\documentclass[12pt,a4paper,notitlepage]{article}
\usepackage[utf8]{inputenc}
\usepackage[english]{babel}
\usepackage[T1]{fontenc}
\usepackage{hyperref}
\usepackage{amsmath}
\usepackage{stmaryrd}
\usepackage{amsfonts}
\usepackage{amssymb}
\usepackage{color} 
\usepackage{caption}
\usepackage{amsmath}
\usepackage{relsize,exscale}
\usepackage{array,multirow,makecell}
\usepackage{graphicx}
\usepackage{multicol}
\usepackage[normalem]{ulem}

\usepackage[toc,page]{appendix}

\setcellgapes{1pt}
\makegapedcells
\newcolumntype{R}[1]{>{\raggedleft\arraybackslash }b{#1}}
\newcolumntype{L}[1]{>{\raggedright\arraybackslash }b{#1}}
\newcolumntype{C}[1]{>{\centering\arraybackslash }b{#1}}
\newcommand{\Tr}{\mathrm{Tr}}

\newtheorem{theorem}{Theorem}
\newtheorem{definition}{Definition}

\newtheorem{proposition}{Proposition}
\newtheorem{remark}{Remark}

\newtheorem{corollary}{Corollary}

\newcommand{\sym}{\mathrm{Sym}}

\newtheorem{lemma}{Lemma}
\usepackage{enumerate}
\usepackage{subfig}
\usepackage{graphicx}

\newcommand{\droit}{\mathrm{d}}

\newcommand{\beq}{\begin{equation}}
\newcommand{\eeq}{\end{equation}}
\newcommand{\bea}{\begin{eqnarray}}
\newcommand{\eea}{\end{eqnarray}}
\definecolor{mygray}{gray}{0.3}

\newcommand{\bes}{\begin{eqnarray}}
\newcommand{\ees}{\end{eqnarray}}

\newcommand\restr[2]{{% we make the whole thing an ordinary symbol
  \left.\kern-\nulldelimiterspace % automatically resize the bar with \right
  #1 % the function
  \vphantom{\big|} % pretend it's a little taller at normal size
  \right|_{#2} % this is the delimiter
  }}

\usepackage{multicol}
\usepackage{amssymb}

\makeatletter
\begin{document}

\begin{center}
\textbf{\Large{
Renormalization group flow of coupled tensorial group field theories:}\\
\medskip
{\large Towards the Ising model on random lattices }}\\
\medskip

\vspace{15pt}

{\large Vincent Lahoche$^a$\footnote{vincent.lahoche@cea.fr}, \,
Dine Ousmane Samary$^{a,b}$\footnote{dine.ousmanesamary@cipma.uac.bj}\, 
and \, Antonio D. Pereira$^{c}$\footnote{adpjunior@id.uff.br}
 }
\vspace{15pt}

a)\,  Commissariat à l'\'Energie Atomique (CEA, LIST),
 8 Avenue de la Vauve, 91120 Palaiseau, France

b)\, Facult\'e des Sciences et Techniques (ICMPA-UNESCO Chair),
Universit\'e d'Abomey-Calavi, 072 BP 50, Benin

c) Instituto de F\'isica, Universidade Federal Fluminense, Campus da Praia Vermelha, Av. Litor\^anea s/n, 24210-346, Niter\'oi, RJ, Brazil
\vspace{0.5cm}
\end{center}
%\vspace{5pt}
\begin{center}
\textbf{Abstract}
\end{center}
We introduce a new family of 
tensorial field theories by coupling 
different fields in a non-trivial way, with a view towards the investigation of the coupling between matter and gravity in the quantum regime. As a first step, we consider the simple case with two tensors of the same rank coupled together, with Dirac like kinetic kernel. We focus especially on rank-$3$ tensors, which lead to a power counting just-renormalizable model, and interpret Feynman graphs as Ising configurations on random lattices. We investigate the renormalization group flow for this model, using two different and complementary tools for approximations, namely, the effective vertex expansion method and finite-dimensional vertex expansion for the flowing action. Due to the complicated structure of the resulting flow equations, we divided the work into two parts. In this first part we only investigate the fundamental aspects on the construction of the model and the different ways to get tractable renormalization group equations, while their numerical analysis will be addressed in a companion paper. 
%\vspace{10cm}
%\pagebreak
%\tableofcontents

%\newpage
\section{Introduction} \label{sec0}
The construction of a fundamental theory which describes the dynamics of spacetime at the quantum level requires a consistent account for the superposition of (pre-)geometric fluctuations \cite{Hawking:1979ig}-\cite{Oriti:2013jga}. However, the naive perturbative path-integral quantization of General Relativity leads to a non-renormalizable theory and therefore lacks predictivity due to the necessity of infinitely many counterterms to absorb ultraviolet (UV) divergences. Consequently, the underlying quantum theory is valid up to some UV cutoff and hence it is not fundamental, i.e., it is not valid up to arbitrarily large energy scales \cite{tHooft:1974toh}-\cite{Goroff:1985th}. This can be viewed as a hint for the departure from the standard perturbative setting or  which geometric degrees of freedom are effective and should be replaced by more fundamental ones. \\

A promising strategy to properly define the quantum gravity path integral is to discretize it in the same way as for lattice gauge theories and look for a suitable continuum limit in this setting. This perspective, implements the ``integral over geometries" by a sum over triangulations which can be tackled by numerical simulations. Such an approach has evolved to the so-called (causal) dynamical triangulation program and quantum Regge calculus, see, e.g., \cite{{Ambjorn:2012jv}}-\cite{Loll:2019rdj} and several non-trivial results as the existence of an extended de Sitter like phase for the emergent geometry were obtained over the last years. Alternatively, in two dimensions, one can construct a sum over random geometries by the well-known matrix models. They correspond to zero-dimensional statistical models whose interactions are dual to elementary polygon-like building blocks, which tessellate two-dimensional surfaces \cite{DiFrancesco:1993cyw}. Feynman diagrams of such models are dual to discretized surfaces, and the perturbative expansion for the partition function of matrix models can be organized by their genus (the so-called $1/N$-expansion, $N$ being the size of the matrices). The  continuum limit can be investigated at different levels of refinement. At the first stage, called single scaling limit, only planar triangulations, with vanishing genus are taken into account. Going beyond the planar sector, and taking into account superpositions of all topologies then requires the so-called double scaling limit, obtained by setting simultaneously the size of the matrix to infinity and the coupling constant to its critical value \cite{Douglas:1989ve}-\cite{Brezin:1992yc}. In two dimensions,  some powerful techniques allow for an analytical probe of such a limit. See also \cite{Duplantier:2009np}-\cite{Duplantier:2009np2} for related discussions. \\

The successful story of matrix models for the description of two-dimensional quantum space-time suggests that a generalization of such models might be a good strategy to set up a theory of quantum gravity in higher dimensions. This led to the introduction of the so-called tensor models, see, e.g. \cite{Ambjorn:1990ge}-\cite{Gurau:2016cjo} (see also              \cite{Rivasseau:2014ima} and references therein), which statistical models for random rank-$d$ tensors, where $d$ is the would be spacetime dimension. In this picture, tensors are dual to $(d-1)$-simplices while the tensor indices are dual to $(d-2)$-simplices. Interactions are given by contractions of tensors, and the Feynman diagrams of the perturbative expansion provide a simplicial decomposition for $d$-dimensional manifold. Historically, the progress on the search for a suitable continuum limit for these models was hampered by the lack of a $1/N$-expansion (with $N$ being the tensor size) similar to the one in matrix models. The situation has changed after the introduction of the so-called colored tensor models in \cite{Gurau:2009tw}-\cite{Gurau:2013pca}. In such models, the interactions are constructed accordingly to an invariance principle under unitary transformations, acting independently on each index of a complex tensor. Generalization to orthogonal transformations for real tensors was successfully considered recently \cite{Carrozza:2015adg}-\cite{Benedetti:2019ikb}. All those models have the nice property of featuring a $1/N$-expansion, a fact that boosted progress in this field. The existence of a well defined power counting, in particular allows to start a renormalization program, which have been strongly developed since the five last years as a promising way to investigate the continuum limit of discrete quantum gravity models. This success does not concern only the tensor models, but a new class of field theories derived from them and called tensorial field theories (TFTs)  \cite{Carrozza:2013wda}-\cite{Lahoche:2015ola}. These field theories differ from the orthodox tensor models due to the introduction of a kinetic operator. Their connection with quantum gravity is inherited from the dual interpretation of their Feynman graphs. The unitary-invariance prescription is broken by the kinetic action while it is preserved for interactions. Moreover, the non-trivial spectrum of the kinetic action provide a canonical path from UV scales (when no fluctuations are integrated) to IR scales (when all the fluctuations are integrated out). For these models, a solid renormalization program has be done, and rigorous BPHZ theorem have been proved, ensuring just-renormalizability for many of them. Moreover, in the same mathematical direction, constructive expansions and Borel summability theorems have been obtained for some super-renormalizable models see   \cite{Rivasseau:2017xbk}-\cite{Erbin:2019zug} and references therein. On the physical counterpart of the investigations, the non-perturbative renormalization group flow has been constructed within some approximations schemes, showing the existence of non-trivial fixed points \cite{Lahoche:2018oeo}-\cite{Eichhorn:2018phj}.  Enriched with group-field theoretic data, the non-trivial combinatorics of such models give rise to group field theories (GFTs) and such fixed points are thought to identify a phase transition from pre-geometric degrees of freedom to a geometric phase due to a ``condensation" mechanism. See, e.g., \cite{Calcagni:2012vb}-\cite{Gielen:2018fqv}.\\

Despite this success, some open questions remain open. Among them, the role and the nature of matter fields in this framework is outstanding. In this paper, we introduce a formalism mixing two (or more) different tensor fields. This can be motivated from different viewpoints. First of all, the two kinds of fields materialize as two states of spin in the Feynman graphs, turning them as random lattice with Ising spins. Ising model on random lattice is not a novel idea \cite{Bonzom:2011ev}-\cite{Sasakura:2014zwa}; and it has been investigated in the beginning of the colored random tensor age \cite{Bonzom:2011ev}. The  new perspective in the proposed approach is the opportunity to implement the renormalization group flow at the lattice level, through a coarse-graining of the lattice, providing an (expected) complementary point of view to the standard block spin approach. Other motivations arise from the difficult question of understanding the influence of matter fields on the quantum pre-geometric spacetime. Indeed, coupling tensors fields of different sizes and ranks may be viewed as a simple way to branch pseudo-manifold discretizations of different dimensions; having their own critical behavior. This may be viewed as a generalization of the multi-critical behavior expected to be closely related to conformal matter field in two dimensional gravity. \\

In this paper, with a view towards the Ising model interpretation, we consider two coupled tensor fields of the same rank; but keeping in mind that the possible generalization and interpretation will be given in forthcoming works. More precisely, we consider a just-renormalizable model, with Dirac-like kinetic kernel, mixing two tensors with different couplings at the interaction level. For this model, we investigate non-perturbative aspects employing two different and complementary ways. First, the focus is  on the leading (melonic) sector, and the use of non-trivial Ward identities arising from the unitary symmetry softly broken by the kinetic term. This method follows the strategy recently developed in a series of paper  \cite{Lahoche:2018oeo}-\cite{Lahoche:2018ggd}, and referred as effective vertex expansion (EVE) method, which, in contrast to the vertex expansion which is a finite-dimensional truncations, cut smoothly in the full theory space, and keep complete infinite dimensional sectors rather than a finite dimensional subspace. Later, we consider finite-dimensional truncations, to investigate regions where EVE remains difficult to be used; especially for disconnected interactions and interactions beyond the melonic sector \cite{Lahoche:2019cxt}-\cite{Lahoche:2019ocf}. Note that this limitation for melonic diagrams does not invalidate the EVE method, which can be extended beyond  melonic sector for the theories from which the Feynman diagrams have the  tree structure representation  \cite{Lahoche:2018oeo}. \\

The outline of this paper is the following. In section \ref{sec1}, we introduce the formalism allowing to couple two or more tensor fields in a coherent way with respect to the renormalization group flow. In section \ref{solving}, we present the general strategies used to solve the exact renormalization group flow equation, the effective vertex method and finite-dimensional truncations. The resulting equations are then numerically investigated  in section \ref{secnum}.  Discussions and conclusion are given in the same section. We provide  three appendices \eqref{appA}, \eqref{appB} and \eqref{appC} on which the power counting theorem, the proof of important propositions and the computation of the sums which are used throughout the paper are given respectively.\\

\section{Preliminaries} \label{sec1}

\subsection{The model}\label{subsec1}
\noindent
In this section we introduce and motivate a new class of models that we will study throughout this work, by mixing different types of tensorial group field theories, that we call  \textit{motley tensorial group field theories} (MTGFT). From this let us briefly give the following definition 
\begin{definition}
Let $\{(\phi_{a(i)}, \,\bar\phi_{a(i)}),\,\, \,i=1,2,\cdots,n\}$ be a set of complex field. We call MTGFT any tensorial group field theory describing two or more different complex tensorial fields interacting together. 
\end{definition}
In this paper we restrict our attention to the case where only \textit{two} complex tensorial fields of the same rank interact together. Let us denote by  $\phi_V$ and $\phi_W$  these two complex fields ($a(1)=V,\,\, a(2)=W$):
\begin{equation}
\phi_V,\phi_W:\mathrm{G}^d \to \mathbb{C}\,,
\end{equation}
where $\mathrm{G}$ denotes an arbitrary compact Lie group and $d$ the number of copies on which the tensors $V$ and $W$ are defined. For our discussion, $d$  will be arbitrary, however $\mathrm{G}$ is fixed to be an Abelian compact Lie group $U(1)$. We denote by $V_{p_1,\cdots,p_d}$ and $W_{p_1,\cdots,p_d}$, or simply $V_{\vec{p}}$ and $W_{\vec{p}}$, $\vec{p}\in\mathbb{Z}^\droit$ the Fourier components of the fields $\phi_V$ and $\phi_W$ such that the vector  $\Phi:=(\phi_V,\phi_W)$  represents the doublet of the fields. \\

\noindent
The fields being defined, the dynamics is given by  the classical action $S[\Phi,\bar{\Phi}]$  which can be represented in standard “rule" for TGFTs as a sum of \textit{tensorial invariants} or \textit{generalized traces}. When we have only a single (complex) tensor field, the trace invariant are obtained as the contractions between the indices of an equal number of the fields  $T$ and $\bar{T}$; such that, for any $T$ and for $i\in\llbracket 1,d\,\rrbracket$\footnote{The notation $\llbracket a,b\,\rrbracket$, used in the rest of the paper refer to the integer interval between $a,b\in\mathbb{N}$.}, the index $p_i$ of $T$ have to be contracted with the index $\bar{p}_i$ of one of the involved $\bar{T}$. As a example a quartic interaction $I_2[T,\bar{T}]$ is:
\begin{equation}\label{preint}
I_2[T,\bar{T}]=\sum_{\vec{p},\vec{p}{\,^\prime}\in\mathbb{Z}^3} \,T_{p_1,p_2,p_3}\bar{T}_{p_1^\prime,p_2,p_3}T_{p_1^\prime,p_2^\prime,p_3^\prime}\bar{T}_{p_1,p_2^\prime,p_3^\prime}.
\end{equation}
Note that the  construction of tensorial invariants quantities are related to a  proper unitary invariance. Restricting the size of the tensors fields to $N$, the  previous interaction \eqref{preint} is invariant under independent unitary transformations $U_1(N)U_2(N)U_3(N)\in\mathbb{U}^{\otimes 3}(N)$, where  the $U_i(N)$ being $N\times N$ unitary matrices. In the same way, for arbitrary dimension $d$, the tensors interactions are invariant under $\mathbb{U}^{\otimes \droit}(N)$. Note that in this paper we will essentially interested to the inductive limit $\mathbb{U}(\infty)$, and like in standard quantum field theory, we assume that, acting on square summables sequences space, the operator $U-\mathbb{I}$ remains of trace or Hilbert-Schmidt class $\forall\,U\in \mathbb{U}(\infty)$\cite{sym} . A general expression of the interaction, involving $2n$-fields,  (where $n$ is a number of field of type $\bar{T}$  coupled with  $n$  number of field of type $T$) may be indexed by a bipartite $d$-colored regular graph $\mathfrak{b}_n$, and denoted as $I_{\mathfrak{b}_n}[T,\bar{T}]$, so that the general interacting part of the action  is written  as:
\begin{equation}
S_{\text{int}}[T,\bar{T}]=\sum_{n>1} \sum_{\mathfrak{b}_n} \,g_{\mathfrak{b}_n} I_{\mathfrak{b}_n}[T,\bar{T}]\,,
\end{equation}
where the relative weights $g_{\mathfrak{b}_n}$ corresponds to the couplings constant. We restrict our attention  to the case of connected graphs $\mathfrak{b}_n$, which are called \textit{bubbles}. We recall that these bubbles are said to be \textit{local} for tensor field theories, the locality principle being inherited from the proper unitary invariance \cite{Carrozza:2012uv}: Now let us given the following definition allowing to point out the connection between bubbles and locality properties.
\begin{definition}\label{locality}
Any connected bubble is said to be local. In the same way, any function expanding as a sum of connected bubbles is said to be local. 
\end{definition}
This property, becoming a standard in the TGFT literature arise from renormalization investigations, and in particular allows to define local counter-terms in the usual sense in quantum field theory. More informations  about this notion can be found in \cite{Carrozza:2013wda}-\cite{Lahoche:2015ola}.  Let's try to quickly give an appropriate method to build a field theory with different fields. For instance let us consider two independent fields denoted by $V$ and $W$.
The rule to build an interacting doublet is not unique, but we will discuss in the end of this section some issues concerning this non-uniqueness. By considering a set of coupling 
constant $\{c_\alpha\}$ for $\alpha$ running from $1$ to $R$, the first step is to define the interacting part of the functional action $S_{\text{int}}[\Phi,\bar{\Phi}]$ such that for  a initial condition $c_\alpha=0,\,\,\forall\alpha$ leads to the following expression:
\begin{equation}
S_{\text{int}}[\Phi,\bar{\Phi}]=S_V[\phi_V,\bar{\phi}_V]+S_W[\phi_W,\bar{\phi}_W]\,.\label{limit}
\end{equation}
In the above relation, the quantities  $S_V$ and $S_W$ are assumed to be built as a sum of connected generalized traces. The index $R$  is  the \textit{degree} of  the mixing tensor i.e.  the number of coupling constants.  From a group symmetry point of view, the action in this  limit  \eqref{limit} is invariant under the  transformations of the form:
\begin{equation}
\prod_{i=1}^\droit\begin{pmatrix} 
U_i^{(V)}(N) & 0 \\
0 & U_i^{(W)}(N)
\end{pmatrix} \in \mathbb{U}^{(V)\,\otimes \droit}(N)\times \mathbb{U}^{(W)\,\otimes \droit}(N)\,,
\end{equation}
where we have to understand that the transformations $U_i^{(V)}(N)$ and $U_i^{(W)}(N)$ act independently on the two components of $\Phi$:
\begin{equation}
\begin{pmatrix} 
\phi_V^\prime  \\
\phi_W^\prime
\end{pmatrix}=
\prod_{i=1}^\droit\begin{pmatrix} 
U_i^{(V)}(N) & 0 \\
0 & U_i^{(W)}(N)
\end{pmatrix} 
\begin{pmatrix} 
\phi_V  \\
\phi_W
\end{pmatrix}=
\begin{pmatrix} 
\prod_iU_i^{(V)}(N) \phi_V  \\
\prod_iU_i^{(W)}(N)\phi_W
\end{pmatrix}\,.
\end{equation}
A first way to build the coupling between these two tensor fields may be given by implementing an  explicit symmetry reduction given by:
\begin{equation}
\mathbb{U}^{(V)\,\otimes \droit}(N)\times \mathbb{U}^{(W)\,\otimes \droit}(N) \to \mathbb{U}^{\otimes \droit}(N)\,,\label{symred}
\end{equation}
simply meaning that we wish to consider mixed interaction with tensorial invariance. Furthermore, we assume that the transformation $U_i(N)$ act together in the same manner  on the two components $\phi_V$ and $\phi_W$. Note that with this definition the limit \eqref{limit} is well defined, and the corresponding interactions are obtained from the bubbles $I_{\mathfrak{b}_n}$ defined  in $S_W$ (or in $S_V$) - up to the substitution of at least one among the number $n$ of the field  $W$ or $\bar{W}$. In order to include this new information in the construction of the interacting two fields, we introduce the integers $m_+$ and $m_-$, such that $m_+,m_-<n$, and the notation $\mathfrak{b}_n^{(m_+,m_-)}$; with the convention that $m_+$ denotes  the number of field $V$ f and $m_-$  the number of  field $\bar{V}$.  Therefore, the \textit{mixing interaction}  denote by $S_{VW}$ will be $S_{VW}:=S_{\text{int}}-S_V-S_W$ and may be given explicitly as follows:
\begin{equation}
S_{VW}=\sum_{n}\sum_{m_++m_-\geq 1}^{n-1} \,c_{\mathfrak{b}_n^{(m_+,m_-)}} I_{\mathfrak{b}_n^{(m_+,m_-)}}[V,\bar{V};W,\bar{W}]\,.\label{def1vW}
\end{equation}
We may adopt a graphical representation  to picture these tensorial invariant as follow. We associate black and white \textit{bulls-nodes} for tensors $V$ and $\bar{V}$ respectively, and black and white \textit{square-nodes} for tensors fields $W$ and $\bar{W}$ respectively. To each nodes  are hooked $d$ colored half edges which are associated to the field variables, and the interaction bubble is built by hooking the edges from black to white nodes following their respective colors.  For instance, if we restrict our attention on the purely quartic case for $d=3$ we have:
\begin{equation}
S_W=\sum_{i=1}^3 g_{W,i}\vcenter{\hbox{\includegraphics[scale=1.1]{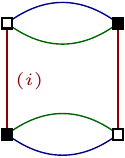} }}\,,\qquad S_V=\sum_{i=1}^3 g_{V,i}\vcenter{\hbox{\includegraphics[scale=1.1]{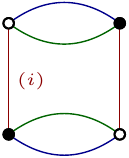} }}\,,
\end{equation}
and :
\begin{align}
\nonumber S_{VW}=\sum_i\Bigg\{c_{i1}\vcenter{\hbox{\includegraphics[scale=1.1]{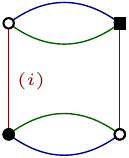} }}+&c_{i2}\vcenter{\hbox{\includegraphics[scale=1.1]{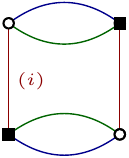} }}+c_{i3}\vcenter{\hbox{\includegraphics[scale=1.1]{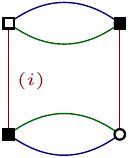} }}+c_{i4}\vcenter{\hbox{\includegraphics[scale=1.1]{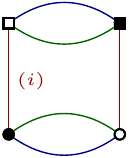} }}\\&+c_{i5}\vcenter{\hbox{\includegraphics[scale=1.1]{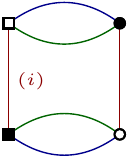} }}+c_{i6}\vcenter{\hbox{\includegraphics[scale=1.1]{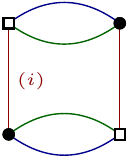} }}+c_{i7}\vcenter{\hbox{\includegraphics[scale=1.1]{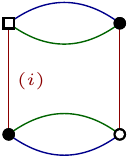} }}\Bigg\}\,,
\end{align}
 the index $i$ on the figure referring to the color of the lonely edge. This construction of the interacting fields  is called \textit{minimal coupling}. Note that  this definition is heuristic  due to the fact that it is not "minimal" with respect to the value of $R$ and on the other hand  It is minimal with respect to the number of additional inputs (the symmetry construction given  below).  Note that the minimal coupling lost \textit{à priori} the reality of the classical action. To recovered it, we impose the invariance with respect to global $U(1)$ symmetry as:
\begin{equation}
\phi_V\to\phi_V^\prime = e^{i\theta_V} \phi_V\,,\qquad \phi_W\to\phi_W^\prime = e^{i\theta_W} \phi_W\,,\qquad \theta_V,\theta_W\in\mathbb{R}\,,
\end{equation}
We denote by ``\textit{reality condition}'' the above relation and obviously we come to
\begin{equation}
I^\prime_{b_n^{(m_+,m_-)}}=e^{i\theta_V(m_+-m_-)-i\theta_W(m_+-m_-)}I_{b_n^{(m_+,m_-)}}.
\end{equation}
Has a result, the reality condition imposes $m_+=m_-=m$, and equation \eqref{def1vW} becomes:
\begin{equation}
S_{VW}=\sum_{n}\sum_{m\geq 1}^{n-1} \,c_{b_n^{(m,m)}} I_{b_n^{(m,m)}}[V,\bar{V};W,\bar{W}]\,,\label{def2vW}
\end{equation}
corresponding to the \textit{real minimal coupling}. For instance by considering  the rank $3$ quartic melonic model, this implies  that $c_{i1}=c_{i2}=c_{i3}=c_{i6}=c_{i7}= 0$ and  therefore $S_{VW}$ becomes 
\begin{equation}
S_{VW}=\sum_i\Bigg\{ c_{i4}\vcenter{\hbox{\includegraphics[scale=1.1]{vert6.pdf} }}+c_{i5}\vcenter{\hbox{\includegraphics[scale=1.1]{vert9.pdf} }}\Bigg\}\,. \label{quarticplus}
\end{equation}
Another important question which need to be addressed at this step is about the  positivity of $S_{VW}$ given in \eqref{quarticplus}. For this let us   fix the index  $i$ to $1$ and provide the following definition:
\begin{equation}
M(n,k):= \sum_{p_2,p_3} V_{n,p_1,p_2}\bar{W}_{k,p_1,p_2}\,,\qquad N(p_1,p_2;p_1^\prime,p_2^\prime):=\sum_{n} V_{n,p_1,p_2}\bar{W}_{n,p_1^\prime,p_2^\prime}\,,
\end{equation}
such that we have:
\begin{equation}
\vcenter{\hbox{\includegraphics[scale=1.1]{vert6.pdf} }}= \sum_{p_i,p_i^\prime}\vert N(p_1,p_2;p_1^\prime,p_2^\prime)\vert^2\geq 0  \,,\qquad \vcenter{\hbox{\includegraphics[scale=1.1]{vert9.pdf} }}=\sum_{k,n} \vert M(n,k)\vert^2\geq 0\,.
\end{equation}
Obviously the construction of the interacting terms allow to recover this positivity  may be given carefully.   From now let us consider the  kinetic part of the action, which have not been discussed above. From our hypothesis given by equation \eqref{limit}, we expect that:
\begin{equation}
S_{\text{kin}}[\Phi,\bar{\Phi}]=S_{\text{kin},V}[V,\bar{V}]+S_{\text{kin},W}[W,\bar{W}]\,,\quad \mu_i=0\,,\forall i
\end{equation}
where we called $\{\mu_i\}$ the set of couplings i.e., the set of weight which involve by mixing $V$ and $W$. Without loss of generality, the kinetic part may be written as:
\begin{equation}
S_{\text{kin}}[\Phi,\bar{\Phi}]= \sum_{\vec{p},\vec{q}\in\mathbb{Z}^d} \bar{\Phi}^T(\vec{p}\,) K(\vec{p},\vec{q}\,) \Phi(\vec{q}\,)\,,
\end{equation}
where $K(\vec{p},\vec{q}\,)$ is a $2\times 2$ matrix whose coefficients depends on $\vec{p}$ and $\vec{q}$. The diagonal part of the matrix correspond to the kinetic kernels of $S_{\text{kin},V}$ and $S_{\text{kin},W}$. For the rest, we assume that $K$ is diagonal i.e. 
\begin{equation}
K(\vec{p},\vec{q}\,)=\mathcal{K}(\vec{p}\,)\delta_{\vec{p},\vec{q}}\,.
\end{equation}
Let us denote as $\mathcal{K}_{ab}$ ($a,b=V,W$) the matrix elements of $\mathcal{K}$. Note that the bubble interactions are not necessarily the same for $V$ and $W$, so that $\mathcal{K}_{VV}$ and $\mathcal{K}_{WW}$ may be different (the families of bubbles could be different in $S_W$ and $S_V$). If we assume the symmetry $\phi_W \leftrightarrow \phi_V$ holds, we have to impose the condition:
\begin{equation}
\mathcal{K}_{VV}=\mathcal{K}_{WW}\,.
\end{equation}
For an example, when $\mathcal{K}$ does not depends on $\vec{p}$, we may have:
\begin{equation}
\mathcal{K}=\begin{pmatrix} 
1 & -\mu \\
-\mu & 1 
\end{pmatrix}\,, \to \mathcal{K}^{-1}=\frac{1}{1-\mu^2}\begin{pmatrix} 
1 & \mu \\
\mu & 1 
\end{pmatrix}\,.
\end{equation}
With this choice, there are no canonical scale on the Feynman graph. In this paper we are interested to the  theories having an autonomous and operational definition of scales. To this end, we consider only momentum-dependent propagators. Other definitions are given  by choosing the Kernel as  the following:
\begin{equation}
\mathcal{K}(\vec{p}\,)=(\vert \vec{p}\,\vert+m)\begin{pmatrix} 
1 & -\mu \\
-\mu & 1 
\end{pmatrix}\,,\qquad \mathcal{K}(\vec{p}\,)=\begin{pmatrix} 
\vert \vec{p}\,\vert+m & -\mu \\
-\mu & \vert \vec{p}\,\vert+m\,, 
\end{pmatrix}
\end{equation}
both having the same limit for $\mu\to 0$, and $\vert \vec{p}\,\vert := \sum_{i=1}^3 \vert p_i\vert$. This choice in particular ensures the \textit{just renormalizability} of the model, as explained in Appendix \ref{appA}.  In general, the presence of the parameter  $\mu$ leads to a  non-zero, Wick contractions between $V$ and $\bar{W}$,  and then this parameter gives their relative weight with respect to $V\bar{V}$ and $W\bar{W}$ contractions. Note that with such terms, we can generate non-real effective couplings. We then expect that the real coupling may be stable only if $\mu=0$. As an example, for the rank $3$ quartic model, we get:
\begin{align}
\nonumber S[\Phi,\bar{\Phi}]=\sum_{\vec{p}\in\mathbb{Z}^3}  (\bar{V}_{\vec{p}},\bar{W}_{\vec{p}})&
\begin{pmatrix} 
\vert \vec{p}\,\vert+m & 0 \\
0 &\vert \vec{p}\,\vert+m 
\end{pmatrix}\begin{pmatrix} 
{V}_{\vec{p} } \\
{W}_{\vec{p}}
\end{pmatrix} +\sum_{i=1}^3 g_{W,i}\vcenter{\hbox{\includegraphics[scale=1.1]{vert2.pdf} }}\\
&+\sum_{i=1}^3 g_{V,i}\vcenter{\hbox{\includegraphics[scale=1.1]{vert1.pdf} }}+ \sum_i\Bigg\{ c_{i4}\vcenter{\hbox{\includegraphics[scale=1.1]{vert6.pdf} }}+c_{i5}\vcenter{\hbox{\includegraphics[scale=1.1]{vert9.pdf} }}\Bigg\}\,.\label{final0}
\end{align}
It can be easily cheeked that the condition $\mu=0$ is stable with respect to the renormalization group flow (in the deep UV sector). At this stage, let us  discuss the existence of another \textit{à priori} stable model with $\mu\neq 0$. For that, by setting  $c_{i5}=0$, it is easy to see that  see Proposition \ref{closed}, and the proof given in Appendix \ref{appB}:
\begin{enumerate}
\item Even if $\mu\neq 0$, the form of the effective action (in the UV)  remains stable, and $c_{i5}$ remains equal to zero. 

\item In the same limit $\mu$ is a non-dynamical parameter, with mass dimension $2$ and flow equation related to this parameter is therefore:
\begin{equation}
\frac{d}{dt}\bar{\mu}=-2\bar{\mu}\,,
\end{equation}
\end{enumerate}
the parameter $t$ designated the standard $\log$-scale parameter along the flow. Note that, the fact that $\mu$ becomes "non-dynamical" might seem strange for a background independent theory. One expect that $\mu$ can be fixed for another way. For instance, it may be corresponds to the non-zero vacuum for an interacting discrete matter model, building on the same way from an additional vector field (which are tensors of order $1$). \\

\noindent
In this paper we focus on non-perturbative renormalization group aspects for the model defined by the classical equation \eqref{final}. More precisely, we choose a slightly  simplified version of the model, where coupling constant are independent of the color index $i$, leading to a classical model without preferred color:
\begin{align}
\nonumber S[\Phi,\bar{\Phi}]=\sum_{\vec{p}\in\mathbb{Z}^3}  (\bar{V}_{\vec{p}},\bar{W}_{\vec{p}})&
\begin{pmatrix} 
\vert \vec{p}\,\vert+m & 0 \\
0 &\vert \vec{p}\,\vert+m 
\end{pmatrix}\begin{pmatrix} 
{V}_{\vec{p} } \\
{W}_{\vec{p}}
\end{pmatrix} +g_{1}\sum_{i=1}^3 \vcenter{\hbox{\includegraphics[scale=1.1]{vert2.pdf} }}\\
&+g_{2}\sum_{i=1}^3 \vcenter{\hbox{\includegraphics[scale=1.1]{vert1.pdf} }}+ \sum_i\Bigg\{ c_{1}\vcenter{\hbox{\includegraphics[scale=1.1]{vert6.pdf} }}+c_{2}\vcenter{\hbox{\includegraphics[scale=1.1]{vert9.pdf} }}\Bigg\}\,.\label{final}
\end{align}
We have essentially two motivations to consider such a kind of theory and to develop nonperturbative renormalization group techniques. The first one arises from the common area of research for TGFT i.e. a pioneering formulation of random geometry throughout the genesis of our classical space-time near the Big Bang theory. Indeed, the MTGFT formalism allows to mix several phases or several critical behaviors, and to add a non-trivial complexity into the primitive or pre-geometric quantum space-time.  The second motivation arises from the similarity between such a model mixing to kind of tensor fields, and the standard Ising model. Indeed, the two fields $V$ and $W$ may be freely interpreted as incarnations of the up and down Ising spins, the interactions and kinetic kernel providing competition between them, as the spin-spin interaction in the sing model. This interpretation, considered initially in \cite{Bonzom:2011ev} may allow considering Ising spin on the random lattice; the interest of a renormalization group analysis, therefore, could be to provide an alternative way to renormalize, based on the lattice itself rather than on the spins as in standard approaches. Note that in this point of view our choices for the model have to be questioned. Indeed, setting $\mu=0$; there is no coupling between up and down spins, except at the interaction level, with $c_1$ and $c_2$ couplings. The corresponding Ising model, therefore, appears to be quite non-conventional. However, we may adopt another point to view to stress the link with the Ising model on random lattice using the intermediate field representation, recalled in Appendix \ref{appA}. To put in a nutshell, an considering a vacuum diagram, the intermediate field representation associate colored edges to vertices and vertices (called loop vertices) to the ordinary loops (see Figure \ref{FigApp1}). An elementary statement is that leading order (melonic) graphs are trees in this representation (see Appendix \ref{appA}, and equation \eqref{treeexample1} for an example). Therefore, setting $c_2=0$ (as we will see in the rest of this paper, this coupling play a marginal role with respect to the other ones), the leading order trees are built with loop vertices of two kinds: the ones with only square nodes, and the ones with bullet nodes. Note that we cannot have loop vertices mixing square and bullet nodes because of $\mu=0$. Then, interpreting the first kind of nodes as down spins, and the other one as up spins, we get that the expansion of the leading order free energy corresponds to Ising spin on random trees. Moreover, the next to leading order diagrams, including loops over trees, may be interpreted as self and long-range interactions. This interpretation may be clearer on the interaction described in remark \ref{remark3} on Appendix \ref{appA}; the coupling $\lambda$ between type $V$ and type $W$ intermediate fields being viewed as the incarnation of the relative Ising weight, and introduce competition between up and down spins. Figure \ref{figureIsing} provides an illustration of the correspondence stressed in this paragraph. \\

\begin{center}
\includegraphics[scale=1]{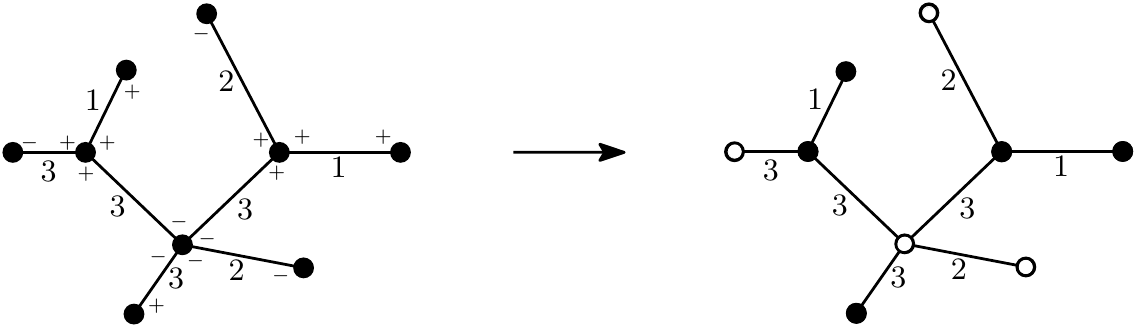}
\captionof{figure}{An illustration of the correspondence with the Ising model on random lattices. On left, a tree in the intermediate field representation, the $+$ and $-$ indicating the nature of the boundary of the colored edges. $+$ for boundaries with squares nodes and $-$ for bullet nodes. On right, the corresponding planar tree with up and down sites, respectively in black and white.}
\end{center}

In the rest of this paper, we do not set $c_2=0$. The reason is that we are essentially interested by the method allowing to extract the RG flow rather than by the specific contacts of the different versions of the model with physics. Then, we remain as general as possible.

\subsection{Non-perturbative renormalization group equation}\label{subsec2}

Among the formulations of the nonperturbative functional renormalization group (FRG), the Polchinski and Wetterich equations are the most popular. Except for very special problems the investigation of the  solutions of  these equations are a difficult challenge which requires approximation. Even if Polchinski and Wetterich equations are essentially two different presentations for the same problem, the Wtterich equation is  showed  to be much more suitable for approximations, and remains the most used to approach the nonperturbative renormalization group. We will focus on the Wetterich approach in this paper, and consider two differents and complementary approximations schemes to solve it, the \textit{vertex expansion} and the \textit{effective vertex expansion} (EVE) method, both associated to a derivative expansion. \\

The model we consider in this paper is described by the action  \eqref{final} on which the dynamic or the evolution of the correlation function is governed by the Gaussian measure $d\mu_C$  from which the propagator is given explicitly by the following definition:
\bea
\int\, d\mu_C\,V_{\vec{p}}\,\bar{V}_{\vec{p}\,^\prime}=\int\, d\mu_C\,W_{\vec{p}}\,\bar{W}_{\vec{p}\,^\prime}=\frac{1}{\vert \vec{p}\,\vert+m}\,\delta_{\vec{p}\,\vec{p}\,^\prime}\,,\quad \int\, d\mu_C\,W_{\vec{p}}\,\bar{V}_{\vec{p}\,^\prime} =0
\eea
The partition function is therefore  defined as
\bea
\mathcal{Z}[J,\bar J]=e^{\mathcal W[J,\bar J]}=\int\, d\mu_{C}\, e^{-S_{\text{int}}+\langle J_W,\bar{W}\rangle+\langle W,\bar{J}_W\rangle+\langle J_V,\bar{V}\rangle+\langle V,\bar{J}_V\rangle}\label{part1}
\eea
where $S_{\text{int}}$ denotes the quartic part of the action \eqref{final}, and: $\langle A,B\rangle:=\sum_{\vec{p}} A_{\vec{p}}\,B_{\vec{p}}$, $J=(J_V,J_W)$ and $\bar J=(\bar J_V,\bar J_W)$ are the external sources and $\mathcal W[J,\bar J]:=\ln\mathcal Z[J,\bar J]$ is the standard free energy. The partition function \eqref{part1} is said to be ``\textit{ \it referent}''. This expression means that  it includes no more ingredients than the quantization principle following with the amplitude of the quantum process\footnote{Strictly speaking, the terminology "quantum" is not appropriate, the amplitude of a real quantum process could be of the form $e^{iS}$. We should speak about "statistical" or "thermal fluctuating" model. We conserve the terminology "quantum" to grant us the conventions of literature.} and  is given from the classical action as $e^{-S}$ up to a normalization factor. Renormalization group provides the partial integration of UV degree of freedom appears to be  a powerful ingredient  in the study of a quantized field theory. To this end, the scale slicing along the path from UV to IR must be considered as a fundamental ingredient to initialize the theory, especially for the cases where we are more interested for the "universality class" rather than  rigorous definition in the UV limit. By "scale slicing", we refer to the way from which the fluctuations are classified as "UV" or "IR" after partial integration over UV modes. In this paper, we   use the  \textit{regulator function} $r_k$ to drive the change of scale e.g. to drag the theory from UV to IR. We state that our quantized model is quite represented scale by scale as a continuous set of models defined with the $k\geq 0$ dependent partition function $\{\mathcal{Z}_k[J,\bar J]\}$,  defined as:
\begin{equation}
\mathcal{Z}_k[J,\bar J]:=\int\, d\mu_{C_k}\, e^{-S_{\text{int}}+\langle J_W,\bar{W}\rangle+\langle W,\bar{J}_W\rangle+\langle J_V,\bar{V}\rangle+\langle V,\bar{J}_V\rangle}\label{part2}\,,
\end{equation}
where $d\mu_{C_k}$ is the Gaussian measure  given with the modified propagator:
\begin{equation}
C_k^{-1}(\vert \vec{p}\,\vert):=(\vert \vec{p}\,\vert+m) \begin{pmatrix}
1&0\\
0&1
\end{pmatrix}+r_k(\vert \vec{p}\,\vert)\begin{pmatrix}
1&0\\
0&1
\end{pmatrix}\,.\label{propnew}
\end{equation}
The index $k$ is the running scale and goes from $\ln \Lambda$ in the deep UV to $0$ in the deep IR, for some fundamental scale $\Lambda$. Among the standard properties of the regulator, we recall the following:
\begin{itemize}
\item $r_k(\vert \vec{p}\,\vert)\geq 0$
\item $r_{k\to 0}(\vert \vec{p}\,\vert) \to 0\quad \forall\,\vert \vec{p}\,\vert\in \mathbb{R}^+$
\item $r_{k\to\infty}(\vert \vec{p}\,\vert)\to+\infty$
\item $r_k(\vert \vec{p}\,\vert\to\infty)\ll 1$\,.
\end{itemize}
From this quantization point of view, we will consider as physically relevant all the conclusions which are independent of the choice of the regulator function. The RG flow in this point of view corresponds to a mapping from $\mathcal{Z}_{k}$ to $\mathcal{Z}_{k+\delta k}$ and  leading to  the following first order equation 
\begin{equation}
\frac{\partial}{\partial k}\Gamma_k=\sum_{\vec{p}} \Tr \left[\frac{\partial r_k}{\partial k}(\vec{p}\,^2)\left(\Gamma^{(2)}_k+r_k\right)^{-1}_{\vec{p},\vec{p}}\right]\,.\label{Wett}
\end{equation}
This equation may be understand as the dynamic respect to the scale of  the average \textit{effective action} $\Gamma_k$ i.e. the modification of $\Gamma_k$ in the range of scale $[k,k+dk]$. Note that the trace $\Tr$  is defined  over the internal indices of the fields  $V$ and $W$. Due to the fact that the Wetterich equation  does not required that couplings remain small, this equation is called  \textit{non perturbative renormalization group} (NPRG) equation. \\

\noindent
We recall that the average effective action is defined as slightly modified Legendre transform of the free energy $\mathcal{W}_k:=\ln\mathcal{Z}_k$ :
\begin{equation}
\Gamma_k[M,\bar{M}]+r_k[M,\bar{M}]=\langle \bar{J}_W,M_W\rangle+\langle \bar{M}_W,J_W\rangle+\langle \bar{J}_V,M_V\rangle+\langle \bar{M}_V,J_V\rangle-\mathcal{W}_k[J,\bar{J}]\,,
\end{equation}
where $M$ (resp. $\bar{M}$) denote the classical field:
\begin{equation}
M_W:= \frac{\partial \mathcal{W}_k}{\partial \bar{J}_W}\,,\quad M_V:= \frac{\partial \mathcal{W}_k}{\partial \bar{J}_V}\,.\label{classical}
\end{equation}
Finally, $\Gamma^{(2)}_k$ denotes the second derivative respect to these classical fields  of the average effective action :
\begin{equation}
\Gamma^{(2)}_k:= \begin{pmatrix}
\Gamma^{(2)}_{k,VV}&\Gamma^{(2)}_{k,VW}\\
\Gamma^{(2)}_{k,WV}&\Gamma^{(2)}_{k,WW}
\end{pmatrix}
\end{equation}
with:
\begin{equation}
\Gamma^{(2)}_{k,IJ}:=\frac{\partial^2 \Gamma_k}{\partial M_I\partial \bar{M}_J}\,,\qquad I,J\in (V,W)\,.
\end{equation}

\noindent
To complete the definition of  our description of  the functional nonperturbative renormalization group formalism, we have to precise the choice of the regulator function $r_k$. The exact solutions of the formal equation \eqref{Wett} does not depend on this choice. However, the approximations used to solve this equation  
introduces a more or less strong dependency on the choice of the regulator. In this paper, we consider only with the (modified) Litim regulator, which has been show to be  optimal in several case and allows to make the computations analytically :
\begin{equation}
r_k(\vec{p}\,):= \begin{pmatrix}
Z_{VV}(k)&0\\
0& Z_{WW}(k)
\end{pmatrix}
(k-\vert \vec{p}\,\vert)\theta(k-\vert \vec{p}\,\vert)\,,
\end{equation} \label{regulator}
where $\theta$ denotes the Heaviside step function and the $Z_{II}$ are the \textit{running field strength renormalization} (see definition \ref{defren} below). Note that the diagonal form of the regulator ensures that we do not generate \textit{nonperturbative anomalies} of the perturbatively stable condition $\mu=0$ (see proposition \ref{closed} or equation \eqref{IJ}).\\

\noindent
Finally, the dimension may be chosen such that the model been just-renormalizable i.e. all the divergences can be subtracted order by order with a finite set of counter-terms. The proof of this important property is given in appendix \ref{appA}, and could be completed with standard references \cite{Carrozza:2013wda}-\cite{BenGeloun:2011rc}. For our model, the  counter-terms related to the mass and quartic couplings, will be denoted by  "bare" notation  following the standard terminology. In addition, a wave-function counter-term will be require for the two components of the bare propagator. We will denote as $Z_{0,II}$ these counter-terms, the index $0$ referring to the value $k=0$ i.e.  to the deep IR limit,which  corresponds to the referring model \eqref{part1}. As a result, taking into account this renormalization, the propagator \eqref{propnew} becomes:
\begin{equation}
C^{-1}_{k}(\vec{p}\,)=\begin{pmatrix}
Z_{0,VV}\vert \vec{p}\,\vert +m &0\\
0&Z_{0,WW}\vert \vec{p}\,\vert +m
\end{pmatrix}+r_{k}(\vec{p}\,) \,.
\end{equation}
\noindent
The path integral quantization allows to define  the Feynman graphs of the theory. The Feynman graphs usually arise from the perturbative expansion of the quantum field theory. However the definitions  we will use here  holds also for "effective graphs" arising in the vertex expansion of the flow equation \ref{Wett}.
Basically, a Feynman graph is a set of vertices, liking together with edges corresponding to Wick contractions. For tensorial field theory however, this definition have to be completed by the concept of \textit{faces}:
\begin{definition}
A face is defined as a maximal and bicolored connected subset of lines, necessarily
including the color 0 which is attributed to the Wick contractions. We distinguish two cases:\\

\noindent
$\bullet$ The closed or internal faces, when the bicolored connected set correspond to a cycle.\\

\noindent
$\bullet$ The open or external faces when the bicolored connected set does not close as a cycle.\\
\end{definition}

\noindent
Usually, we attribute the color $0$ for the Wick contraction, and picture them as dotted edges. The boundary of a given face is then the subset of its $0$-colored edges, and its length is defined as the number of internal dotted edges on its boundary. Note that for external faces, the external edges are not included into the boundary set. Moreover, we call color of the face $f$ denotes as $c(f)$ the color of the edges building  with the corresponding cycle together with the $0$-colored edges. 
To complete this definition let us provide the following:
\begin{definition}
On a given Feynman graph, the set of edges split into internal and external edges. External edges come from Wick contraction with external fields and  internal edges comes from Wick contractions between vertex fields. Moreover, a vertex is said to be a boundary vertex if at least one of the external edges is hooked to him, or an interior vertex otherwise. Finally, we define the interior of a Feynman diagram as the set of internal vertices and dotted edges. \label{definitionTomTom}
\end{definition}
Now let us define other important notion: \textit{ the boundary graphs}:
\begin{definition}
Let $\mathcal{G}$ be a connected Feynman graph amputated of its external edges. Let $\partial n$ the set of external nodes (i.e. nodes hooked to external edges), $F^\prime=\{f^\prime\}$ the set of external faces and $\partial f^\prime$ the boundary of $f^\prime$. The boundary graph $\partial \mathcal{G}$ of $\mathcal{G}$ is the $d$-colored bipartite regular graph build as follows:
\begin{enumerate}
\item set $(n,\bar{n})\in \partial n$ two boundary nodes respectively black and white. 
\item set $\partial_{n\bar{n}} F^\prime\subset \partial F^\prime$ the subset of boundaries having the couple $(n,\bar{n})$ as boundaries. 
\item For each path $\partial f^\prime\in \partial_{n\bar{n}} F^\prime$ we create an edge of color $c(f^\prime)$ between $n$ and $\bar{n}$.
\item and so one for each pair $(n,\bar{n})\in \partial n$ .
\end{enumerate}
\end{definition}
Note that the boundary graph is not necessarily connected, and corresponds generally to a set of bubble. The number of bubble corresponds to the number of (external) face-connected components of the graph $\mathcal{G}$. 

\section{Solving the NPRG equation}\label{solving}
As explained  above, the NPRG equation \eqref{Wett} cannot be exactly solved appart for  a  very special cases. Approximations are then necessary to track nonperturbative aspects of the renormalization group flow. We focus on two approaches in this section, the standard vertex expansion method and the EVE method.  Let us briefly recall the mean ingredients of this two point of views.

 \begin{itemize}

\item \textit{The standard vertex expansion} consist in a systematic "crude" projection into a reduced phase space of finite dimension, so that a big information on the observable, beyond this reduced phase space, is lost. Especially, for a truncation around interactions of valence $n$, $\Gamma^{n+1} \approx 0$. The vertex expansion method is a powerful formalism to deal with non-local interactions, especially for tensor field theory, as shown by the number of papers on the subject \cite{Carrozza:2014rba}-\cite{Eichhorn:2017xhy}. In particular, the formalism offers great flexibility in the type of interactions  which are considered. Moreover, this truncation scheme seems to be more appropriate to investigate intermediate regimes between deep UV and deep IR.

\item \textit{The EVE method}, in contrast cut "smoothly" into the full phase space, and select "sectors" (that is, infinite sets of observables) rather than a finite dimensional subset of interactions. The results about fixed points are similar with vertex expansion in the melonic sector. However, the methods differ on their philosophy. With the EVE method, the phase space is build of an infinite set of interactions, parametrized with a finite set of couplings, and the full momentum dependence of the effective vertices is took into account. Moreover, some singularities occurring in the vertex expansion method disappear in the EVE, in that sens, EVE extend maximally the domain of investigation of the phase space \cite{Lahoche:2018oeo}. However, the formalism seems to be less flexible than truncations via vertex expansion. In particular, it is difficult to nest "sectors with sectors", as explained in \cite{Lahoche:2018oeo},\cite{Lahoche:2019cxt}. 

 \end{itemize}
 Both methods will be discussed in the same section as complementary ways of investigation.

\subsection{The effective vertex expansion method}
In this section we investigate the EVE method. This point of view to derive the flow equations was introduced recently in \cite{Lahoche:2018oeo} and extensively discussed  in \cite{Lahoche:2019cxt}. Then in this paper  we will focus only on the major modifications due to the specificity of the two tensor models that we consider. As remark in this section, we reintroduce the rank of the tensor fields to arbitrary dimension $d$ to clarify some results. In details, in this section we discuss successively points:

\begin{itemize}
\item The local expansion, investigating the structure of the leading order graph to get relations between effective vertices holding to all orders of the perturbative expansion. 
\item The Ward identities, used to express the derivative of the effective vertices with respect to the external momenta, playing an important role in the definition of the anomalous dimension. The Ward identities moreover introduce a non-trivial relation between $\beta$-functions that we call Ward constraint.
\item We deduce the flow equations in the EVE approximation, taking into account the Ward constraint in a second time, defining the constrained melonic flow. Finally, we investigate the vertex expansion as a complementary way, allowing to track effects depending on the disconnected diagrams. 
\end{itemize}

\subsubsection{Local expansion for Hierarchical RG equations} \label{sectionEVE}

Taking successive functional derivations, the first order differential flow equation \eqref{Wett} split as an infinite hierarchical system, expressing the derivative of $\Gamma^{(n)}$ in terms of $\Gamma^{(n+2)}$, and so on. The basic strategy of the EVE is to close this hierarchical system with the  a set of rigid relations called \textit{structure equations}, inherited from the structure of the leading order graphs in the deep UV. These equations express all the irrelevant effective vertex in terms of a restricted set of parameter: the mass, and all the just-renormalizable couplings. More precisely, we consider a special domain for $k$, so far from the very deep UV limit, and so far from the deep IR limit: $1 \ll k \ll \Lambda$. In this regime, one expect that revelant and marginal couplings dominate the flow, and are sufficient to drag the irrelevant observables. 
We will   investigate  the boundary configuration for the effective vertices and fix some conditions about this investigation. First, we focus our investigation in a phase where the expansion around vanishing classical fields is assumed to be a good vacuum. As explained in \cite{Lahoche:2018vun}, in this region of the global phase space, the two-point functions have to be diagonal, and all the odd vertex functions have to vanish. More precisely:
\begin{corollary}
Into the symmetric phase, the $2$-point functions $G_{IJ}(\vec{p},\vec{q}\,)$, $I,J\in (V,W)$ are diagonals in the momentum space:
\begin{equation}
G_{IJ}(\vec{p},\vec{q}\,):=G_{IJ}(\vec{p}\,)\,\delta_{\vec{p},\vec{q}}\,,
\end{equation}
and all the odd vertex functions vanish, as well as vertex functions which does not involve the same number of derivatives with respect to $\bar{M}_I$ and $M_I$. 
\end{corollary}

The properties of the effective vertex in the symmetric phases are inherited of the behavior of the Feynman diagrams contributing to their perturbative expansion. This has been proved for quartic models with a single field (see \cite{Lahoche:2018vun}). In fact, this is not mysterious, because vanishing classical field is precisely the good vacuum for the perturbative expansion, and one expect that the symmetric phase is contained, or coincide with the analyticity domain of the effective vertex function. For leading order graphs, the definition of the melonic diagrams  is recalled in appendix \ref{appA}  and the series have been explicitly ressumed for purely quartic models. Note that these well analytic properties are a direct consequence of the tree structure of the melonic diagrams (see appendix \ref{appA}), trees increasing less faster than ordinary Feynman graphs with the number of vertices, and are easy to count. \\

\noindent
Following the argument explained with full details in \cite{Lahoche:2018vun}, the boundary structure of the Feynman graphs contributing to the perturbative expansion of the effective vertex reduces them to a set of $d\times n$ matrix valued functions $\pi_2^{(i,n)} : \mathbb{Z}^2\to \mathbb{R}$, the index $n$ labeled the type of interaction that we consider (see equation \eqref{defvertex} of appendix \ref{appA}) and $i\in\llbracket 1,d\rrbracket$. Indeed, from the proposition \ref{cormelons} (appendix \ref{appA}),  it follows that all melonic quartic graphs (except for the first term of the perturbative expansion) contributing to the effective $4$-points vertex function $\Gamma^{(4)}$ have two boundary vertices, two external faces of the same color connected to boundary nodes of two different boundary vertices, and $(d-1)$ short external faces of length $1$ per boundary vertex. Finally, the stability of the condition $\mu=0$  given in the proposition \ref{closed}, ensures the stability of the flow in the UV-limit. As a result, there are only four allowed boundaries, corresponding to the quartic graphs involved in the classical action and the effective $4$-point function $\Gamma^{(4)}_{\vec{p}_1,\vec{p}_2,\vec{p}_3,\vec{p}_4}$ may be decomposed in term of $d\times n$ functions $\Gamma^{(4),(i,n)}_{\vec{p}_1,\vec{p}_2,\vec{p}_3,\vec{p}_4}$ as:
\begin{equation}
\Gamma^{(4),n}_{k,\,\vec{p}_1,\vec{p}_2,\vec{p}_3,\vec{p}_4}=\sum_{i=1}^d \Gamma^{(4),(i,n)}_{k,\,\vec{p}_1,\vec{p}_2,\vec{p}_3,\vec{p}_4}\,.
\end{equation}
Graphically, up to the permutations coming from Wick theorem, the four effective vertex functions $\Gamma^{(4),(i,n)}$ are  the following, up to permutations of the external indices :\\
\begin{align}
\nonumber{\Gamma^{(4),(i,1)}_{k,\,\vec{p}_1,\vec{p}_2,\vec{p}_3,\vec{p}_4}}&\equiv{\vcenter{\hbox{\includegraphics[scale=1]{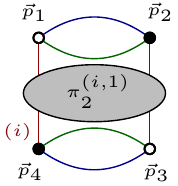} }}}\,,\quad {\Gamma^{(4),(i,2)}_{k,\,\vec{p}_1,\vec{p}_2,\vec{p}_3,\vec{p}_4}}\equiv {\vcenter{\hbox{\includegraphics[scale=1]{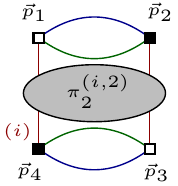} }}}\,\\
{\Gamma^{(4),(i,3)}_{k,\,\vec{p}_1,\vec{p}_2,\vec{p}_3,\vec{p}_4}}&\equiv {\vcenter{\hbox{\includegraphics[scale=1]{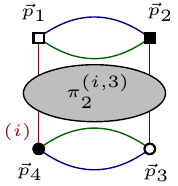} }}}\,,\quad {\Gamma^{(4),(i,4)}_{k,\,\vec{p}_1,\vec{p}_2,\vec{p}_3,\vec{p}_4}}\equiv {\vcenter{\hbox{\includegraphics[scale=1]{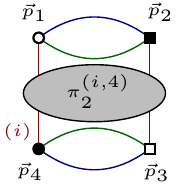} }}}\,.
\end{align}
Explicitly :
\begin{equation}
\Gamma^{(4),(i,n)}_{k,\,\vec{p}_1,\vec{p}_2,\vec{p}_3,\vec{p}_4}=\pi_{2,p_{1i}p_{3i}}^{(i,n)} \sym_n\, \mathcal{W}_{\vec{p}_1,\vec{p}_2,\vec{p}_3,\vec{p}_4}\,,
\end{equation}
where $\mathcal{W}_{\vec{p}_1,\vec{p}_2,\vec{p}_3,\vec{p}_4}$  is a product of Kronecker delta building the general melonic quartic interaction:
\begin{equation}
\sum_{\{\vec{p}_k\}}\mathcal{W}_{\vec{p}_1,\vec{p}_2,\vec{p}_3,\vec{p}_4} \,\phi_{I,\vec{p}_1}\bar{\phi}_{J,\vec{p}_2}\phi_{K,\vec{p}_3}\bar\phi_{L,\vec{p}_4}
\end{equation}
with $,J,K,L \in (V,W)$. The operator $\sym_n$ takes into account all the allowed contractions due to Wick contractions with external edges, and its action depends on the type of boundary that we consider. Indeed, for $n=3,4$, there is only a single configuration of the external edges, each type of field occurring only one time. In contrast, there are four allowed contractions for $n=1,2$, because there are two white and two black nodes of the same type (bull or square). Explicitly:
\begin{equation}
\sym_n\, \mathcal{W}_{\vec{p}_1,\vec{p}_2,\vec{p}_3,\vec{p}_4} = \left\{
    \begin{array}{ll}
       2(\mathcal{W}_{\vec{p}_1,\vec{p}_2,\vec{p}_3,\vec{p}_4} +\mathcal{W}_{\vec{p}_3,\vec{p}_2,\vec{p}_1,\vec{p}_4})\quad & \mbox{if} \,n=1,2 \\
       \mathcal{W}_{\vec{p}_1,\vec{p}_2,\vec{p}_3,\vec{p}_4} & \mbox{if}\, n=3,4
    \end{array}
\right.
\end{equation}
\noindent
Deriving the Wetterich equation \eqref{Wett} with respect to $M_I$ and $\bar{M}_J$ in  the deep UV, one get, graphically
\begin{equation}
\dot{\Gamma}_{k,\,IJ}^{(2)}(\vec{p}\,)= 0 \,\quad \mbox{for\,} I\neq J\,, \label{IJ}
\end{equation}
and:
\begin{align}
\dot{\Gamma}_{k,\,VV}^{(2)}(\vec{p}\,)= - \sum_{i=1}^d \left\{ 2\vcenter{\hbox{\includegraphics[scale=0.8]{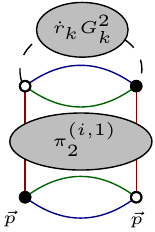} }} \,+\, \vcenter{\hbox{\includegraphics[scale=0.8]{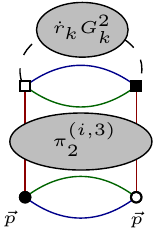} }}   \right\}\,,\quad \dot{\Gamma}_{k,\,WW}^{(2)}(\vec{p}\,)= - \sum_{i=1}^d \left\{ 2\vcenter{\hbox{\includegraphics[scale=0.8]{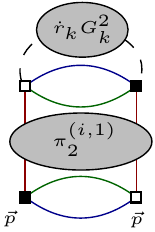} }} \,+\, \vcenter{\hbox{\includegraphics[scale=0.8]{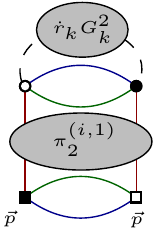} }}   \right\}\,,\label{II}
\end{align}
where we keep only the leading order contractions in the large $k$ limit, the contraction with dotted edges building the effective loop with propagator $\dot{r}_k G^2_k$, $G_k$ denoting the effective $2$-point function :
\begin{equation}
(G_k^{-1})_{IJ}=(\Gamma^{(2)}_k+r_k)_{IJ}\,.
\end{equation}
Note that the factor $2$ in front of some effective vertex is at least abusive since all the combinatorics is in fact included in the definition of the effective vertex itself. Then, we can think this notation as a simple way to count the symmetry factors, i.e. the number of equivalent diagrams without permutation of the external indices. The first equation \eqref{IJ} ensures the stability of the condition $\mu=0$, and provides a non-perturbative version of the first point of the proposition \ref{closed}. The second equation \eqref{IJ} allows to extract the flow equations for the effective mass parameters $m_{II}$ and wave functions renormalization $Z_{II}$. \\

\noindent
In order to get the flow equations, we have to fix the \textit{renormalization conditions} by giving the effective coupling constant at the scale $k$:
\begin{definition} \label{defren}\textbf{Renormalization condition in the symmetric phase.} 
The effective mass parameters $m_{II}(k)$ for $I\in(V,W)$ is defined as the zero-momenta value of the $2$-point function $\Gamma^{(2)}_{k,II}$:
\begin{equation}
m_{II}(k):= \Gamma^{(2)}_{k,II}(\vec{p}=\vec{0}\,)\,.
\end{equation}
In the same way, the effective couplings $g_I(k)$ and $c_I(k)$ are defined from the zero momenta $4$-point functions as:
\begin{align}
\Gamma^{(4),(i,n)}_{k,\,\vec{0},\vec{0},\vec{0},\vec{0}}&=4\pi_{2,00}^{(i,n)}=:4g_n(k)\,, \,\mbox{for\,\,} n=1,2\\
\Gamma^{(4),(i,n)}_{k,\,\vec{0},\vec{0},\vec{0},\vec{0}}&=\pi_{2,00}^{(i,n)}=:c_n(k)\,, \quad\mbox{for\,\,} n=3,4\,.
\end{align}
Finally, in the symmetric phase, the wave functions renormalization factors $Z_{II}(k)$ are defined as :
\begin{equation}
Z_{II}(k):=\frac{\partial}{\partial \vert p_i\vert } \Gamma_{k,\,II}^{(2)}(\vec{p}=\vec{0}\,)\,.\label{equationZ}
\end{equation}
\end{definition}
Note that we must distinguish between the effective and the bare couplings (i.e. the couplings involved in the classical action) by the  evolution parameter $k$. Moreover, note that our initial conditions for mass are such that $m_{VV}=m_{WW}=m$ in the limit $k\to \Lambda$. \\

\noindent
Our strategy will be to close the hierarchical system of coupled flow equations for local couplings by expressing the effective vertices $\Gamma^{(6)}_k$ in terms of $\Gamma^{(4)}_k$ and $\Gamma^{(2)}_k$. We recall that local couplings correspond to couplings for local observable, in the sense of the definition \ref{locality}. For $2$ and $4$-points melonic observables, these local couplings are defined by the renormalization conditions given by definition \ref{defren}. However, strictly local couplings are not sufficient to express the flow of the local couplings. In particular, this requires the full momentum dependence of the $2$-point function. In the melonic sector, this function is fixed by a system of closed equations from the following statement (For more detail see Appendix \ref{appB}):

\begin{proposition} \textbf{Melonic closed equations.} \label{closed}
In the symmetric phase, and in the deep UV limit, we have the following two statements about the leading order self energy:

\begin{itemize}
\item The components $\Sigma_{VW}$ and $\Sigma_{WV}$ of the self energy $\Sigma$ have to vanish. As a consequence, the coupling constant $\mu$ does not receives radiative corrections, and the condition $\mu=0$ is stable.

\item The leading order remaining components $\Sigma_{VV}$ and $\Sigma_{WW}$ splits into $d$ functions $\sigma_{II}^{(i)}$, $I\in (V,W)\,, i\in\llbracket 1,d\rrbracket$, depending on a single component of the external momentum:
\begin{equation}
\Sigma_{II}(\vec{p}\,):=\sum_{i=1}^d \sigma_{II}^{(i)}(p_i)\,,
\end{equation}
and the components $\sigma_{II}^{(i)}(p_i)$ satisfies to the coupled system of closed equations:
\begin{align}
\sigma_{VV}^{(i)}(p)&= -2g_1 \sum_{\vec{q}\in \mathbb{Z}^d} \delta_{q_ip}G_{VV}(\vec{q}\,)-c_1 \sum_{\vec{q}\in \mathbb{Z}^d} \delta_{q_ip}G_{WW}(\vec{q}\,)\,,\\
\sigma_{WW}^{(i)}(p)&=-2g_2 \sum_{\vec{q}\in \mathbb{Z}^d} \delta_{q_ip}G_{WW}(\vec{q}\,)-c_1 \sum_{\vec{q}\in \mathbb{Z}^d} \delta_{q_ip}G_{VV}(\vec{q}\,)\,.
\end{align}
\end{itemize}
\end{proposition}

\noindent
Note that the couplings involved in these equations are the bare couplings  without $k$ dependence. Remark that these melonic closed equations are reputed to be difficult to solve  
  \cite{Samary:2014tja}-\cite{Samary:2014oya}, and we do not focus on the solution of this system. In spite of all the difficulties to provide the explicitly solution, we remind the reader that the resolution of these equations is deserved for a forthcoming investigation. Our strategy in this note is to use the flow equation itself to get an approximation of the exact solution. More concretely, we will choose an ansatz for $\Gamma^{(2)}$, and follows its evolution along the RG flow. Because we are in the symmetric phase, $\Gamma^{(2)}$ have to be independent of the classical mean fields, and following references \cite{Lahoche:2018oeo}-\cite{Lahoche:2018ggd}, we will keep only the first terms of the derivative expansion (that is to say, we keep only marginal and essential terms in the expansion of $\Gamma^{(2)}(\vec{p}\,)$ with respect to $\vert \vec{p}\,\vert$). Our ansatz is then the following:
\begin{equation}
\Gamma^{(2)}_{II}(\vec{p}\,):=Z_{II}(k)\vert \vec{p}\,\vert+m_{II}(k)\,. \label{derivativeexp}
\end{equation}

\noindent
Note that this approximation hold only in the symmetric phase. From the renormalization conditions, by setting the external momenta to zero, i.e.  $\vec{p}=\vec{0}$  in equation \eqref{II} and in the first derivative of the same equation, one  can deduce the equations for $\dot{m}_{II}$ and $\dot{Z}_{II}$. 
Defining the \textit{anomalous dimension} $\eta_I$ as:
\begin{equation}
\eta_I:= \frac{1}{Z_{II}}\frac{d}{dt} Z_{II}\,;
\end{equation}
one get:
\begin{equation}
\dot{m}_{II}:=-6g_{i(I)}(k) I_{2\,,II}(0) -3c_1(k) I_{2\,,\hat{I}\hat{I}}(0) \label{floweq1}
\end{equation}
\begin{equation}
Z_{II}\,\eta_I:= -2\frac{\partial \pi_{2,\,00}^{(1,n(I))}}{\partial \vert p_1 \vert } I_{2\,,II}(0)-\frac{\partial \pi_{2,\,00}^{(1,3)}}{\partial \vert p_1 \vert } I_{2\,,\hat{I}\hat{I}}(0)-2g_{n(I)}(k) I_{2\,,II}^\prime(0) -c_1(k) I_{2\,,\hat{I}\hat{I}}^\prime(0)\label{floweq2}
\end{equation}
where, $n(V)=1$, $n(W)=2$; and as in section \ref{truncation} the "dot" refer to the derivative with respect to the log-scale $t:=\ln(k)$, and we defined the one-loop sums $I_{n\,,II}$ as:
\begin{equation}
I_{n\,,II}(\vert q \vert ):=\sum_{\vec{p}} \delta_{p_1q} \frac{\dot{r}_k(\vec{p}\,)}{(Z_{II} \vert \vec{p}\,\vert+m_{II}+(r_k)_{II}(\vec{p}\,))^n}\,.
\end{equation}
Finally, the notation $\hat{I}$ is defined as: For $I=V$, $\hat{I}=W$ and for $I=W$, $\hat{I}=V$. Using the explicit expression of the regulator \eqref{regulator}, the sums $I_{n\,,II}(\vert q \vert )$ writes as:
\begin{equation}
I_{n\,,II}(\vert q \vert )=\frac{Z_{II}}{(Z_{II} k+m_{II})^n} \sum_{\vec{p}} \delta_{p_1q}\, \theta \left(k-\vert \vec{p}\,\vert\right) \left[\eta_{I}(k-\vert \vec{p}\,\vert)+k\,\right]\,.
\end{equation}
In the deep UV limit, the sums can be approached by integrals of the type:
\begin{equation}
J_n(\vert q \vert):= \int d^2x \vert \vec{x} \,\vert^n \theta(R-\vert x_1 \vert -\vert x_2 \vert)\,
\end{equation}
for the continuous variables $x_i:=p_i/k$; more detail about the explicit computation of this sum can be found in appendix \ref{appC}. Then we get:
\begin{equation}
I_{n\,,II}(\vert q \vert )=Z_{II}\frac{J_0(\vert q \vert)[\eta_I(k-\vert q \vert)+k]-J_1(\vert q \vert)\eta_I}{(Z_{II} k+m_{II})^n} \,.
\end{equation}
This expression allows to compute each terms in the equations \eqref{floweq1} and \eqref{floweq2}, except the terms $\frac{\partial \pi_{2,\,00}^{(1,n)}}{\partial \vert p_1 \vert } I_{2\,,II}(0)$. These terms take into account an additional information about the momentum dependence of the effective vertices with respect to the finite local vertex expansion. We compute them in the next section using Ward identities. \\

\noindent
The equations for $\dot{g}_i$ and $\dot{c}_i$ may be obtained from the same principle by deriving twice with respect to $M$ and $\bar{M}$, before setting $M=\bar{M}=0$. Formally, if we forget temporally the index structure $V,W$, we get an equation schematically of the form:
\begin{align}
\nonumber\dot\Gamma^{(4)}_{k,\vec 0,\vec 0,\vec 0,\vec 0}=-\sum_{\vec p}\Tr\,\dot r_k(\vec p\,) G^2_k(\vec p\,)\Big[\Gamma^{(6)}_{k,\vec p,\vec 0,\vec 0,\vec p,\vec 0,\vec 0}&-2\sum_{\vec p\,'}\Gamma^{(4)}_{k,\vec p,\vec 0,\vec p\,',\vec 0}G_k(\vec p\,')\Gamma^{(4)}_{k,\vec p\,',\vec 0,\vec p,\vec 0}\Big]\\
&+2\sum_{\vec p}\Tr\,\dot r_k(\vec p\,) G^3_k(\vec p\,)[\Gamma^{(4)}_{k,\vec p,\vec 0,\vec p,\vec 0}]^2\,, \label{flowfour}
\end{align} 
the trace $\Tr$ running through the $V$-$W$ indices. This equation involve $6$-point functions, and therefore in the hope to  isolate the contributions of each beta functions, we have to  provide the structure of the effective vertices $\Gamma^{(6)}_k$, as in the case of  the $4$-point vertices. \\  

\noindent
From proposition \ref{cormelons} of appendix \ref{appA}, we keep that the boundary of a 1PI melonic $6$-point graph have to be made of three external vertices, each of them sharing $d-1$ colored faces of length one. Moreover, there must exist three external faces of the same color, whose boundaries link pairwise each external vertex (i.e. their end nodes are one different vertices). However, all the boundary configurations are not allowed, and we have the following constraint ( the proof is given in appendix \ref{appB}):

\begin{proposition} \label{coro1}
Let us consider a 1PI non-vacuum diagram having $\bar{V}_4$ external vertices of type $(i,4)$ among the $\bar{V}$ external vertices and $2N$ external edges. In the limit $\mu \to 0$, all the leading order connected $1PI$ graphs with odd $\bar{V}_4$ vanish except for the special case $2N=4$ and $\bar{V}_4=\bar{V}=1$.
\end{proposition}
In particular, for 1PI $6$-points graphs we get the following statement:
\begin{corollary}
The $6$-point vertex functions with $3$ and $1$ boundary vertex of type $(i,4)$ vanish in the limit $\mu\to 0$. 
\end{corollary}
As a result, the full melonic function $\Gamma_k^{(6)}$ splits into $d$ components, labeled by the color of the external face running through the interior of the diagrams contributing to its perturbative expansion. Moreover, each of these components split as $9$ components, labeled by all the allowed configurations for the boundary vertices:
\begin{equation}
\Gamma_{k,\vec{p}_1,\vec{p}_2,\vec{p}_3,\vec{p}_4,\vec{p}_5,\vec{p}_6}^{(6)}=\sum_{i=1}^d\sum_{a,b,c\in(1,2,4)}\Gamma_{k,\vec{p}_1,\vec{p}_2,\vec{p}_3,\vec{p}_4,\vec{p}_5,\vec{p}_6}^{(6),(i,abc)}\,,
\end{equation}
where the allowed structures for the functions $\Gamma_{k,\vec{p}_1,\vec{p}_2,\vec{p}_3,\vec{p}_4,\vec{p}_5,\vec{p}_6}^{(6),(i,abc)}$ are listed on Figure \ref{figlist6} below. Note that, as for $4$-point functions, we took  into account the permutations of the external momenta.

\begin{center}
\begin{align*}
&\Gamma_{k,\vec{p}_1,\vec{p}_2,\vec{p}_3,\vec{p}_4,\vec{p}_5,\vec{p}_6}^{(6),(i,111)}\,\equiv\,\vcenter{\hbox{\includegraphics[scale=0.6]{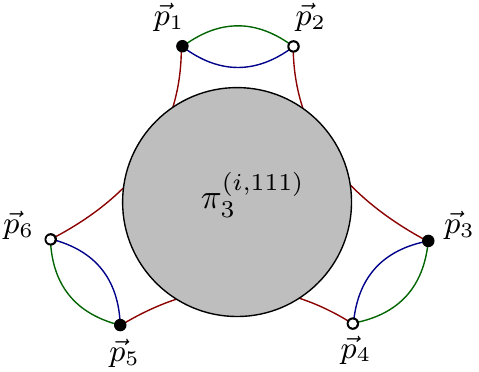} }}\,,\quad \Gamma_{k,\vec{p}_1,\vec{p}_2,\vec{p}_3,\vec{p}_4,\vec{p}_5,\vec{p}_6}^{(6),(i,112)}\,\equiv\,\vcenter{\hbox{\includegraphics[scale=0.6]{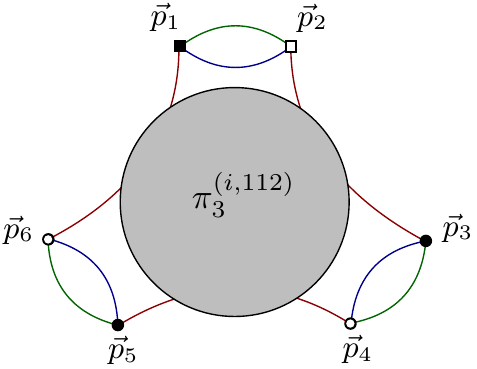} }}\\  
&\Gamma_{k,\vec{p}_1,\vec{p}_2,\vec{p}_3,\vec{p}_4,\vec{p}_5,\vec{p}_6}^{(6),(i,122)}\,\equiv\,\vcenter{\hbox{\includegraphics[scale=0.6]{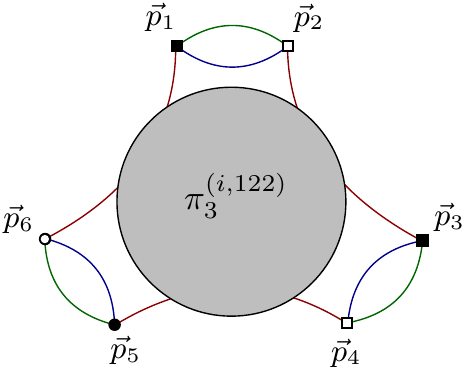} }}\,,\quad
\Gamma_{k,\vec{p}_1,\vec{p}_2,\vec{p}_3,\vec{p}_4,\vec{p}_5,\vec{p}_6}^{(6),(i,222)}\,\equiv\,\vcenter{\hbox{\includegraphics[scale=0.6]{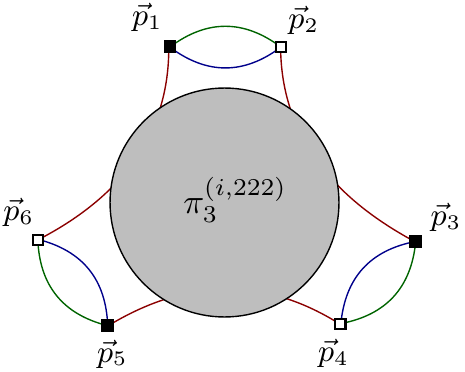} }}\\
&\Gamma_{k,\vec{p}_1,\vec{p}_2,\vec{p}_3,\vec{p}_4,\vec{p}_5,\vec{p}_6}^{(6),(i,144)}\,\equiv\,\vcenter{\hbox{\includegraphics[scale=0.6]{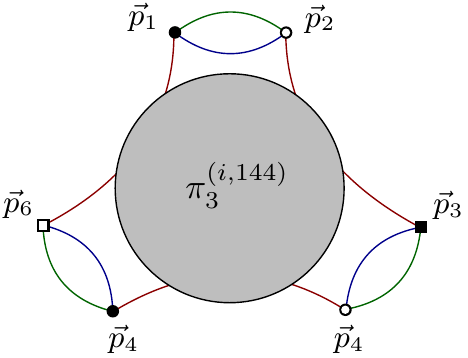} }} \,,\quad \Gamma_{k,\vec{p}_1,\vec{p}_2,\vec{p}_3,\vec{p}_4,\vec{p}_5,\vec{p}_6}^{(6),(i,244)}\,\equiv\, \vcenter{\hbox{\includegraphics[scale=0.6]{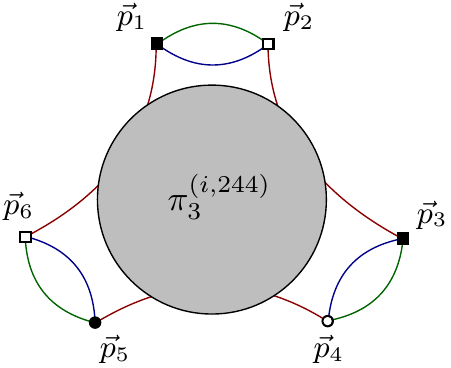} }}
\end{align*}
\captionof{figure}{The allowed boundary structures for the vertex functions $\Gamma_{k,\vec{p}_1,\vec{p}_2,\vec{p}_3,\vec{p}_4,\vec{p}_5,\vec{p}_6}^{(6),(i,abc)}$. All the other combinations of the indices $a$, $b$, $c$ labels  are vanished functions.}\label{figlist6}
\end{center}

\noindent
Note that all the reduced vertex functions $\pi_3^{(i,abc)}:\mathbb{Z}^3\to \mathbb{R}$ will be fixed latter to close the hierarchy. \\

\noindent
Moving on to the flow equation \eqref{flowfour}, we are now in position to investigate the beta functions for each couplings from the renormalization conditions, identifying their contributions on the right hand side from the boundary structure of the resulting effective obervables. Let us consider $\dot{g}_1$. From the definition \ref{defren}, one get:
\begin{equation}
4\dot{g}_1=-4\left(3\vcenter{\hbox{\includegraphics[scale=0.5]{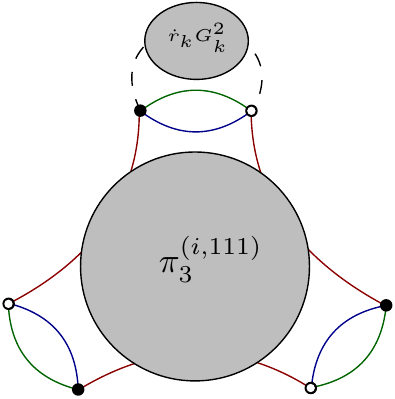}}}+\vcenter{\hbox{\includegraphics[scale=0.5]{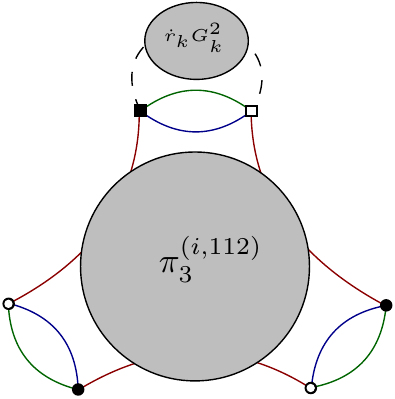}}} \right)+4 \left(4\,\vcenter{\hbox{\includegraphics[scale=0.6]{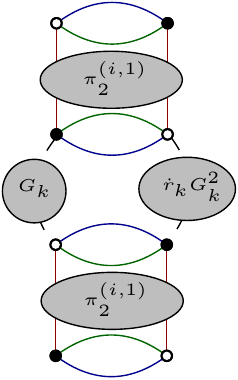}}}+\vcenter{\hbox{\includegraphics[scale=0.6]{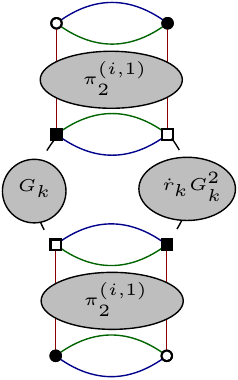}}}\right)\,.\label{flowspe}
\end{equation}
More explictly we write: 
\begin{equation}
\dot{g}_1= -3\pi_3^{(1,111)} I_{2,VV}(0)-\pi_3^{(1,112)} I_{2,WW}(0)+4 g_1^2 I_{3,VV}(0)+c_1^2I_{3,WW}(0)\,.\label{flow2}
\end{equation}
The same computation may be given  for $\dot{g}_2$, $\dot{c}_1$ and $\dot{c}_2$, we get:
\begin{align}
\dot{g}_2&= -3\pi_3^{(1,222)} I_{2,WW}(0)-\pi_3^{(1,122)} I_{2,VV}(0)+4 g_2^2 I_{3,WW}(0)+c_1^2I_{3,VV}(0)\,,\\
\dot{c}_1&= -2\pi_3^{(1,112)} I_{2,VV}(0)-2\pi_3^{(1,122)} I_{2,WW}(0)+4c_1\left(g_1 I_{3,VV}(0)+g_2 I_{3,WW}(0)\right) \,,\\
\dot{c}_2&=-\pi_3^{(1,144)}I_{2,VV}(0)-\pi_3^{(1,244)}I_{2,WW}(0)+c_2^2 I_{3,VW}(0) \,,\label{flow3}
\end{align}
where for the last equation we introduce the mixed loop $I_{3,VW}(0)$, defined from:
\begin{equation}
I_{3,VW}(\vert q \vert):=\sum_{\vec{p}} \delta_{p_1q}\left(\dot{r}_{VV}G^2_{k,VV}G_{k,WW}+\dot{r}_{WW}G^2_{k,WW}G_{k,VV}\right)\,,\label{hetero}
\end{equation}
and we used of the notation $\pi_{3}^{(i,abc)}\equiv \pi_{3,000}^{(i,abc)}$. As explained before, to close the hierarchy we have to find a relation between $6$, $4$ and $2$ points functions. As we will see, there are two nonequivalent ways to do this. The first one follows the strategy explained in \cite{Lahoche:2019cxt}, and use the structure of the melonic diagram to find explicitly the structure of the leading order $6$ point functions. This strategy have to be finally completed with a constraint coming from Ward-identity, and provides a complicate description of the effective physical phase space. Another way is to constraint the expression of the $6$-point functions themselves using Ward identities. We will discuss these two strategies in the two next sections.

\subsubsection{Structure equations and Ward identities}
The unitary symmetry of the classical interactions \eqref{classical} induce a non-trivial Ward-Takahashi identity at the quantum level, due to the translation invariance of the Lebesgue measure $\prod_I d\phi_I d\bar{\phi}_I$ involved in the definition of the partition function \eqref{part1}:
\begin{equation}
\int d\mu_C:= \int \prod_{I=V,W} d\phi_I d\bar{\phi}_I e^{-S_{\text{kin}}[\phi_I ,\bar{\phi}_I]}\,.
\end{equation}
In the first section \ref{sec1}, we defined as $\mathbb{U}(N)$ the set of unitary symmetries of size $N$, admitting an inductive limit for arbitrary large $N$. A transformation is then a set of $d$ independent elements of $\mathbb{U}(N)$, $\mathcal{U}:=(U_1,\cdots, U_d)\in\mathbb{U}(N)^d$, one per index of the tensor fields, and the transformation rule is explicitly the following:
\begin{equation}
\mathcal{U}[\phi_I]_{\vec{p}}\to(\phi_I^\prime)_{\vec{p}} :=\sum_{q_1,\cdots,q_d} (U_1)_{p_1q_1}\cdots (U_d)_{p_dq_d} (\phi_I)_{\vec{q}}\,.
\end{equation}
Note that, as explained in the section \ref{sec1}, the same transformation act on $V$ and $W$ fields. The global translation invariance of the Lebesgue measure ensure the invariance of the partition functions \eqref{part2}:
\begin{equation}
\mathcal{U}[\mathcal{Z}_k[J,\bar{J}]]=\mathcal{Z}_k[J,\bar{J}]\,.
\end{equation}
Focusing on an infinitesimal transformation : $\delta_1:=(\mathrm{id}+\epsilon,\mathrm{id},\cdots,\mathrm{id})$ acting non-trivially only on the color $1$ for some infinitesimal anti-Hermitian transformations, and keeping only the variation terms of order $\epsilon$, one get the following statement:
\begin{theorem}
The partition function $\mathcal{Z}_k[J,\bar{J}]=:e^{\mathcal{W}_k[J,\bar{J}]}$ of the theory defined by the action \ref{classical} satisfy the following relation (we sum over the index $I$):
\bea\label{Ward0}
\nonumber\sum_{\vec{p}_\bot, \vec{p}_\bot\,^{\prime}}  \delta_{\vec{p}_\bot\vec{p}_\bot\,^{\prime}}  \bigg\{\big[\Delta C_s(\vec p,\vec p\,')\big]_{II}\left[\frac{\partial^2 \mathcal{W}_k}{\partial \bar{J}_{I,\vec{p}\,^\prime}\,\partial {J}_{I,\vec{p}}}+\bar{M}_{I,\vec{p}}M_{I,\vec{p}\,^\prime}\right]-\bar{J} _{I,\vec{p}}\,M_{I,\vec{p}\,^\prime}+{J} _{I,\vec{p}\,^\prime}\bar{M}_{I,\vec{p}}\bigg\}=0\,,
\eea
with $\vec{p}_\bot:=(0,p_2,\cdots ,p_d)\in \mathbb{Z}^{d}$ and:
\begin{equation}
\big[\Delta C_s(\vec p,\vec p\,')\big]_{II}:=C_{k,II}^{-1}(\vec{p}\,)-C_{k,II}^{-1}(\vec{p}\,^{\prime})
\end{equation}
\end{theorem}
The proof  can be found in  \cite{Lahoche:2018oeo}-\cite{Lahoche:2019ocf}. The interest of the Ward identities is to connect $n$ and $n+2$ effective functions, due to the specific non-invariance of the kinetic term. In particular, we will extensively interested by the connection between $4$ and $2$ point function on one hand, and between $6$ and $4$ point functions on a second hand. Taking successive derivative with respect to the external sources, $J_V$ and $\bar{J}_V$, setting $\vec{p}\to\vec{p}\,^\prime$ on one hand, and finally $\vec{p}\to \vec{0}$, one get the relation:
\begin{equation}
2\,\vcenter{\hbox{\includegraphics[scale=1]{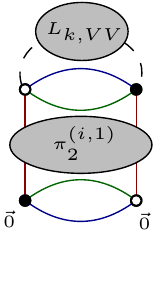} }}+\vcenter{\hbox{\includegraphics[scale=1]{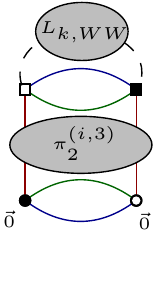} }}=-\frac{\partial}{\partial \vert p_1\vert} \left(\Gamma_{k,VV}^{(2)}(\vec{p}_{\bot})-Z_{0,VV}\vert\vec{p}\,\vert \right)\bigg\vert_{\vec{p}=0}\,,
\end{equation}
where:
\begin{equation}
L_{k,I}(\vec{p}\,):=\left(Z_{0,II}+\frac{\partial r_{k,II}}{\partial \vert p_1\vert}(\vec{p}\,)\right)G^2_{k,I}(\vec{p}\,)\,.
\end{equation}
The same relation hold in the replacement $V\to W$ for the external points, and because of the renormalization conditions, definition \ref{defren}, one get the following statement:

\begin{corollary}\label{Ward1}
In the symmetric phase, and in the deep UV sector, the effective local quartic couplings and the field strength renormalization obey to the following set of coupled equations:
\begin{equation}
 \left\{
    \begin{array}{ll}
        2g_1(k) \mathcal{L}_{k,V}+c_1(k)  \mathcal{L}_{k,W}=Z_{0,VV}-Z_{VV}(k)\\
       2g_2(k) \mathcal{L}_{k,W}+c_1(k)  \mathcal{L}_{k,V}=Z_{0,WW}-Z_{WW}(k)\,,
    \end{array}
\right.
\end{equation}
where:
\begin{equation}
\mathcal{L}_{k,I}:= \sum_{\vec{p}_\bot} L_{k,I}(\vec{p}_\bot)\,,\quad I\in(V,W)\,.
\end{equation}
\end{corollary}
Note that the coupling $c_2(k)$ does not appear in these equations, and it is not constrained by the Ward identity. We will discuss extensively this point at the end of this section and in the next one. \\

\noindent
Now, let us consider the $6$-point function. Deriving twice with respect to the external sources the Ward identity given by theorem \ref{Ward0}, setting $J=\bar{J}=0$ at the end of the computation, and following the same strategy as explained in references \cite{Lahoche:2018oeo}-\cite{Lahoche:2018vun}, we obtain a relation between the derivative of the reduced vertex functions $\pi_2$ with respect to the external momenta and the reduced functions $\pi_3$. Graphically:
\begin{equation}
6\vcenter{\hbox{\includegraphics[scale=0.6]{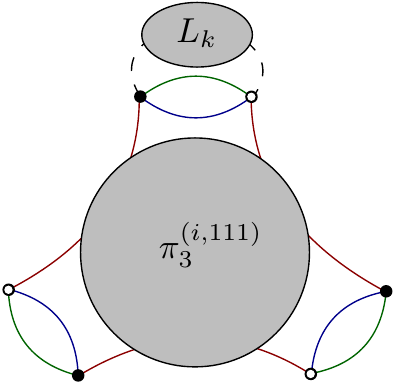} }}+2\vcenter{\hbox{\includegraphics[scale=0.6]{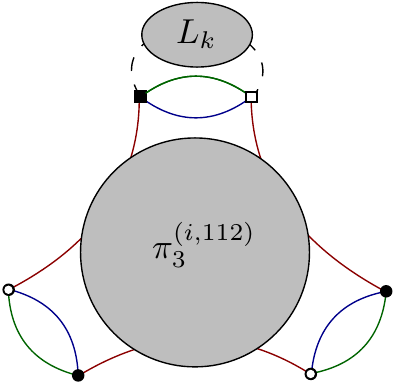} }}-\,8\,\vcenter{\hbox{\includegraphics[scale=0.6]{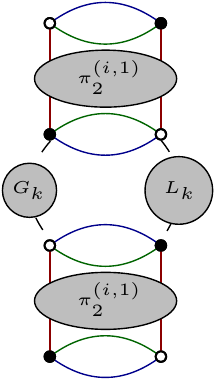} }}-\,2\,\vcenter{\hbox{\includegraphics[scale=0.6]{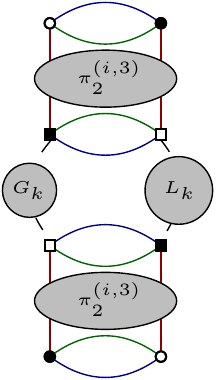} }}=-\frac{\partial \pi_{2,00}^{(1,1)}}{\partial \vert p_1\vert} \,, \label{Wardspe}
\end{equation}
\begin{equation}
6\vcenter{\hbox{\includegraphics[scale=0.6]{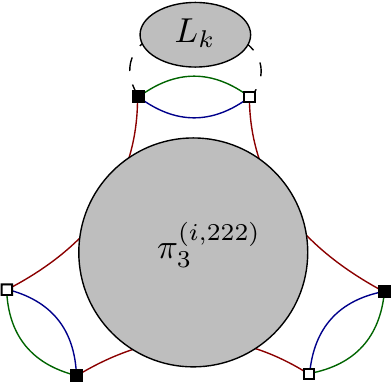} }}+2\vcenter{\hbox{\includegraphics[scale=0.6]{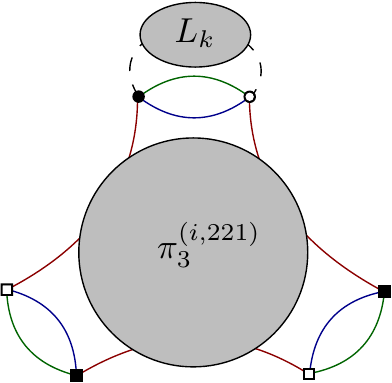} }}-\,8\,\vcenter{\hbox{\includegraphics[scale=0.6]{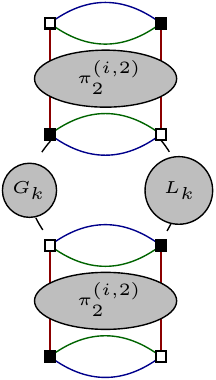} }}-\,2\,\vcenter{\hbox{\includegraphics[scale=0.6]{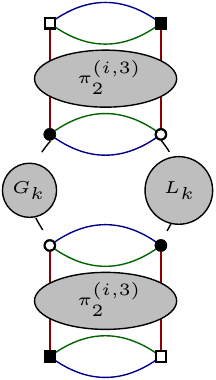} }}=-\frac{\partial \pi_{2,00}^{(1,2)}}{\partial \vert p_1\vert} \,,
\end{equation}
\begin{equation}
2\vcenter{\hbox{\includegraphics[scale=0.6]{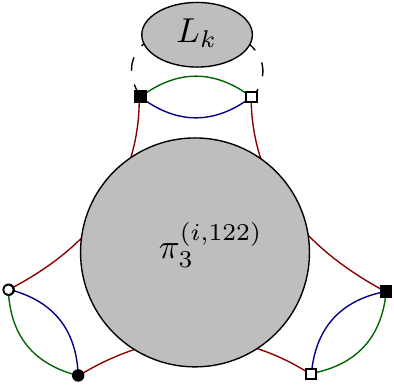} }}+2\vcenter{\hbox{\includegraphics[scale=0.6]{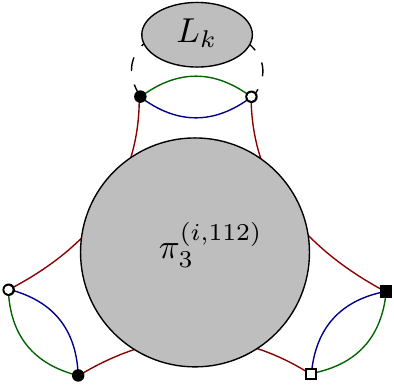} }}-\,4\,\vcenter{\hbox{\includegraphics[scale=0.6]{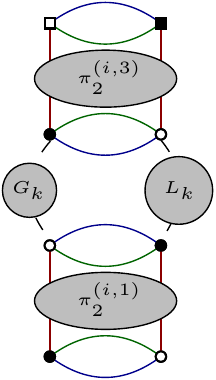} }}-\,4\,\vcenter{\hbox{\includegraphics[scale=0.6]{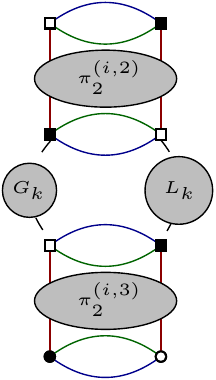} }}=-\frac{\partial \pi_{2,00}^{(1,3)}}{\partial \vert p_1\vert} \,,
\end{equation}
\begin{equation}
\vcenter{\hbox{\includegraphics[scale=0.6]{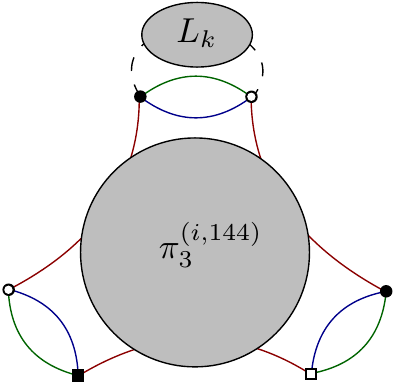} }}+\vcenter{\hbox{\includegraphics[scale=0.6]{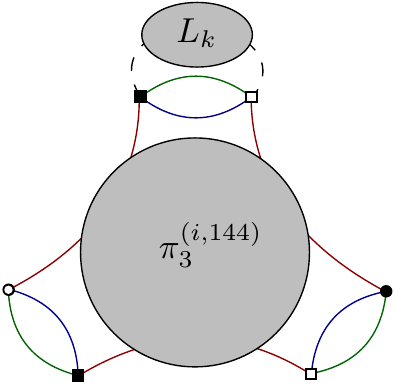} }}-\,\,\vcenter{\hbox{\includegraphics[scale=0.6]{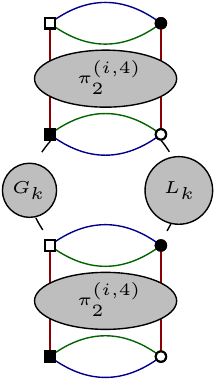} }}-\,\,\vcenter{\hbox{\includegraphics[scale=0.6]{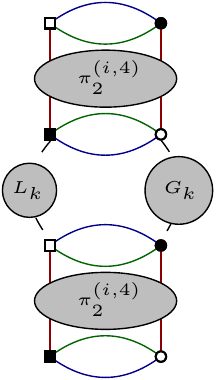} }}=-\frac{\partial \pi_{2,00}^{(1,4)}}{\partial \vert p_1\vert} \,.
\end{equation}
Note that for each of these equations:
\begin{equation}
\frac{\partial \pi_{2,00}^{(1,n)}}{\partial \vert p_1\vert} \equiv \frac{\partial \pi_{2,p_1p_1}^{(1,n)}}{\partial \vert p_1\vert} \bigg\vert _{p_1=0}\,.
\end{equation}
Translating each of these equations into formula, we get, to complete the the Corollary \ref{Ward1}:
\begin{corollary}\label{Ward3}
In the symmetric phase, and in the deep UV limit, the first derivatives of each reduced effective quartic functions with respect to the external momenta are related to $6$ and $4$ point effective vertices as:
\begin{align}
6 \pi_3^{(1,111)} \mathcal{L}_{k,V}+2\pi_3^{(1,112)} \mathcal{L}_{k,W}-8g_1^2(k) \mathcal{U}_{k,V}-2c_1^2(k) \mathcal{U}_{k,W}&=-\frac{\partial \pi_{2,00}^{(1,1)}}{\partial \vert p_1\vert}\,,\\
6\pi_3^{(1,222)} \mathcal{L}_{k,W}+2\pi_3^{(1,122)} \mathcal{L}_{k,V}-8g_2^2(k) \mathcal{U}_{k,W}-2c_1^2(k) \mathcal{U}_{k,V}&=-\frac{\partial \pi_{2,00}^{(1,2)}}{\partial \vert p_1\vert}\,,\\
2 \pi_3^{(1,112)} \mathcal{L}_{k,V}+2\pi_3^{(1,122)} \mathcal{L}_{k,W}-4c_1(k)\left(g_1(k)\mathcal{U}_{k,V}+g_2(k) \mathcal{U}_{k,W}\right)&=-\frac{\partial \pi_{2,00}^{(1,3)}}{\partial \vert p_1\vert}\,,\\
 \pi_3^{(1,144)} \mathcal{L}_{k,V}+\pi_3^{(1,244)} \mathcal{L}_{k,W}-c_2^2\left(\mathcal{U}_{k,VVW}+\mathcal{U}_{k,WWV}\right)&=-\frac{\partial \pi_{2,00}^{(1,4)}}{\partial \vert p_1\vert} \,,
\end{align} 
where we introduced $\mathcal{U}_{k,I}$ and $\mathcal{U}_{k,IIJ}$ defined as:
\begin{equation}
\mathcal{U}_{k,I}= \sum_{\vec{p}_\bot} L_{k,I}(\vec{p}_\bot)G_{k,I}(\vec{p}_\bot)\,,\quad \mathcal{U}_{k,IIJ}= \sum_{\vec{p}_\bot} L_{k,I}(\vec{p}_\bot)G_{k,J}(\vec{p}_\bot)\,.
\end{equation}
\end{corollary}

\noindent
These equations allows to express the derivatives on the right hand side of the expression \eqref{floweq2} in term of the local observables $\pi_2$ and $\pi_3$. At this stage, $\pi_3$ is the only inconvenience to close the hierarchy. \\

\noindent
The recursive definition of the melonic diagram provide a set of solid relations called \textit{structure equations} in  \cite{Lahoche:2018oeo}-\cite{Lahoche:2018ggd}, and allowing to express all the melonic effective vertex functions in terms of the finite set of them corresponding to just-renormalizable interactions. We only provide the main step of the proof, referring to the cited papers for details. \\

\noindent
Let us consider a Feynman diagram $\mathcal{G}$ contributing to perturbative expansion of the reduced melonic effective vertex $\pi_{3}^{(1,111)}$. We use of the intermediate field representation, the reader unfamiliar with this formalism may consult the Appendix \ref{appA}. As a leading order graph, $\mathcal{G}$ corresponds to a tree $\mathcal{T}_{\mathcal{G}}$ in the intermediate field representation, whose an example is pictured in Figure \ref{figinter} below. The structure of the diagram follows the proposition  \ref{cormelons} of appendix \ref{appA}. The diagram has three cilia corresponding to the external (dotted) edges to which the three external vertex are hooked. Each of these vertices share $d-1$ short external faces of length one, and are the end points of $3$ monocolored external faces running through the interior of the diagram. These three internal faces connect two-by-two the boundary vertex, and then ensure the existence a path of color $1$ between each pair of boundary vertices. Let $\mathcal{P}_{c_1c_2}^{(1)}$, $\mathcal{P}_{c_2c_3}^{(1)}$ and $\mathcal{P}_{c_1c_3}^{(1)}$. Merging together these paths $\mathcal{P}_{c_1c_2}^{(1)}\cup \mathcal{P}_{c_2c_3}^{(1)}\cup \mathcal{P}_{c_1c_3}^{(1)}$ build a monocilored tree made of three monocolored arms $\mathcal{P}_{c_i}^{(1)}$. Topologically, these arms have to be hooked to a common loop-vertex $v$. Some other connected components may be hooked to this vertex, has $\tau_1$ and $\tau_2$ between the arc $(\mathcal{P}_{c_3}^{(1)},\mathcal{P}_{c_1}^{(1)})$:
\begin{equation}
\vcenter{\hbox{\includegraphics[scale=1]{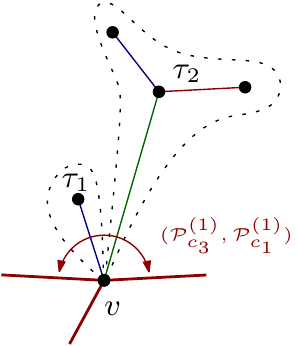}}}
\end{equation}
Each of these components $\tau_i$ are sub-trees with two external points, and then they contribute to the perturbative expansion of the leading order self-energy. As a result, order by order, connected components hooked to the arc $(\mathcal{P}_{c_3}^{(1)},\mathcal{P}_{c_1}^{(1)})$ build nothing but the effective $2$-point function $G_k$, between each arcs $(\mathcal{P}_{c_i}^{(1)},\mathcal{P}_{c_j}^{(1)})$. In the same way, each arm $\mathcal{P}_{c_i}^{(1)}$ are nothing but contributions to the perturbative expansion of the effective $4$-point functions. More precisely, due to the boundary conditions, the effective functions may be homogeneous and corresponds to $\Gamma_k^{(4,1)}$ or heteroclyte and corresponds to $\Gamma_k^{(4, 3)}$. Indeed, because of propositions \ref{closed} and \ref{coro1}, only type $1$, $2$ or $3$ vertices may be hooked to the global node $v$. Moreover, to ensure its closure, all the arms hooked to $v$ have to be of the same type. In other worlds, all of them have to be a contribution of the perturbative expansion of $\Gamma_k^{(4,1)}$ or $\Gamma_k^{(4,3)}$, but mixed contributions are not allowed. As a result, one expect the following structure (up to permutations of the external momenta):
\begin{equation}
\Gamma_{k,\vec{p}_1,\vec{p}_2,\vec{p}_3,\vec{p}_4,\vec{p}_5,\vec{p}_6}^{(6),(1,111)}=\alpha\vcenter{\hbox{\includegraphics[scale=0.8]{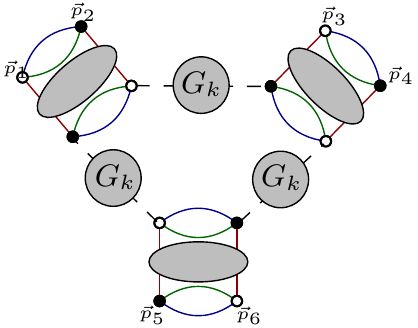}}}+\beta\vcenter{\hbox{\includegraphics[scale=0.8]{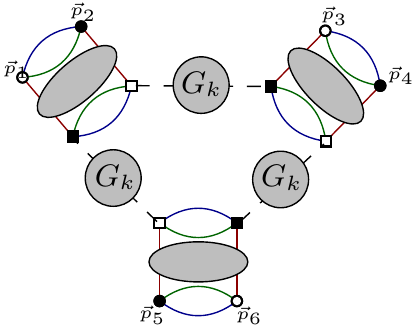}}}\,,\label{equation6}
\end{equation}
and it easy to check that any tree of the type of Figure \ref{figinter} is a term of the perturbative expansion of one among the two configurations pictured above. In this sense, the identification of both sides of the previous equation make sense only in the analyticity domain of $\Gamma_k^{(6)}$. The remaining coefficients $\alpha$ and $\beta$ may be computed by considering the first term of the pertubative expansion of $\Gamma_k^{(6), (1,111)}$. The diagrams are exactly of the form of those pictured on the right hand side of the equation \eqref{equation6}, and from Wick theorem one get straightforwardly:
\begin{equation}
\alpha=16\,,\qquad \beta=2\,.
\end{equation}
By reexpressing the diagramicall  equation \eqref{equation6} and setting to zero all the external momenta, one get:
\begin{equation}
\pi_{3}^{(1,111)}=16g_1^3(k) \mathcal{A}_{3,VVV} + 2c_1^3(k) \mathcal{A}_{3,WWW}\,, \label{six0}
\end{equation}
where we introduced $\mathcal{A}_{n,IJ\cdots K} $ as:
\begin{equation}
\mathcal{A}_{n,IJ\cdots K} := \sum_{\vec{p}_\bot\in\mathbb{Z}^{d-1}}\,G_{k,I}(\vec{p}_\bot)G_{k,J}(\vec{p}_\bot)\cdots G_{k,K}(\vec{p}_\bot)\,.
\end{equation}
In the same way, one get, for the remaining reduced effective vertices:
\begin{align}
\pi_{3}^{(1,222)}&=16g_2^3(k) \mathcal{A}_{3,WWW} + 2c_1^3(k) \mathcal{A}_{3,VVV}\,,\label{sixset1}\\
\pi_{3}^{(1,112)}&=2g_2(k)c_1^2(k) \mathcal{A}_{3,VVW}+4g_1^2(k)c_1(k) \mathcal{A}_{3,VVW}\,,\\
\pi_{3}^{(1,221)}&=2g_1(k)c_1^2(k) \mathcal{A}_{3,WWV}+4g_2^2(k)c_1(k) \mathcal{A}_{3,WWV}\,,\label{sixset2}\\
\pi_{3}^{(1,144)}&=2c_2^2(k)\left(2g_1(k)\mathcal{A}_{3,VVW}+c_1(k)\mathcal{A}_{3,WWV}\right)\,,\label{sixset3}\\
\pi_{3}^{(1,244)}&=2c_2^2(k)\left(2g_2(k)\mathcal{A}_{3,WWV}+c_1(k)\mathcal{A}_{3,VVW}\right)\,.\label{sixset4}
\end{align}

The set of effective melonic functions \eqref{sixset1},\eqref{sixset2}, \eqref{sixset3} and \eqref{sixset4} allows to close our hierarchical system. Indeed, from Ward identities of the proposition \eqref{Ward3}, we be  able to express the lacking derivatives $\partial \pi_2^{(i,a)}/\partial \vert p_1\vert$ involved in the expression of $\eta_I$ (equation \eqref{floweq2}) and express all the equations \eqref{flow3} in terms of local couplings only (thanks to the approximation \eqref{derivativeexp}). However in this way we lack the contribution of the Ward identities provided by corollary \ref{Ward1}. As discussed in \cite{Lahoche:2018oeo}, and as we will see in the next section, the relations provided by the corollary may be translated locally as a constraint between beta functions, and finally on the renormalization group flow as well. Instead of this constraint approach, ensuring that all melonic structure equations remains true along the flow, we will adopt another strategy in the next section, defining the six (and higher) effective vertices along the "physical" flow itself. 

\begin{center}
\includegraphics[scale=1.1]{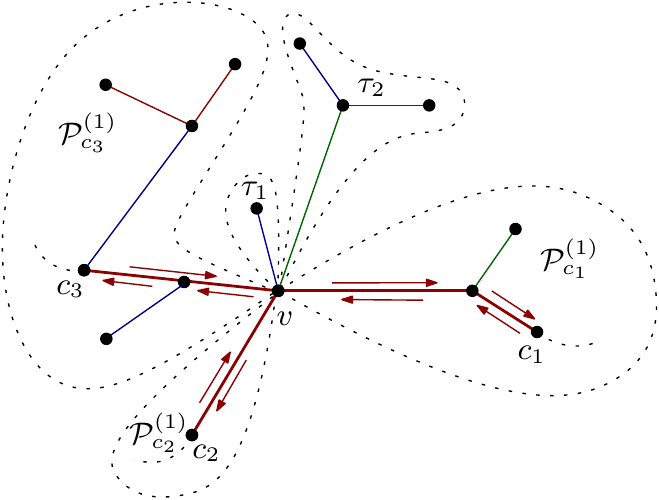} 
\captionof{figure}{Intermediate field representation of a typical 1PI melonic diagram contributing to $\pi_{3}^{(1,111)}$. The dotted cilia correspond to external edges, and some connected components are surrounded with dotted lines: the three monocolored arms $\mathcal{P}_{c_i}^{(1)}$ ending with ciliated vertex and the two connected components $\tau_i$. The red arrows picture the external monocolored faces running through the internior of the diagram from the boundary vertices.} \label{figinter}
\end{center}

\subsubsection{Description of the constrained melonic flow}

Let us consider the two relations of corollary \ref{Ward1}. Setting $k=0$, the regulator $r_k$ disappears, and the relations reduces to:
\begin{equation}
 \left\{
    \begin{array}{ll}
        2g_1^RZ_{0,VV}\mathcal{A}_{2,VV}(k=0)+c_1^R Z_{0,WW} \mathcal{A}_{2,WW}(k=0)=Z_{0,VV}-1\\
       2g_2^RZ_{0,WW} \mathcal{A}_{2,WW}(k=0)+c_1^R Z_{0,VV} \mathcal{A}_{2,VV}(k=0)=Z_{0,WW}-1\,,
    \end{array}
\right.\label{equationsWardIR}
\end{equation}
where the \textit{renormalized couplings} $g_{i}^R$ and $c_i^{R}$ are defined as:
\begin{equation}
g_i^R:=g_i(k=0)\,,\qquad c_i^R:=c_i(k=0)\,,
\end{equation}
and we chose $Z_{II}(k=0)=1$. The last condition ensuring the following behavior for small momenta in the deep IR limit:
\begin{equation}
G_{k=0,II}(\vec{p}\,)\sim \frac{1}{\vert \vec{p}\,\vert +m_{II}^R}\,, \qquad \vert \vec{p} \,\vert \ll 1\,,
\end{equation}
assuming the definition $m_{II}^R:=m_{II}(k=0)$. Note that the condition $Z_{II}(k=0)=1$\footnote{Note that, we have not to confuse $Z_{II}(k=0)$ and $Z_{0,II}$. The first one correspond to the end point of the RG trajectory, the second one is an initial condition ensuring the model to be well defined in the UV.} do not make sens necessarily if the flow is attracted toward a non-trivial fixed point. In this case, we can choose a referent scale $k_0$ far away from the deep UV sector to fix the renormalization condition. We left this irrelevant difficulty for our purpose. Interestingly, the set of coupled equations given by proposition \ref{Ward1} have to be compared with the closed system given by proposition \ref{closed}. Indeed, deriving them with respect to $p$, setting $p=0$ and taking into account the definition:
\begin{equation}
\frac{\partial \sigma^{(1)}_{II}}{\partial \vert p\,\vert }(p=0)=Z_{0,II}-Z_{II}(k)\,,
\end{equation}
we get:
\begin{equation}
 \left\{
    \begin{array}{ll}
        2g_1^B Z_{VV}(k) \mathcal{A}_{2,VV}(k)+c_1^B Z_{WW}(k)  \mathcal{A}_{2,WW}(k)=Z_{0,VV}-Z_{VV}(k)\\
       2g_2^B Z_{WW}(k) \mathcal{A}_{2,WW}(k)+c_1^BZ_{VV}(k)  \mathcal{A}_{2,VV}(k)=Z_{0,WW}-Z_{WW}(k)\,,
    \end{array}
\right. \label{contraint2}
\end{equation}
where we introduced the subscript $B$ for "bare" to clearly distinguish between bare and renormaized quantities. The equations \eqref{contraint2} and the corollary \ref{Ward1} provide two \textit{à priori} nonequivalent expressions for the difference $Z_{0,II}-Z_{II}(k)$. Equaling them together, one get (and the same exchanging $V\leftrightarrow W$):
\begin{equation}
 2g_1^B Z_{VV}(k) \mathcal{A}_{2,VV}(k)+c_1^B Z_{WW}(k)  \mathcal{A}_{2,WW}(k)= 2g_1(k) \mathcal{L}_{k,V}+c_1(k)  \mathcal{L}_{k,W}\,.
\end{equation}
Especially, setting $k=0$:
\begin{equation}
 \left\{
    \begin{array}{ll}
        2(g_1^B-g_1^RZ_{0,VV})  \mathcal{A}_{2,VV}(k=0)+(c_1^B-c_1^R Z_{0,WW})  \mathcal{A}_{2,WW}(k=0)=0\\
       2(g_2^B-g_2^RZ_{0,BB})  \mathcal{A}_{2,WW}(k=0)+(c_1^B-c_1^R Z_{0,VV})  \mathcal{A}_{2,VV}(k=0)=0\,,
    \end{array}
\right.\label{conservation}
\end{equation}
In a sense, these equations relate the end points of the RG flow history, the bare quantities being for the classical model in the UV and the renormalized ones for the classical model in the IR. They simply means that something is conserved along the flow, a conservation that we will translate locally, in the infinitesimal interval $[k,k+\delta k]$. Let us return on the pair of equations \eqref{equationsWardIR}. Solving in terms of $Z_{0,VV}$ and $Z_{0,WW}$, we get:
\begin{equation}
 \left\{
    \begin{array}{ll}
      Z_{0,VV}^{-1}=\frac{1- 2g_1^R\mathcal{A}_{2,VV}}{1+c_1^R Z_{0,WW} \mathcal{A}_{2,WW}}\\
      Z_{0,WW}^{-1}=\frac{1-2g_2^R\mathcal{A}_{2,WW}- \frac{(c_1^R)^2\mathcal{A}_{2,VV}\mathcal{A}_{2,WW}}{1- 2g_1^R\mathcal{A}_{2,VV}}}{1+ \frac{c_1^R\mathcal{A}_{2,VV}}{1- 2g_1^R\mathcal{A}_{2,VV}}}\,,
    \end{array}
\right.\label{equationsWardIRsolve}
\end{equation}
Expanding the coefficients $\mathcal{A}_{2,VV}$ and $\mathcal{A}_{2,WW}$ in power of couplings, we generate the log-divergent wave function counter-terms. For some fundamental cut-off $\Lambda$, one then expect that $\mathcal{A}_{2,II}\sim \ln(\Lambda)$, and a direct inspection of the equations \eqref{equationsWardIRsolve} show that $ Z_{0,VV}$ and $ Z_{0,WW}$ vanish as $1/\ln(\Lambda)$ in the \textit{continuum limit} $\Lambda\to \infty$, except for the special configurations:
\begin{equation}
4g_1^Rg_2^R=(c_1^R)^2\,. \label{regionbreak}
\end{equation}
In our consideration, we will consider a region so far from the region defined by \eqref{regionbreak}. \\

\noindent
Let us return on the equations given by the corollary \ref{Ward1}. Deriving with respect to $t=\ln(k)$, one get
\begin{equation}
2 \dot{g}_1  \mathcal{L}_{k,V}+ 2g_1 \left(Z_{0,VV} \dot{\mathcal{A}}_{2,VV}+\dot{\Delta}_{2,V}\right)+\dot{c}_1  \mathcal{L}_{k,W}+c_1\left(Z_{0,VV} \dot{\mathcal{A}}_{2,WW}+\dot{\Delta}_{2,W}\right)=0\,,\label{localconst1}
\end{equation}
and the same equation exchanging $V\leftrightarrow W$, the quantity ${\Delta}_{2,I}$ being defined as:
\begin{equation}
\Delta_{n,I}:= \sum_{\vec{p}_\bot}\frac{\partial r_{k,II}}{\partial \vert p_1\vert}(\vec{p}_\bot) G_I^n(\vec{p}_\bot)\,.
\end{equation}
There are some obstacle to simplify the equation \eqref{localconst1}. The big one is the computation of $\mathcal{A}_{k,II}$, which is a divergent quantity. One could expect that we can use of the approximation \eqref{derivativeexp}. However, such an approximation has to be showed very bad for quartic models with a single field   \cite{Lahoche:2018oeo}-\cite{Lahoche:2018vun}. In particular, the universal one loop asymptotic freedom is lost. The expected reason is that the approximation \eqref{derivativeexp}, used to solve the flow equation is valid only in the interior of the windows of momenta corresponding to the support of the distribution $\dot{r}_{k,II}$, whereas $\mathcal{A}_{k,II}$ involves an integration over all the momenta, from the very deep UV scale $\vert \vec{p}\,\vert \sim \Lambda$ to $\vert \vec{p}\,\vert \sim 0$, so far from the domain in which the approximation \eqref{derivativeexp} make sens. In the same paper the authors show that for superficially divergence free quantities like $\mathcal{A}_{n,I,J,\cdots,K}$ for $n>2$, however, the approximation \eqref{derivativeexp} is trusty. \\

\noindent
To skirt  the difficulty, we express $\mathcal{L}_{k,I}$ in terms of $Z_{II}$ and the effective couplings $g_1$, $g_2$ and $c_1$ from the corollary \ref{Ward1}. In the continuum limit, we obtain straightforwardly :
\begin{align}
\mathcal{L}_{k,V}&=\frac{c_1Z_{WW}-2g_2 Z_{VV}}{4g_2g_1-c_1^2}\,,\label{cond1}\\
\mathcal{L}_{k,W}&=\frac{c_1Z_{VV}-2g_1 Z_{WW}}{4g_2g_1-c_1^2}\,. \label{cond2}
\end{align}
In the continuum limit, we seen that $Z_{0,II}\to 0$. Moreover, $\dot{\mathcal{A}}_{2,II}$ is essentially the difference of two logarithmic divergent quantity, and because the integrated functions have the same behavior in the deep UV limit,  $\dot{\mathcal{A}}_{2,II}$ has to be finite in the continuum limit \footnote{This point has to be explicitly cheeked in  \cite{Lahoche:2018oeo}-\cite{Lahoche:2018hou}.}. Therefore: $Z_{0,II}\dot{\mathcal{A}}_{2,II} \to 0$, and from \eqref{cond1} and \eqref{cond2}:
\begin{equation}
 \left\{
    \begin{array}{ll}
       2 \dot{g}_1  \mathcal{L}_{k,V}+ 2g_1 \dot{\Delta}_{2,V}+\dot{c}_1  \mathcal{L}_{k,W}+c_1\dot{\Delta}_{2,W}=0\,,\\
       2 \dot{g}_2  \mathcal{L}_{k,W}+ 2g_2 \dot{\Delta}_{2,W}+\dot{c}_1  \mathcal{L}_{k,V}+c_1\dot{\Delta}_{2,V}=0\,.
    \end{array}
\right.\label{localconst2}
\end{equation}
Because the allowed windows of momenta for $\frac{\partial r_{k,II}}{\partial \vert p_1\vert}$ and $\dot{r}_{k,II}$ are essentially the same, the approximation \eqref{derivativeexp} used to solve the flow equation may be used to compute $\Delta_{2,I}$. The explicit computation is given on appendix \eqref{appC}. Finally, we are not really interested by the velocities of the effective couplings, but by the velocity of the dimensionless and renomalized quantities, that is to say, quantities for which we retain only the intrinsic part of the dynamic. For our model, the notion playing the role of dimension is called \textit{canonical dimension}, and refers to the optimal scaling of the quantum corrections for each couplings. From the results of appendix \ref{appA}, all the quartic couplings have zero canonical dimension, and the mass have canonical dimension equal to $1$. Note these dimension reflect nothing but the UV behavior of the RG flow, and are the direct consequence of the just-renormalizability of the model that we consider. The renormalized couplings, finally, are obtained by subtracting the contribution coming from the wave function renormalization. To summarize:
\begin{definition} \textbf{Dimensionless renormalized couplings}
In the deep UV, the dimensionless and renormalized couplings $\bar{g}_i$, $\bar{c}_i$ and $\bar{m}_{II}$ are defined as:
\begin{equation}
g_i=:Z_{II}^2 \bar{g}_i\,,\qquad c_i=:Z_{VV}Z_{WW} \bar{c}_i \,,\qquad m_{II}=kZ_{II} \bar{m}_{II}\,.
\end{equation}
\end{definition}
The velocity of the dimensionless renormalized couplings are the standard beta functions, that is to say : $\beta_{g_i}=\dot{\bar{g}}_i$, $\beta_{c_1}:=\dot{\bar{c}}_i$ and $\beta_{m_{II}}:=\dot{\bar{m}}_{II}$; and we have the following statement:
\begin{corollary}\label{constmelo}
In the continuum limit, and in the symmetric phase, the beta function of the renormalized couplings along the flow are related as:
\begin{align*}
       2 \beta_{g_1} \frac{\bar c_1-2\bar g_2}{4\bar g_2\bar g_1-\bar c_1^2}+{12\bar g_1}\left(\frac{1}{1+\bar{m}_{VV}}\right)^4\beta_{m_{VV}}&+\beta_{c_1}  \frac{\bar c_1-2\bar g_1 }{4\bar g_2\bar g_1-\bar c_1^2}+{6\bar c_1}\left(\frac{1}{1+\bar{m}_{WW}}\right)^4\beta_{m_{WW}}\\      
       &+\frac{\eta_{V}(5\bar c_1-2(4\bar g_1+\bar g_2))+\eta_W(\bar c_1-2\bar g_1)}{4\bar g_2\bar g_1-\bar c_1^2}=0\,,
\end{align*}  
\begin{align*}
       2 \beta_{g_2}  \frac{\bar c_1-2\bar g_1 }{4\bar g_2\bar g_1-\bar c_1^2}+ {12\bar g_2}\left(\frac{1}{1+\bar{m}_{WW}}\right)^4\beta_{m_{WW}}&+\beta_{c_1} \frac{\bar c_1-2\bar g_2 }{4\bar g_2\bar g_1-\bar c_1^2}+{6\bar c_1}\left(\frac{1}{1+\bar{m}_{VV}}\right)^4\beta_{m_{VV}}\\
       &+ \frac{\eta_{W}(5\bar c_1-2(4\bar g_2+\bar g_1))+\eta_V(\bar c_1-2\bar g_2)}{4\bar g_2\bar g_1-\bar c_1^2}=0\,.
\end{align*}
\end{corollary}
Note that the singularity \eqref{regionbreak} is the end version of the singularity $4\bar g_2(k)\bar g_1(k)-\bar c_1^2(k)$ occurring in each of these equations. They define a breakdown region, where our method do not make sense. In fact, we have to distinguish two region $4\bar g_2(k)\bar g_1(k)-\bar c_1^2(k) >0$ and $4\bar g_2(k)\bar g_1(k)-\bar c_1^2(k)<0$, we will return on this point in the next section. \\

\noindent
These equations introduce a non-trivial constraint along the flow, and have to be solved simultaneously with the flow equations. Indeed, they come from the Ward identity, which have to treated on the same footing as the flow equations. Indeed, they provide two complementary description of the global phase space, and to see why, it suffice to compare the equations \eqref{Wardspe} and \eqref{flowspe}. The same diagrams appears in two case. For the Ward identity \eqref{Wardspe}, the contraction involves a variation of the propagator with respect to the momentum, and the equation express the variation of the quantity $\pi_2^{(i,1)}$ with respect to $\vert p\vert$. In the same way, the flow equation \eqref{flowspe} 
 describes the evolution of $\pi_2^{(i,1)}$ with the change of scale $k$, and the contractions involves the variation of the propagator with respect to $k$. In both case, this is the non-trivial variation of the propagator which generate the change; and there are no reason to discard the Ward identity, especially because as it is easy to cheek, the two variations, with respect to $p$ and $k$ are not independent (see equation \eqref{floweq2})\footnote{This point is extensively discussed in \cite{Lahoche:2019cxt}-\cite{Lahoche:2019vzy}.}. \\

To close this section, therefore, we will provide two description of the melonic flow. The first one discarding the constraint, and the second one taking into account the constraint, and describing what we will call \textit{physical melonic phase space}. \\

\noindent
$\bullet$ \textbf{Unconstrained melonic flow.} As announced we start by a description of the unconstrained melonic flow. From the Ward equations given by the corollary \ref{Ward3}, we may obtain the derivative of the effective vertex and then complete the equation \eqref{floweq2}. The functions $\mathcal{L}_{k,I}$ have been computed before, and we have to compute the functions $\mathcal{U}_{k,I}$. Because it is a superficially convergent quantity, and from our previous discussion, it is suitable to use of the equation \eqref{derivativeexp}. In the continuum limit, we get from Appendix \ref{appC}:
\begin{equation}
\mathcal{U}_{k,I}\to\Delta_{3,I}=-2\frac{1}{Z_{II}^{2}}\frac{1}{k} \left(\frac{1}{1+\bar{m}_{II}}\right)^3\,. 
\end{equation}
Ultimately we are interested by the beta functions, that is to say, by dimensionless and renormalized quantity. To this end, we extend the notation "bar" for any function $X(\{g_i, c_i, m_I\})$, extracting of them their explicit dependence in $k$ and $Z_{I}$:
\begin{equation}
X(\{g_i, c_i, m_I\})=: k^{d_X} Z_{VV}^{\partial^{(V)}_X}Z_{WW}^{\partial^{(W)}_X} \bar{X}(\{\bar g_i, \bar c_i, \bar m_I\})\,.
\end{equation}
As for the couplings constants, we call \textit{canonical dimension} the exponent $d_X$, whereas we call \textit{heteroclite degree} the quantities $\partial^{(V)}_X$ and $\partial^{(W)}_X$. As an example:
\begin{equation}
\bar{\mathcal{U}}_{k,I}=-2\left(\frac{1}{1+\bar{m}_{II}}\right)^3\,.
\end{equation}
We have to proceed in the same way for the reduced effective vertices $\pi_{3}^{(1,abc)}$ given by equations \eqref{six0} and \eqref{sixset1}, \eqref{sixset2}, \eqref{sixset3} and \eqref{sixset4}. The computation of the dimensionless quantities $\bar{\mathcal{A}}_{3,IJK}$ are given in Appendix \ref{appC}, the result is:
\begin{align}
\bar{\mathcal{A}}_{3,IIJ}=\frac{2}{(1+\bar{m}_{II})^2}\frac{1}{1+\bar{m}_{JJ}}+\frac{4}{\bar{m}_{JJ}-\bar{m}_{II}} \left(\frac{\bar{m}_{JJ}}{\bar{m}_{JJ}-\bar{m}_{II} }\ln \left(\frac{1+\bar{m}_{JJ}}{1+\bar{m}_{II}}\right)-\frac{\bar{m}_{II}}{1+\bar{m}_{II}}\right)\,,
\end{align}
and:
\begin{equation}
\bar{\mathcal{A}}_{3,III}=\frac{2}{(1+\bar{m}_{II})^3}+4\left(-\frac{1}{1+\bar{m}_{II}}+2\frac{1}{(1+\bar{m}_{II})^2}\right)\,.
\end{equation}
As a result, the dimensionless effective $6$-points vertex functions writes as:
\begin{align}
\bar \pi_{3}^{(1,111)}&=16\bar g_1^3(k) \bar{\mathcal{A}}_{3,VVV} + 2\bar c_1^3(k) \bar{\mathcal{A}}_{3,WWW}\label{sixset20}\\
\bar\pi_{3}^{(1,222)}&=16\bar g_2^3(k) \bar{\mathcal{A}}_{3,WWW} + 2\bar c_1^3(k) \bar{\mathcal{A}}_{3,VVV}\,,\label{sixset21}\\
\bar\pi_{3}^{(1,112)}&=2\bar g_2(k)\bar c_1^2(k) \bar{\mathcal{A}}_{3,VVW}+4\bar g_1^2(k)\bar c_1(k) \bar{\mathcal{A}}_{3,VVW}\,,\\
\bar\pi_{3}^{(1,221)}&=2\bar g_1(k)\bar c_1^2(k) \bar{\mathcal{A}}_{3,WWV}+4\bar g_2^2(k)\bar c_1(k) \bar{\mathcal{A}}_{3,WWV}\,,\label{sixset22}\\
\bar\pi_{3}^{(1,144)}&=2\bar c_2^2(k)\left(2\bar g_1(k)\bar{\mathcal{A}}_{3,VVW}+\bar c_1(k)\bar{\mathcal{A}}_{3,WWV}\right)\,,\label{sixset23}\\
\bar\pi_{3}^{(1,244)}&=2\bar c_2^2(k)\left(2\bar g_2(k)\bar{\mathcal{A}}_{3,WWV}+\bar c_1(k)\bar{\mathcal{A}}_{3,VVW}\right)\,.\label{sixset24}
\end{align}
and from the flow equations \eqref{floweq1}, \eqref{floweq2}, \eqref{flow2} and \eqref{flow3}, we obtain, up to straightforward manipulations:
\begin{align}
\beta_{m_{VV}}&=-(1+\eta_V)\bar{m}_{VV}-6\bar g_{1} \bar I_{2\,,VV} -3\bar c_1 \bar I_{2\,,WW}\,,\\
\beta_{m_{WW}}&=-(1+\eta_W)\bar{m}_{WW}-6\bar g_{2} \bar I_{2\,,WW} -3\bar c_1 \bar I_{2\,,VV}\,,\\
\beta_{g_1}&=-2\eta_V \bar g_1 -3\bar\pi_3^{(1,111)} \bar I_{2,VV}-\bar\pi_3^{(1,112)} \bar I_{2,WW}+4\bar g_1^2 \bar I_{3,VV}+\bar c_1^2\bar I_{3,WW}\,,\\
\beta_{g_2}&=-2\eta_W \bar{g}_2 -3\bar\pi_3^{(1,222)} \bar I_{2,WW}-\bar\pi_3^{(1,122)}\bar I_{2,VV}+4 \bar g_2^2 \bar I_{3,WW}+\bar c_1^2\bar I_{3,VV}\,,\\
\beta_{c_1}&=-(\eta_V+\eta_W)\bar c_1 -2\bar \pi_3^{(1,112)} \bar I_{2,VV}-2\bar\pi_3^{(1,122)} \bar I_{2,WW}+4\bar c_1\left(\bar g_1 \bar I_{3,VV}+\bar g_2 \bar I_{3,WW}\right) \,,\\
\beta_{c_2}&=-(\eta_V+\eta_W) \bar c_2-\bar\pi_3^{(1,144)}\bar I_{2,VV}-\bar\pi_3^{(1,244)}\bar I_{2,WW}+\bar c_2^2 \bar I_{3,VW} \,,\label{flow4}
\end{align}
where the relevant effective vertices are given by equations \eqref{sixset20}--\eqref{sixset24}, and, from Appendix \ref{appC} ($\bar I_{n\,,II}\equiv \bar I_{n\,,II}(0)$):
\begin{equation}
\bar I_{n\,,II}(\vert q \vert )=\frac{\bar{J}_0(\vert q \vert)[\eta_I(1-\vert q \vert/k)+1]-\bar{J}_1(\vert q \vert)\eta_I}{( 1+\bar m_{II})^n} \,, \qquad \bar{J}_n(\vert q\vert )=2^2 \frac{(1-\vert q \vert/k)^{n+2}}{n+2}\,, \label{equationI}
\end{equation}
and, from definition \eqref{hetero}:
\begin{equation}
\bar I_{3,VW}(\vert q \vert)=\frac{\bar{J}_0(\vert q \vert)[\eta_V(1-\vert q \vert/k)+1]-\bar{J}_1(\vert q \vert)\eta_V}{( 1+\bar m_{VV})^2( 1+\bar m_{WW})} +\frac{\bar{J}_0(\vert q \vert)[\eta_W(1-\vert q \vert/k)+1]-\bar{J}_1(\vert q \vert)\eta_W}{( 1+\bar m_{WW})^2( 1+\bar m_{VV})} \,.
\end{equation}
To get the explicit expression for $\eta_I$ from equation \eqref{floweq2}, we have to compute $\bar I_{2\,,II}^\prime(\vert q \vert =0)$:
\begin{equation}
I_{2\,,II}^\prime(\vert q \vert =0)=\frac{\bar J_0^\prime(0)(\eta_I+1)-(\bar J_0(0)+\bar J_1^\prime(0))\eta_I}{(1+\bar m_{II})^2} \,,\quad \bar{J}^\prime_n(0)=(2+n)\bar{J}_n(0)\,. 
\end{equation}
Solving in $\eta_I$, the equation \eqref{floweq2}, we get:
\begin{equation}
\eta_I=\frac{C_IB_{\hat{I}}-AC_{\hat{I}}}{B_VB_W-A^2}\,,\label{eta}
\end{equation}
where :
\begin{align}
A&:=\left(\bar\pi_{2,\,00}^{(1,3)}\right)^\prime\left(\bar{J}_1-\bar{J}_0\right)+\bar{c}_1\left[(\bar{J}_0+\bar{J}_1^\prime)-\bar{J}_0^\prime\right]\,,\\
B_I&:= \left(\bar\pi_{2,\,00}^{(1,n(I))}\right)^\prime \left(\bar{J}_1-\bar{J}_0\right)+2g_{n(I)}\left[(\bar{J}_0+\bar{J}_1^\prime)-\bar{J}_0^\prime\right]-(1+\bar{m}_{II})^3\,,\\
C_I&:=\left( 2\left(\bar\pi_{2,\,00}^{(1,n(I))}\right)^\prime+ \left(\bar\pi_{2,\,00}^{(1,3)}\right)^\prime \right)\bar J_0+\left(2g_{n(I)}+c_1\right)\bar J^\prime_0\,.
\end{align}

\noindent
$\bullet$ \textbf{Constrained melonic flow.}
The set of equations derived above describe the behavior of the minimal relevant couplings in the UV sector, describing the renormalization group flow toward the infinite melonic subset. However, they completely ignore the corollary \ref{constmelo}. As explained, a solution to take into account this corollary is to consider the relations between beta functions as constraint along the RG flow, and to keep only the solutions of the flow equations which satisfy them. This restriction define a subset of the full melonic phase space corresponding to that we called physical melonic phase space. However, the description of the physical phase space is not the more practical. Indeed, even with a single field like in \cite{Lahoche:2018hou}, taking into account the constraint in this way provide an inelegant description. But we can remark that in this description, what we conserve is precisely the $6$-point melonic equations. We impose that the set of equations \eqref{six0}-\eqref{sixset4} remain true along the flow. The description that we propose in this section relax this unessential constraint. More precisely, we use of the two relations provided by corollary \ref{constmelo} to define $\beta_{g_1}$ and $\beta_{g_2}$:
\begin{align}
   \nonumber    \beta_{g_1} =-{6\bar g_1}\frac{4\bar g_2\bar g_1-\bar c_1^2}{\bar c_1-2\bar g_2}\left(\frac{1}{1+\bar{m}_{VV}}\right)^4\beta_{m_{VV}}&-\frac{\beta_{c_1}}{2}  \frac{\bar c_1-2\bar g_1 }{\bar c_1-2\bar g_2}\\
       -{3\bar c_1}\frac{4\bar g_2\bar g_1-\bar c_1^2}{\bar c_1-2\bar g_2}\left(\frac{1}{1+\bar{m}_{WW}}\right)^4\beta_{m_{WW}}      
       &-\frac{\eta_{V}(5\bar c_1-2(4\bar g_1+\bar g_2))+\eta_W(\bar c_1-2\bar g_1)}{2(\bar c_1-2\bar g_2)}\,, \label{contrainte1}
\end{align}  

\begin{align}
   \nonumber    \beta_{g_2} =-{6\bar g_2}\frac{4\bar g_2\bar g_1-\bar c_1^2}{\bar c_1-2\bar g_1}\left(\frac{1}{1+\bar{m}_{WW}}\right)^4\beta_{m_{WW}}&-\frac{\beta_{c_1}}{2}  \frac{\bar c_1-2\bar g_2 }{\bar c_1-2\bar g_1}\\
       -{3\bar c_1}\frac{4\bar g_2\bar g_1-\bar c_1^2}{\bar c_1-2\bar g_1}\left(\frac{1}{1+\bar{m}_{VV}}\right)^4\beta_{m_{VV}}      
       &-\frac{\eta_{W}(5\bar c_1-2(4\bar g_2+\bar g_1))+\eta_V(\bar c_1-2\bar g_2)}{2(\bar c_1-2\bar g_1)}\,.\label{contrainte2}
\end{align}  
Then, we keep all the equations of the set \eqref{flow4}, except the ones given $\beta_{g_1}$ and $\beta_{g_2}$ which are now given by \eqref{contrainte1} and \eqref{contrainte2}. The old  equations for $\beta_{g_1}$ and $\beta_{g_2}$ become the constraint, but not on the flow of the relevant coupling constant. The constraint define the effective vertices $\pi_3^{(1,111)}$ and $\pi_3^{(1,222)}$ along the physical flow. In other word, we replaced the constraint along the flow of the relevant couplings \textit{with} structure equations by a constrained flow relaxing the structure equations for two effective vertices, which are now defined by the flow itself. Moreover, it is easy to cheek that this construction does not break the closure of the flow equations ! The physical effective vertices that we call $\Pi_3^{(1,111)}$ and $\Pi_3^{(1,222)}$ to distinguish them from the unconstrained ones \eqref{six0} and \eqref{sixset1} being fixed, there beta function may be found taking the derivation with respect to $k$. Equaling the result with the corresponding flow equation obtained from the Wetterich equation \eqref{Wett}, we fix some $8$-point effective vertices, and so one.  We will investigate numerically the unconstrained and constrained melonic flows in the next section. Note that, even two close this section that our equations have some singularities. First of all, a singularity for $\bar{m}_{II}=-1$, inherited from the restriction to the symmetric phase. Moreover, corollary \ref{constmelo} add the condition $4\bar g_2\bar g_1-\bar c_1^2\neq 0$, which define two regions $I_{\pm}$ respectively for $4\bar g_2\bar g_1-\bar c_1^2> 0$ and $4\bar g_2\bar g_1-\bar c_1^2< 0$. Finally, the definition of the anomalous dimension \eqref{eta} provides a new restriction :
\begin{equation}
B_VB_W-A^2>0\,,
\end{equation}
assuming that the physical region has to contain the Gaussian fixed point, i.e. the point at which $\bar{g}_i=\bar{c}_i=\bar{m}_{II}=0$.

\subsection{Finite-dimensional vertex expansion and multi-trace interactions}\label{truncation}

In this section, we obtain the system of $\beta$-functions for the heteroclite tensorial group field theory within the simplest truncation containing all the symmetric quartic interactions mixing the tensor fields $V$ and $W$. As recalled before, the vertex expansion method is finite dimensional; and assume a systematic projection into the finite dimensional phase space parametrized by the truncated average action. This approach has proved its powerful, especially for treating branching and multi-trace interactions; which lack to form a closed sector, as required by the EVE method. Nevertheless, those terms will be generated along the renormalization group flow unavoidably, and with this respect, the understanding of how the couplings associated to these interactions couple to the full system of $\beta$-functions is of uttermost importance. Note that as in the previous section, we focus on the deep UV region ($k\gg 1$), ensuring that flow equations must be rewritten as an autonomous system for dimensionless couplings. 

\subsubsection{Setting up the vertex expansion}

In order to benefit of the efficiency of the vertex expansion method, we consider the following quartic truncation $\Gamma_k$, taking into account disconnected contributions:

\begin{eqnarray}
\Gamma_k =\Gamma_{k,\text{kin}}+\Gamma^{(V)}_{k}+\Gamma^{(W)}_{k}+\Gamma^{(V,W)}_{k}\,,
\label{trunc1}
\end{eqnarray}
where $\Gamma_{k,\text{kin}}$ is a diagonal matrix in the $\{V$-$W\}$-space, with entries $\Gamma^{(2)}_{II}$ given by equation \eqref{derivativeexp}, and the quartic interactions $\Gamma^{(V)}_{k}$,$\Gamma^{(W)}_{k}$ and $\Gamma^{(V,W)}_{k}$, including mixing couplings, are graphically defined as:
\begin{equation}
\Gamma^{(V)}_{k}=g_1 \left\{ \sum_{i=1}^d \vcenter{\hbox{\includegraphics[scale=1]{vert1.pdf}  }}  \right\}+ {\kappa}_{1} \left\{\vcenter{\hbox{\includegraphics[scale=1]{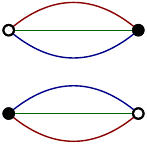}  }} \right\}\,,
\end{equation}
\begin{equation}
\Gamma^{(W)}_{k}=g_2 \left\{ \sum_{i=1}^d \vcenter{\hbox{\includegraphics[scale=1]{vert2.pdf}  }}  \right\}+ {\kappa}_{2} \left\{\vcenter{\hbox{\includegraphics[scale=1]{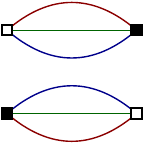}  }} \right\}\,,
\end{equation}
and :
\begin{equation}
\Gamma^{(VW)}_{k}=c_1 \left\{ \sum_{i=1}^d \vcenter{\hbox{\includegraphics[scale=1]{vert6.pdf}  }}  \right\}+c_2\left\{ \sum_{i=1}^d \vcenter{\hbox{\includegraphics[scale=1]{vert9.pdf}  }}  \right\}+c_0 \left\{ \vcenter{\hbox{\includegraphics[scale=1]{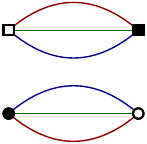}  }} \right\}+c_3 \left\{ \vcenter{\hbox{\includegraphics[scale=1]{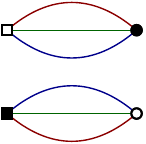}  }} \right\}\,. 
\end{equation}
The truncated effective action \eqref{trunc1} provides a parametrization of the theory space going beyond the connected melonic sector that we considered in the previous sections with the EVE method. Note that the list of quartic interaction is then complete for tensor fields of rank three. We use the same regulator as in the section \ref{sectionEVE}, that is, the Litim-like regulator, building as a diagonal in the $V-W$ coordinates. More precisely, in matrix form, the regulator kernel $\mathbb{R}_k (\vec{p},\vec{q}\,)$ must be expressed as:
\begin{equation}
\mathbb{R}_k (\vec{p},\vec{q}\,) = \delta_{\vec{p},\vec{q}}\begin{pmatrix}
R_{V,k} & 0\\ 
0 &R_{W,k}
\end{pmatrix}\,,\nonumber
\label{trunc5}
\end{equation}
where the functions $R_{I,k} $ are defined as:
\begin{equation}
R_{I,k} := Z_{II}\left[k - \sum^3_{i=1}|p_i|\right]\theta\left(k - \sum^3_{i=1}|p_i|\right)
\label{trunc5a}
\end{equation}
and the scale-derivative of the regulator yields
\begin{equation}
\dot{R}_{I,k} = \dot{Z}_{II}\left[k - \sum^3_{i=1}|p_i|\right]\theta\left(k - \sum^3_{i=1}|p_i|\right)+k Z_{II}\,\theta\left(k - \sum^3_{i=1}|p_i|\right)\,,
\label{trunc6}
\end{equation}
with once again $I$ take values $I=V,W$. Following the projection method, we demand that the truncated action \eqref{trunc1} satisfies the full Wetterich equation \eqref{Wett}. 
An alternative way to the successive functional derivatives considered in the previous section is to expand the right hand side of the flow equation as a power series in fields (vertex expansion), namely:
\begin{equation}
\partial_t \Gamma_k = \mathrm{Tr}\left[(\partial_t \mathbb{R}_k)P^{-1}_k+\sum^{\infty}_{n=1}(-1)^{n}(\partial_t\mathbb{R}_{k})P^{-1}_{k}(\mathbb{H}P^{-1}_{k})^n\right]\,,
\label{trunc7}
\end{equation}
the big matrix $\mathbb{H} $ writing in the $V-W$ space as :
\begin{equation}
\mathbb{H} = \begin{pmatrix}
\Gamma^{\mathrm{int}}_{{V}V} & \Gamma^{\mathrm{int}}_{{V}W}\\ 
\Gamma^{\mathrm{int}}_{{W}V} & \Gamma^{\mathrm{int}}_{{W}W} 
\end{pmatrix}\,,
\label{trunc8}
\end{equation}
where we defined $\Gamma^{\mathrm{int}}_{k}$ as  $\Gamma^{\mathrm{int}}_{k} = \Gamma^{(V)}_k+\Gamma^{(W)}_k+\Gamma^{(V,W)}_k$, $\Gamma^{\mathrm{int}}_{IJ}$ denoting the derivative of $\Gamma^{\mathrm{int}}_{k}$ with respect to $I$ and $J$ fields, and where the matrix valued regulated propagator $P^{-1}_k$ is defined for vanishing classical fields as
\begin{equation}
P_k := \Gamma^{(2)}_k\Big|_{(\Phi_V,\Phi_W)=(0,0)} + \mathbb{R}_{k}\,.
\label{trunc9}
\end{equation}
Explicitly, using matrix notation in the $V-W$ space, we get:
\begin{equation}
P_k (\vec{p},\vec{q}\,)= \delta_{\vec{p},\vec{q}}\begin{pmatrix}
\,k Z_{VV}+{m}_{VV}& 0\\ 
0 & \,k Z_{WW}+{m}_{WW} 
\end{pmatrix}
\label{trunc10}
\end{equation}
The first term in eq.\eqref{trunc7} is field independent and therefore it is just a vacuum contribution which can be disregarded. Within our truncation, we must consider the expansion up to $n=2$ order since this will generate quartic terms in the fields. The projection of the results at $n=1$ to the left-hand side of the flow equation \eqref{trunc7} will provide the running of the masses as well as the anomalous dimensions while at $n=2$ we can extract the beta functions of the couplings for each interaction class. In the following, we present the analysis for each order $n=1$ and $n=2$ separately.

\subsubsection{Running of masses and anomalous dimensions}   

Computing the $\beta$-functions for masses $(m_{VV},m_{WW})$ and running wave function normalization $(Z_{VV},Z_{WW})$ follows the same strategy as in section \ref{sectionEVE}. In the vertex expansion formalism described just above, we have to compute the $n=1$ contribution of equation.\eqref{trunc7}, that is to say:
\begin{equation}
\partial_t \Gamma_k\Big|_{n=1}=-\mathrm{Tr}\left[(\partial_t\mathbb{R}_{k})P^{-1}_{k}\mathbb{H}P^{-1}_{k}\right]\,.
\label{trunc11}
\end{equation}
Due to the diagonal structure of the regulated propagators, eq.\eqref{trunc11} can be factorized as
\begin{equation}
\partial_t \Gamma_k\Big|_{n=1} = - \sum_{I=V,W}\mathrm{Tr}\left[\partial_{t}R_{I,k}(k Z_{II}+{m}_{II})^{-2}\Gamma^{\mathrm{int}}_{II}(\vec{p}\,^\prime,\vec{p})\right]\,,
\label{trunc12}
\end{equation}
or explicitly:
\begin{equation}
\partial_t \Gamma_k\Big|_{n=1} = - \sum_{I=V,W}\mathrm{Tr}\left[\left(\beta^{I}+\alpha^{I}K\right)\theta\left(k-\sum^{3}_{i=1}|p_i|\right)\Gamma^{\mathrm{int}}_{II}(\vec{p}\,^\prime,\vec{p})\right]\,,
\label{trunc13}
\end{equation}
with $K \equiv \sum^{3}_{i=1}|p_i|$,
\begin{equation}
\alpha^{I} = - Z_{II}\frac{\eta_I}{[ k  Z_{II}+{m}_{II}]^2}\,,
\label{trunc14}
\end{equation}
and
\begin{equation}
\beta^{I} = k  Z_{II}\frac{1+\eta_I}{[kZ_{II}+{m}_{II}]^2}\,,
\label{trunc15}
\end{equation}
where we used of the definition of the anomalous dimension. The explicit evaluation of the traces over the tensor indices can be performed directly employing the techniques described in the previous section, i.e. taking second derivative with respect to the classical field, and setting them to zero at the end of the computation. In regard to the connected contributions, the result is exactly the same as for equations \eqref{flow4}. For the contributions involving disconnected couplings however, the computation have to be done, and it is not so hard to check that only the four diagrams pictured on Figure \ref{discomass} must provide a relevant contribution in the large $k$ limit for the flows of $\bar{m}_{VV}$ and $\bar{m}_{WW}$. What is unclear for now is the canonical dimension of the disconnected couplings. The canonical dimension may be fixed from the general argument stressed in \cite{Lahoche:2018oeo}, considering the behavior of the RG flow in the vicinity of the Gaussian fixed point. Another and consistent way to fix the canonical dimension is to impose the scaling of the disconnected couplings with respect to $k$ in such a way that the flow equations become an autonomous system in the large-$k$ limit. 

\begin{center}
\begin{equation*}
\vcenter{\hbox{\includegraphics[scale=1.2]{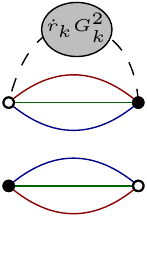} }}\quad \vcenter{\hbox{\includegraphics[scale=1.2]{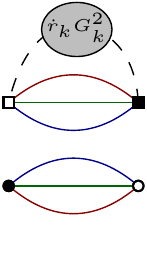} }} \qquad \vcenter{\hbox{\includegraphics[scale=1.2]{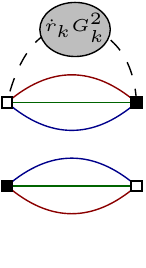} }}\quad \vcenter{\hbox{\includegraphics[scale=1.2]{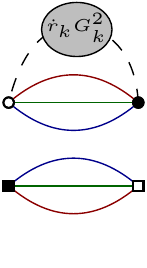} }}
\end{equation*}
\captionof{figure}{The four relevant contractions contributing to the mass RG flow in the large $k$ limit. The two contractions on right contribute to $\dot{\bar{m}}_{VV}$ whereas the two lat ones contribute to $\dot{\bar{m}}_{WW}$.  } \label{discomass}
\end{center}

The computation of the new loop integrals is a straightforward variation from the computation of section \ref{sectionEVE}. The only difference with respect to the computation of $J_n$ is that we have one additional internal face, and therefore one additional integration. For instance, up to numerical factors, the weight associated to the sum on the first diagrams of Figure \ref{discomass} is, in the large $k$ limit:
\begin{equation}
\vcenter{\hbox{\includegraphics[scale=1]{vertdecVcont.pdf} }} \sim \sum_{\vec{p}\in \mathbb{Z}^d}  \frac{\dot{r}_k(\vec{p}\,)}{(Z_{VV} \vert \vec{p}\,\vert+m_{VV}+(r_k)_{VV}(\vec{p}\,))^2} \,,
\end{equation}
which motivate the definition of the integral kernel:
\begin{equation}
K_{n\,,II}:=Z_{II}\frac{G_0[\eta_I+1]k-G_1\eta_I}{(Z_{II} k+m_{II})^n} \,,
\end{equation}
with, in terms of the continuous variables $x_i:= p_i/k$:
\begin{equation}
G_n:= \int d^3x \vert \vec{x} \,\vert^n \theta(k-\vert x_1 \vert -\vert x_2 \vert-\vert x_3 \vert)=\frac{8}{n+3}\,k^{n+3}\,,
\end{equation}
where we used the same strategy described in Appendix \ref{appC} to compute the integral. From the behavior of this integral, it is in fact no hard to check that the canonical dimension for \textit{all} disconnected interactions must be fixed to $-1$. As a result, we are now in position to write down the flow equations for masses. Let us consider for instance the $\beta$-function for $m_{VV}$:
\begin{equation}
\beta_{m_{VV}}=-(1+\eta_V)\bar{m}_{VV}-6\bar g_{1} \bar I_{2\,,VV} -3\bar c_1 \bar I_{2\,,WW}-2\bar{\kappa}_1 \bar{K}_{2\,,VV}-\bar{c}_0 \bar{K}_{2\,,WW}\,,
\end{equation}
where the three first terms are nothing but the ones that we get in the previous section, and the last two terms provide the disconnected contributions; the numerical factors may be easily computed: There are two kind of independent contractions for the interaction corresponding to the coupling $\bar{\kappa}_V$, and a single one for $\bar{c}_0$. Moreover, there are no summation over colors for disconnected interactions. Note that the accordance with the canonical dimension expected for disconnected couplings may be checked directly on this expression. Indeed, while $I_{2\,,VV}$ scale as $k^2$, $ \bar{K}_{2\,,VV}$ scale as $k^3$, then, disconnected couplings must be scale as $k^{-1}$ with respect to the connected quartic couplings. Obviously, due to the $V\leftrightarrow W$ symmetry of the model, the equation for $ \beta_{m_{WW}}$ may be easily deduced:
\begin{equation}
\beta_{m_{WW}}=-(1+\eta_W)\bar{m}_{WW}-6\bar g_{1} \bar I_{2\,,WW} -3\bar c_1 \bar I_{2\,,VV}-2\bar{\kappa}_2 \bar{K}_{2\,,WW}-\bar{c}_0 \bar{K}_{2\,,VV}\,.
\end{equation}
The anomalous dimension may be computed in the same way. From equation \eqref{equationZ}, it is no hard to deduce that:
\begin{equation}
Z_{II}\,\eta_I:= -2g_{n(I)}(k) I_{2\,,II}^\prime(0) -c_1(k) I_{2\,,\hat{I}\hat{I}}^\prime(0)\,.\label{floweq2prime}
\end{equation}
Note that there are no contribution coming from disconnected interactions, because any external momenta flows through the effective loop (this is why $K_{n,II}$ does not depends on $q$). From the explicit expression of $I_{2\,,II}^\prime(0)$, equation \eqref{equationI}, it is straightforward to solve in $\eta_I$, leading to:
\begin{equation}
\eta_I=\frac{\tilde{C}_I\tilde{B}_{\hat{I}}-\tilde{A}\tilde{C}_{\hat{I}}}{\tilde{B}_V\tilde{B}_W-\tilde{A}^2}\,,\label{eta2}
\end{equation}
where :
\begin{align}
\tilde{A}&:=\bar{c}_1\left[(\bar{J}_0+\bar{J}_1^\prime)-\bar{J}_0^\prime\right]\,,\\
\tilde{B}_I&:= 2g_{n(I)}\left[(\bar{J}_0+\bar{J}_1^\prime)-\bar{J}_0^\prime\right]-(1+\bar{m}_{II})^3\,,\\
\tilde{C}_I&:=\left(2g_{n(I)}+c_1\right)\bar J^\prime_0\,.
\end{align}
which is nothing but the equation \eqref{eta} with vanishing vertex derivatives. We have then obtained all the flow equations for two-point relevant parameters. The computation of the beta functions of the couplings of the quartic interactions requires the computation of the $n=2$ contributions from equation \eqref{trunc7}. Those will be reported in the next subsection.

\subsubsection{Running of quartic couplings}
The computation of the running of quartic couplings requires the expansion of \eqref{trunc7} up to $n=2$, the reason being that due to the fact that the truncation is limited to quartic truncations, two insertions of $2$-point functions are necessary to do a relevant contribution with four tensors fields. Explicitly, this is achieved by equation \eqref{flowfour}, setting $\Gamma^{(6)}_k=0$. In the vertex expansion approach, at order $n=2$, the contribution we want to evaluate is given by :
\begin{equation}
\partial_t \Gamma_k\Big|_{n=2}=\mathrm{Tr}\left[(\partial_t\mathbb{R}_{k})P^{-1}_{k}\mathbb{H}P^{-1}_{k}\mathbb{H}P^{-1}_k\right]\,.
\label{n2cont1}
\end{equation}
As before, the diagonal structure of the regulator leads to the following expression,
\begin{eqnarray}
\partial_t \Gamma_k\Big|_{n=2} &=& \mathrm{Tr}\left[(\partial_t R^{(V)}_{k})P^{-1}_{k,V}\Gamma^{\mathrm{int}}_{VV}P^{-1}_{k,V}\Gamma^{\mathrm{int}}_{VV}P^{-1}_{k,V}\right]\nonumber\\
&+&\mathrm{Tr}\left[(\partial_t R^{(W)}_{k})P^{-1}_{k,W}\Gamma^{\mathrm{int}}_{WW}P^{-1}_{k,W}\Gamma^{\mathrm{int}}_{WW}P^{-1}_{k,W}\right]\nonumber\\
&+&\mathrm{Tr}\left[(\partial_t R^{(V)}_{k})P^{-1}_{k,V}\Gamma^{\mathrm{int}}_{VW}P^{-1}_{k,W}\Gamma^{\mathrm{int}}_{WV}P^{-1}_{k,V}\right]\nonumber\\
&+& \mathrm{Tr}\left[(\partial_t R^{(W)}_{k})P^{-1}_{k,W}\Gamma^{\mathrm{int}}_{WV}P^{-1}_{k,V}\Gamma^{\mathrm{int}}_{VW}P^{-1}_{k,W}\right]\,.
\end{eqnarray}
The explicit evaluation of the traces as well as a suitable projection into the different combinatorial structures yield systematically all the $\beta$-functions for each of the couplings involved in the truncation. First of all, the equations for connected couplings must have no additional contributions arising from disconnected interactions with respect to the EVE formula \eqref{flow4}, vanishing the $\Gamma^{(6)}_k$ contribution (for instance, the relevant diagrams for $\dot{g}_1$ remain the ones given by equation \eqref{flowspe}). Therefore, we must have:

\begin{align}
\beta_{g_1}&=-2\eta_V \bar g_1 +4\bar g_1^2 \bar I_{3,VV}+\bar c_1^2\bar I_{3,WW}\,,\\
\beta_{g_2}&=-2\eta_W \bar{g}_2 +4 \bar g_2^2 \bar I_{3,WW}+\bar c_1^2\bar I_{3,VV}\,,\\
\beta_{c_1}&=-(\eta_V+\eta_W)\bar c_1 +4\bar c_1\left(\bar g_1 \bar I_{3,VV}+\bar g_2 \bar I_{3,WW}\right) \,,\\
\beta_{c_2}&=-(\eta_V+\eta_W) \bar c_2+\bar c_2^2 \bar I_{3,VW} \,.\label{flownew}
\end{align}
Therefore, we have only to check the $\beta$-functions for $\kappa_I$, $c_0$ and $c_3$. Let us consider $\beta_{\kappa_V}$ for instance. A direct inspection of the different allowed contractions leads to the following diagrammatic equation:
\begin{align}
\nonumber 4\dot{\kappa}_1&=4\left\{ 4\kappa_1^2\, \vcenter{\hbox{\includegraphics[scale=0.8]{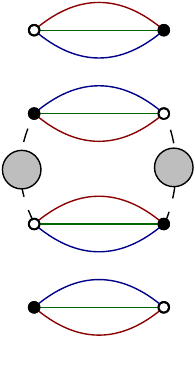} }}   + c_0^2\, \vcenter{\hbox{\includegraphics[scale=0.8]{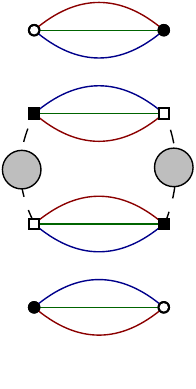} }}   \right\}+4\frac{d(d-1)}{2}\left\{ 4g_1^2\, \vcenter{\hbox{\includegraphics[scale=0.8]{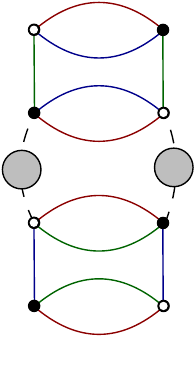} }}   + c_1^2\, \vcenter{\hbox{\includegraphics[scale=0.8]{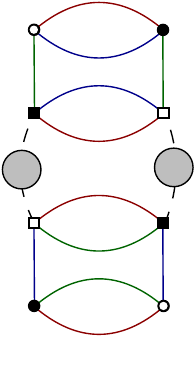} }}   \right\}\\
&\qquad  4d \left\{ 2g_1\kappa_1\, \vcenter{\hbox{\includegraphics[scale=0.8]{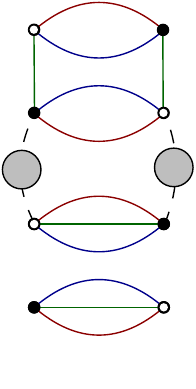} }}   + c_0c_1\, \vcenter{\hbox{\includegraphics[scale=0.8]{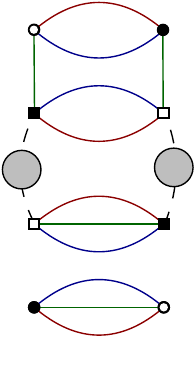} }}   \right\}\,,
\end{align}
where we keep the notation $d$ for the rank of the tensor, to highlight the nature of the numerical coefficient. Taking into account the canonical dimension and the $[{\kappa}_1]=-1$, we easily get the equation for $\beta_{\kappa_1}$:
\begin{align}
\nonumber \beta_{\kappa_1}=&(1-2\eta_V)\bar{\kappa}_1+ 4 \bar{\kappa}_1^2 \bar{K}_{3,VV}+\bar{c}_0^2 \bar{K}_{3,WW} +12 \bar{g}_1^2 \bar{Q}_{3,VV}(0)+3\bar{c}_1^2  \bar{Q}_{3,WW}(0)\\
&\qquad +6\bar{g}_1\bar{\kappa}_1  \bar{I}_{3,VV}(0)+3\bar{c}_0\bar{c}_1  \bar{I}_{3,WW}(0)\,.
\end{align}
The equation for $\beta_{\kappa_2}$ may be immediately deduced from the $W\leftrightarrow V$ symmetry:
\begin{align}
\nonumber \beta_{\kappa_2}=&(1-2\eta_W)\bar{\kappa}_2+ 4 \bar{\kappa}_2^2 \bar{K}_{3,WW}+\bar{c}_0^2 \bar{K}_{3,VV} +12 \bar{g}_2^2 \bar{Q}_{3,WW}(0)+3\bar{c}_1^2  \bar{Q}_{3,VV}(0)\\
&\qquad +6\bar{g}_2\bar{\kappa}_2  \bar{I}_{3,WW}(0)+3\bar{c}_0\bar{c}_1  \bar{I}_{3,VV}(0)\,.
\end{align}
where the new loop kernel $Q_n$ is defined as:
\begin{equation}
Q_{n\,,II}:=Z_{II}\frac{H_0[\eta_I+1]k-H_1\eta_I}{(Z_{II} k+m_{II})^n} \,,
\end{equation}
with, in terms of the continuous variables $x_i:= p_i/k$:
\begin{equation}
H_n:= \int dx \vert \vec{x} \,\vert^n \theta(k-\vert x_1 \vert )=\frac{2}{n+1}\,k^{n+1}\,,
\end{equation}
In the same way, counting the number of relevant contractions following their respective combinatorics, we deduce the flow equations for $c_0$ and $c_3$:
\begin{align}
\nonumber \beta_{c_0}&=(1-\eta_V-\eta_W)\bar{c}_0+4\bar{c}_0\bar{\kappa}_1 \bar{K}_{3,VV}+4\bar{c}_0\bar{\kappa}_2 \bar{K}_{3,WW}+12\bar{c}_1\left(\bar{\kappa}_1 I_{3,VV}+\bar{\kappa}_2 I_{3,WW} \right)\\
& +12\bar{c}_0\left(\bar{g}_1 I_{3,VV}+\bar{g}_2 I_{3,WW} \right)+12 \bar{c}_1\left( \bar{g}_1 Q_{3,VV}  + \bar{g}_2 Q_{3,WW}  \right)\,,
\end{align}
\begin{align}
\nonumber \beta_{c_3}&=(1-\eta_V-\eta_W)\bar{c}_3+2\bar{c}_3^2 K_{3,VW}+12\bar{c}_3\bar{c}_2 I_{3,VW} +12\bar{c}_2^2 Q_{3,VW}\,,
\end{align}
where $I_{3,VW}$ has been defined in equation \eqref{hetero}, and:
\begin{equation}
K_{3,VW}:=\sum_{\vec{p}} \left(\dot{r}_{VV}G^2_{k,VV}G_{k,WW}+\dot{r}_{WW}G^2_{k,WW}G_{k,VV}\right)(\vec{p}\,)\,,\label{hetero2}
\end{equation}
\begin{equation}
Q_{3,VW}:=\sum_{\vec{p}} \delta_{p_10}\delta_{p_20}\left(\dot{r}_{VV}G^2_{k,VV}G_{k,WW}+\dot{r}_{WW}G^2_{k,WW}G_{k,VV}\right)(\vec{p}\,)\,,\label{hetero3}
\end{equation}
We have now all the theoretical material to address the numerical analysis of the RG flow. The next section provides a first look on these difficult investigations, which, as announced in the introduction will be reported for an incoming work.

\section{Preliminary numerical  investigations, discussion and conclusion}\label{secnum}

As announced, this section is devoted to a first look at the numerical investigations of the flow equations obtained in the previous sections. But before to come on this analysis, we briefly summarize the results of the previous sections:
\begin{itemize}
\item We defined a non conventional TGFT, mixing two complex tensors interacting together. For this field theory, we will investigate nonperturbative aspects of the renormalization group flow, following two different approximation schemes: The EVE method and the standard vertex expansion.
\item For the EVE methods, we showed that the recursive structure of the melonic diagrams provides non trivial relation between effective vertex functions, providing a way to close the hierarchical renormalization group equation expanded around local interactions.
\item The non-trivial Ward identities, arising from the Unitary symmetry breaking by the kinetic action provide relations between local effective couplings and the derivative of the effective vertices with respect to the external momenta. This was especially useful for the computation of the anomalous dimension. Moreover, Ward identities provide a relation between $\beta$-functions. Projecting the flow along with this constraint, we obtained a set of equations describing the constrained melonic flow.
\item In a complementary way, we investigate the renormalization group flow from a more standard strategy, using vertex expansion, exploiting the flexibility of the formalism to explore regions of the phase space where the EVE break-down (for instance for the disconnected diagrams, or more generally as soon as we left the melonic sector).
\end{itemize}
Then, moving on to the numerical analysis, we only consider the unconstrained flow, discarding the constraints \eqref{contrainte1} and \eqref{contrainte2} coming from Ward identities, and investigate separately the equations obtained from the EVE and finite-dimensional truncations. Due to the complicated structure of the flow equations, the numerical search for the fixed becomes challenging. In fact, here we provide  the  search for the fixed point in a  systematic inspection of some regions in the vicinity of the Gaussian fixed point. In this way, the first relevant fixed points, closer to the Gaussian one, are given in the table \ref{table1}; their respective critical exponents, computed as the opposite of the eigenvalues of the stability matrix are given in table \ref{table2}. Note that all these fixed points are obtained from the EVE equations, no fixed point being obtained from vertex expansion, keeping disconnected interactions or not within our limited preliminary search - fixed points with the mixing interactions turned off are found, of course. 

\begin{center}
\begin{tabular}{|l|l|l|l|l|l|l|l|l|}
\hline Fixed points /  & $\bar{m}^*_{VV}$&$\bar{m}^*_{WW}$&$\bar{g}^*_1$ &  $\bar{g}^*_2$ &  $\bar{c}^*_1$&$\bar{c}^*_2$&$\eta_V$&$\eta_W$\\
\hline $FP_1$ &-0.36& 0& -0.051&0&0&0&-2.16&0\\
\hline $FP_2$ &0.91& -0.41& -0.15&0.01&-0.011&-0.13&-0.27&-0.19\\
\hline
\end{tabular}
\captionof{figure}{The list of the first non-Gaussian fixed points around the Gaussian one keeping only connected interactions with the EVE method. }\label{table1}
\end{center}

\begin{center}
\begin{tabular}{|l|l|l|l|l|l|l|l|||}
\hline Fixed points /  & $\theta_1$ & $\theta_2$ & $\theta_3$ & $\theta_4$ &  $\theta_5$ & $\theta_6$ \\
\hline $FP_1 $ &-8 &-5.1 &-2.1 &-1.3 &1 &$7.10^{-9}$ \\
\hline $FP_2 $ &3.3-1.4i &3.3+1.4i &-0.066-1.2i &-0.066+1.2 i & 0.46&-0.12 \\
\hline
\end{tabular}
\captionof{figure}{The critical exponents corresponding to fixed points keeping only connected interactions.}\label{table2}
\end{center}

Despite the strong incompleteness of this analysis, these two fixed points require some comments. First of all, due to the $V\leftrightarrow W$ symmetry of the $\beta$-functions, each of these fixed point admits a symmetric image coming from the $W$-$V$ exchange. Moreover, the fixed point FP2 appears essentially as an UV attractor, in contrast to the first fixed point FP1. It has four relevant directions in the IR, and then appears as an IR fixed point, spanning a basin of attraction of dimension four, where only the couplings of type $V$ survive. It is tempting to interpret this situation as a Ising-type phase transition, where only spins "up" survives, below the Curie point. \\

%Interestingly, this result seems to be in conflict with the analysis provided in \cite{Bonzom:2011ev}, and may be an indication of the role played by the nontrivial propagator of TGFTs. 

Another question can be raised. How can we explain that we do not find fixed points with truncated equations? How can we explain the strong disagreement that we observe here? One way to explain this feature is to argue that the coupled between $V$ and $W$ fields enhanced the dependence on the terms that vertex expansion do not take into account. This can be tracked for instance, on the expressions \eqref{cond1} and \eqref{cond2}. The expression that we found for $\mathcal{L}_{k,V}$ and $\mathcal{L}_{k,W}$ exhibit a line of singularity for $4g_1g_2-c_1^2=0$, where  $\mathcal{L}_{k,V}$ and $\mathcal{L}_{k,W}$ are not defined. In contrast, the singularity behaves like $1/2g$ for a single coupling, and is compensated by the factors $g$ which provides a suitable limit for $g\to 0$. The occurrence of such a singularity line is expected to be the main source of enhancement of EVE contributions, and seems to have a strong impact on the fixed point structure. If this interpretation holds, this simply means that the understanding of the fixed point structure cannot be well found from a too small finite-dimensional truncations, and requires a deep investigation of the theory space, including more and more operators, which is especially what EVE intends to do. However, this conclusion may be confirmed/improved by taking into account disconnected interaction at the vertex expansion level, with a deeper numerical analysis. In a first time, we can interpret our result as a pathology of the vertex expansion, and as a hint about the role of higher interactions to correctly understand the physical theory space and the fixed point structure. This was already pointed out in \cite{BenGeloun:2018ekd} for disconnected interactions, and confirmed the necessity of finding methods allowing to explore more sophisticated theory spaces. There is no version of the EVE for disconnected interactions, but this result should be viewed as a strong motivation for future investigations. Note that, the emergence of the disconnected interaction seems to be unavoidable from a RG point of view, even starting with connected interactions, and it is a direct consequence of the fact that many connected interactions exist. This remark shows that a simple way to circumvent the problem of disconnected interaction may be to consider a single melonic interaction, rather than a model symmetric under color permutation. One can for instance replace the equation \eqref{final0} by:
\begin{align}
\nonumber S[\Phi,\bar{\Phi}]=\sum_{\vec{p}\in\mathbb{Z}^3}  (\bar{V}_{\vec{p}},\bar{W}_{\vec{p}})&
\begin{pmatrix} 
\vert \vec{p}\,\vert+m_1 & 0 \\
0 &\vert \vec{p}\,\vert+m_2 
\end{pmatrix}\begin{pmatrix} 
{V}_{\vec{p} } \\
{W}_{\vec{p}}
\end{pmatrix} + g_{2}\vcenter{\hbox{\includegraphics[scale=1.1]{vert2.pdf} }}\\
&+ g_{1}\vcenter{\hbox{\includegraphics[scale=1.1]{vert1.pdf} }}+\Bigg\{ c_{1}\vcenter{\hbox{\includegraphics[scale=1.1]{vert6.pdf} }}+c_{2}\vcenter{\hbox{\includegraphics[scale=1.1]{vert9.pdf} }}\Bigg\}\,,\label{final0bis}
\end{align}
which discard all the disagreements expected to come from disconnected interactions but keep the disagreements between EVE and vertex truncations. All these numerical investigations will be addressed on the companion paper. 

\section*{Acknowledgments}

ADP acknowledges ACRI under the Young Investigator Training Program, INFN and FAPERJ for partial financial support and SISSA for hospitality during the ACRI training program.

\pagebreak
\begin{center}
\textbf{{\Large Appendix}}
\end{center}

\appendix
\section{Leading order, just-renormalizability and canonical  dimension} \label{appA}

In this section we briefly investigate the leading order graphs of the theory and provide an argument for just-renormalizability. For more extensive developments, the reader may consult references \cite{Carrozza:2013wda}-\cite{BenGeloun:2011rc}. We introduce an alternative representation of the theory, called \textit{intermediate field representation}, in which the properties of the leading sector become very nice. Usually, intermediate field representation is introduced as a ‘‘trick" coming from the properties of the Gaussian integration, and allowing to break a quartic interaction for a single field into a three body interaction for two fields. To simplify the presentation, we introduce the intermediate field decomposition as a one-to-one correspondence between Feynman graphs \cite{Lahoche:2018oeo}. We will prove the following result:
\begin{theorem}
The 1PI leading order vacuum graphs are trees in the intermediate field representation. In the original representation, these trees are called melonic diagrams. \label{theoremtreees}
\end{theorem}
The rule are the following. First of all, we picture as dotted edges the Wick contractions in the original representation. The graphs one the left hand side of the Figure \ref{FigApp1} provides some examples of vacuum Feynman graphs. To introduce the intermediate field decomposition, remark that all our vertices can be labeled with a pair of numbers $(i,n)$ where $i$ run from $1$ to $3$ whereas $n$ run from $1$ to $4$. The first index $i$ is a color label, corresponding for a given melonic vertex to the color of the single edges; and $n$ label the four configurations. Explicitly:
\begin{equation}
(i,1)\equiv \vcenter{\hbox{\includegraphics[scale=1]{vert1.pdf} }} \,,\quad (i,2)\equiv \vcenter{\hbox{\includegraphics[scale=1]{vert2.pdf} }}\,,\quad (i,3)\equiv \vcenter{\hbox{\includegraphics[scale=1]{vert6.pdf} }} \,,\quad (i,4)\equiv \vcenter{\hbox{\includegraphics[scale=1]{vert9.pdf} }} \label{defvertex}
\end{equation}
The correspondence from original to intermediate field representation is then as follow. For a given vacuum Feynman graph, we associate an edge of color $i$, labeled with an index $n$ to each vertex of type $(i,n)$. In the same way, to each loop made of doted edges, we associate an effective black vertex; the number of \textit{corners} corresponding to the length of the original loop. Figure \ref{FigApp1} provides some examples. To distinguish this representation with the standard Feynman one, we call \textit{colored edges} the lines of an intermediate field graph, and \textit{loop-vertices} its vertices. We can now moving on to the proof of theorem \ref{theoremtreees}. \\

\noindent
Let us consider the following lemma (power-counting):
\begin{lemma}
For any Feynman graph $\mathcal{G}$ in the original representation, with $L$ internal (dotted) edges, $F$ closed faces and $V$ vertices, the divergent degree $\omega(\mathcal{G})$ is given by:
\begin{equation}
\omega(\mathcal{G})=-L(\mathcal{G})+F(\mathcal{G})\,.
\end{equation}
\end{lemma}
Standard proofs using multi-scale decomposition may be found in \cite{BenGeloun:2012pu}, the presence of two tensor fields introduce only minor modifications of the standard proofs. \\

\noindent
\textit{Proof of theorem \ref{theoremtreees} (sketched).}  We only sketch the proof, focusing on aspects depending on the specificity of the model. We proceed recursively on the number of intermediate field edges. Let $\ell$ the number of colored edges. A tree with $\ell$ edges has $c=2\ell$ corners, and $F=(d-1)(\ell+1) +1 $ faces\footnote{We introduce $d$, the rank of the tensor field to clarify the origin of the different factors.}, since each colored edge glues two faces. As a result, for trees: $\omega_T=-2\ell+(d-1)(\ell+1)+1=(d-3)\ell+d$, the index $T$ being for "trees".  \\

\noindent
$\bullet$ For $\ell=1$, there are essentially two configurations, independently of the choice of the integer $n$:
\begin{equation}
\vcenter{\hbox{\includegraphics[scale=1]{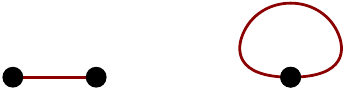} }}
\end{equation}
and so one for each choices of colors for the intermediate field edges. From direct computation, the divergent degrees are respectively, from left to right: $\omega_L=-2+5=3$ and $\omega_R=\omega_L-1=2$; then the leading order graph if this one on the left, and it is a tree. \\

\noindent
$\bullet$ Now, we assume that we have a tree for arbitrary $\ell$, and we investigate the way to build a graph with $\ell=1$ colored edges. From the typical tree
\begin{equation}
\vcenter{\hbox{\includegraphics[scale=1]{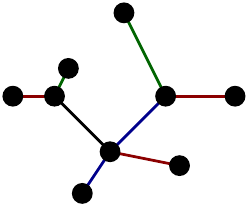} }}\,, \label{treeexample1}
\end{equation}
we have four possible moves:
\begin{equation}
\vcenter{\hbox{\includegraphics[scale=0.8]{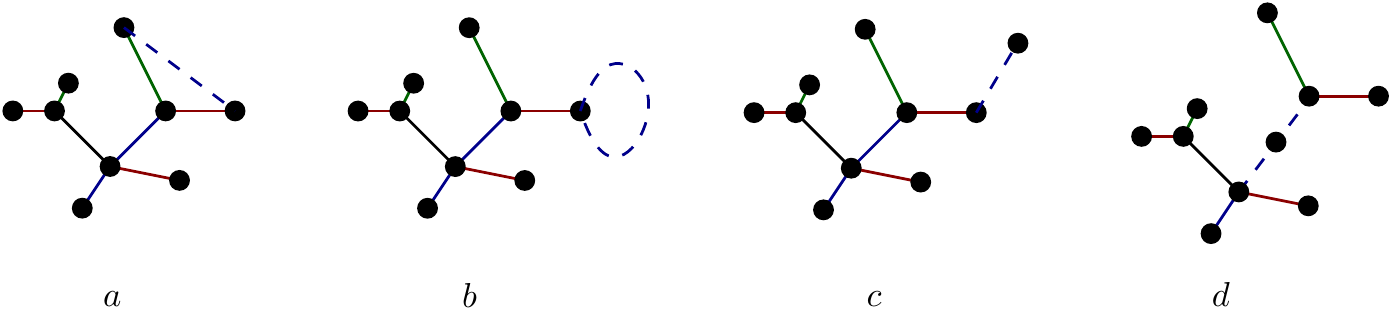} }} \label{recurrence}
\end{equation}

\noindent
where the moves are represented with dotted edges. The two moves one the right, (c) and (d) preserve the tree structure, then, the power counting is the expected for a tree, and setting $d=3$: $\omega_T=3$, such that the variation of the power counting vanish: $\delta \omega=0$. The two moves (a) and (b) on left however introduces a loop. For the first one (a), we introduce at least a single face and we create two corners. The variation for power counting is then optimally : $\delta\omega=-2+1=-1$. Obviously this bounds hold for the second move (b) on the left which introduce a tadpole edge. Then for this two moves, the divergent degree decreases from its value for trees, and the theorem is proved.
\begin{flushright}
$\square$
\end{flushright}

\begin{center}
\includegraphics[scale=0.8]{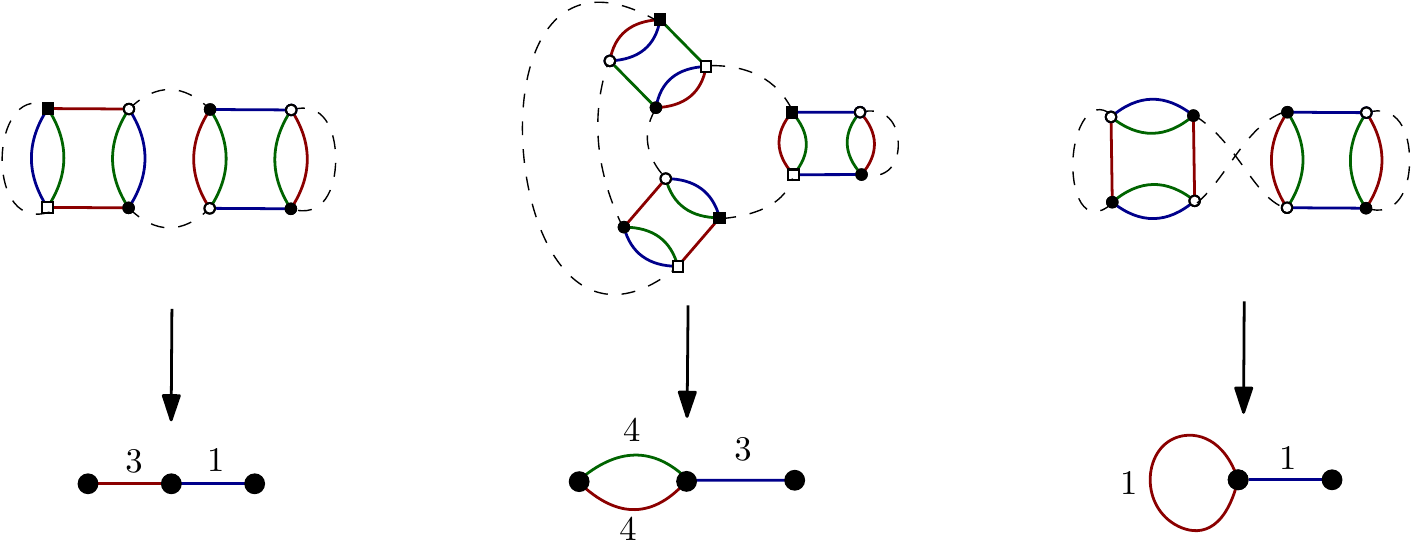} 
\captionof{figure}{Example of the correspondence between original representation (on left) and intermediate field representation (on right).}\label{FigApp1}
\end{center}

This definition extends for non-vacuum diagrams:

\begin{definition}
All the leading order diagrams (vacuum or not) are said to be \textit{melonic}. 
\end{definition}
 \noindent
 \begin{remark}
Note that in this proof we assumed that all the interactions have the same "scaling", that we justify at the end of this section. 
 \end{remark}
 
\noindent
The leading order non-vacuum graphs can be obtained following a recursive procedure as well, opening some internal loops from a vacuum graphs. When a single field is involved, this procedure allows to obtain the structure of the leading order graph very quickly. Indeed, consider a "pure" vacuum, made only of vertices of type $(i,1)$ or $(i,2)$. We obtain a $2$-point graph by cutting one of the dotted edge in a corner. Because of the structure of melonic diagrams, it is clear that if we cut a dotted edge which is not a tadpole line (i.e. a line in a loop of length upper than one), we get a 1PR diagrams. Therefore, we have to cut only a tadpole edge to get a 1PI diagram. Cutting the first one, we delete $d$ faces, and obtain a 1PI $2$-points melonic diagram. To obtain a $4$-points melonic diagram, we have to cut another tadpole edge on the same diagram. However, it is clear that such a cutting could be deleted $d$ internal faces, \textit{except} if the chosen vertex share the opened heart external face. Indeed, in this case, the cutting cost $d-1$ faces (which become boundary external lines) for the same "cost" in dotted edge, and the power counting is clearly optimal.\\

\noindent
It is easy to cheek that interactions of type $(i,3)$ do not modify the argument. However, a difficulty occurs for interactions of type $(i,4)$. Indeed, we have no Wick contractions between $V$ and $W$ fields, and there are no melonic contractions build of a single vertex of the type $(i,4)$. For the same reason, there are no leaf hooked to an edge of color $(i,4)$ in the intermediate field representation. As a consequence, non-vacuum leading order graphs having boundary vertices of type $(i,4)$ cannot be obtained from a leading order graph vacuum graph following the procedure described above. To solve this difficulty, we propose the following "trick". We temporarily modify the bare propagator in a way to allow all Wick contractions:
\begin{equation}
C_\mu^{-1}(\vec{p}\,):= (\vert \vec{p}\,\vert+m) \begin{pmatrix}
1&\mu\\
\mu& 1
\end{pmatrix}\,.
\end{equation}
In this way, the procedure described before holds, and at the end of the cutting procedure:
\begin{itemize}
\item We delete all the graphs involving a contraction between $V$ and $W$ fields.
\item We set the coupling $\mu$ to zero.
\end{itemize}

\noindent
As a result, we proved the following statement:
\noindent
\begin{proposition} \label{cormelons}
A 1PI melonic diagram with $2N$ external lines has $N(d-1)$ external faces of length $1$ shared by external vertices and $N$ heart external faces of the same color running through the internal vertices and/or internal edges (i.e. through the heart graph). 
\end{proposition}
To complete these definitions, and of interest for our incoming results, we have the following proposition:
\begin{corollary}
The perturbative expansion of the model is power-counting just renormalizable.
\end{corollary}
Indeed, we showed that the power counting is bounded by the melonic diagrams. Moreover, it is easy to see, from the recursive definition of melons that $F=(d-1)(L-V+1)$. Indeed, contracting a tree line does not change the divergent degree and the number of faces. Then, contracting all the line over a spanning tree, we get $L-V+1$ lines contracted over a single vertex. Now, we delete the lines optimally. We have some external lines, but we know from the definition of melons that no more one 
heart external face pass through one of them. Then, an optimal cutting is for a line which is on the boundary of one external face. As a result, the cutting remove $4$ internal lines. Processing the operation until the last line have been contracted, we find the desired counting for faces.. Therefore, taking into account the topological relation $2L=4V-N_{ext}$ coming because our model is quartic, and where $N_{ext}$ denotes the number of external lines; we get:
\begin{align}\label{melocountinplus}
\omega=-L+F= -2V+N_{ext}/2+2(2V-N_{ext}/2-V+1)=2-\frac{1}{2}N_{ext}\,,
\end{align}
which does not depend on the number of vertices. Moreover, it is positive for $N_{ext}=2,4$, and then negative for $N_{ext}>4$. \\

\noindent
To conclude this appendix, let us briefly discuss the problem of the dimension. Standard field theory defined over space-time are delivered with a canonical notion of scale, provided by the background space-time itself. However, as background independent field theories, there are no canonical notion of scale for GFTs. Except if we impose a physical contact with ordinary space-time, all the quantity involved in the classical actions for GFTs are dimensionless. The renormalization group however provides a specific notion of scale from the behavior of the running coupling constant. For instance, it is tempting to attribute the dimension zero for couplings which behaves like $\ln(\Lambda)$ -- for some UV cut-off $\Lambda$. Generally, recognizing that radiative corrections behaves like a power $\Lambda^n$ of the cut-off, we define the \textit{canonical dimension} as the optimal $n$, that is, following the behavior of the leading order quantum corrections. Because the power counting \eqref{melocountinplus} does not depend on the number of vertex, it follows that for leading order graphs, adding an elementary melon has no cost, and then their canonical dimension have to be zero:
\begin{equation}
[g_i]=[c_i]=0\,,\quad \mbox{for}\, i=1,2\,.
\end{equation}
In the same way because $2$-point melonic graphs diverge like $\Lambda$ ($\omega=1$), we have to fix at $1$ the dimension of the parameter $m$:
\begin{equation}
[m]=1\,.
\end{equation}

\begin{remark}
We proved that power counting is bounded by the melonic one. However, we do not proved that all divergences are localized into the melonic sector, and a direct inspection show that it is not the case. For instance, the following non-melonic diagram:
\begin{equation}
\includegraphics[scale=1]{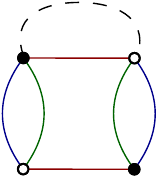} 
\end{equation}
diverge logarithmically and therefore require counter-terms to make the theory divergence free. 
\end{remark}

\begin{remark} \label{remark3}
As a remark, note that if we discard the couplings $(i,4)$ (i.e. if we set $c_2=0$), the effective action in the intermediate field representation takes an interesting form. Indeed, in this case the original action \eqref{final} can be broken with the help of $\droit$ doublets of matrix-like hermitian fields $\sigma_i:=(\sigma_{Vi},\sigma_{Wi})$ as:
\begin{equation}
 S[\Phi,\bar{\Phi},\sigma]=\bar{\Phi}^T\mathcal{K}\Phi + \frac{1}{2}\sum_{i} \Tr \sigma_i^T  C^{-1}_{VW} \sigma_i- \sqrt{-2} \sum_{i=1}^\droit\Tr \Psi_i^T \sigma_i \,,
\end{equation}
with:
\begin{equation}
\Psi_i=(\sqrt{g_{1}}\psi_{Vi},\sqrt{g_{2}}\psi_{Wi})\,,\qquad
\begin{matrix}
(\psi_{V1})_{pq}:= \sum_{p_1,p_2} \bar{V}_{p,p_1,p_2}\,V_{q,p_1,p_2}\\
 (\psi_{W1})_{pq}:= \sum_{p_1,p_2} \bar{W}_{p,p_1,p_2}\,W_{q,p_1,p_2}
\end{matrix}
\end{equation}
and:
\begin{equation}
C_{VW}=
\begin{pmatrix}
1& \lambda\\
\lambda & 1
\end{pmatrix}\,,\quad \text{with:}\quad  \lambda = \frac{1}{2} \frac{c_{1}}{\sqrt{g_{1}g_{2}}} \,.
\end{equation}
\end{remark}

\section{Proofs of propositions \ref{closed} and \ref{coro1}} \label{appB}

In this section we prove the proposition \ref{closed}, providing the closure relations defining formally the self energy, and the corollary \ref{coro1}. \\

\noindent
\textit{Proof of proposition \ref{closed}.} To begin, let us focus on the first part of the proposition, i.e. the components $\Sigma_{VW}$ and $\Sigma_{WV}$ of the self energy have to vanish, so that the coupling constant $\mu$ has no radiative corrections in the deep UV. \\

\noindent
We prove this statement by recurrence on the number of vertices. To simplify the proof, we use of the intermediate field representation recalled in appendix \ref{appA}. In this representation, vertices become edges, and we build our recurrence on the number of edges. \\

\noindent
Let $\mathcal{G}$ a melonic $2$-point diagram contributing to the perturbative expansion of $\Sigma_{VW}$, and $\mathcal{T}_{\mathcal{G}}$ the corresponding tree in the intermediate field representation. Let $n$ be the number of colored edges and $n_4$ the number of vertices of type $4$. For $n=1$, $n_4=1$, and it is easy to cheek that there are no leading order diagram (there is no melonic tadpole with vertex of type $(i,4)$). Now, we assume $n>1$. Because we cannot build leafs with edges of type $(i,4)$, we have necessarily $n>n_4$. Assuming $n_4=1$, we can investigate all the way to build a tree with $n_4=2$. All the different moves are given on equation \eqref{recurrence}, figures (c) and (d). The configuration (c) is impossible for vertex of type 4. Moreover, the configuration (d) is impossible as well. Indeed, closing the boundary loop vertex require the same number of bull and square nodes, and therefore another edge of type $(i,4)$ hooked to him, in contradiction with the recursion hypothesis $n_4=1$.  From the same argument, it is impossible to build a tree with $n_4+1$ nodes from a tree with $n_4>1$ nodes. \\

\noindent
We now move on to the second part of the proposition. The proof follows the proposition \ref{cormelons} of appendix \ref{appA}. Let us consider the component $\Sigma_{VV}$ having boundary made of bull nodes. From proposition \ref{cormelons}, we know that all the graphs contributing to the perturbative expansion of $\Sigma_{VV}$ can be obtained from a vacuum graph by opening a leaf. As a result, the two external (dotted) edges have to be hooked to the same vertex. The color of the edges hooked to this leaf split $\Sigma_{VV}$ into $d$ components $\sigma_{VV}^{(i)}:\mathbb{Z}\to \mathbb{R}$, labeled with a color index, corresponding to the color of the long opening face, running through the interior of the diagram. Let us consider the component $i$. Deleting the edge of color $i$ from its interior loop vertex, we get a $2$-point graph, and we have to distinguish two cases:
\begin{enumerate}
\item  If the deleted colored edge was of type $(i,1)$, this $2$-point graph has to be an element of the perturbative expansion of $G_{VV}$.

\item If the deleted colored edge was of type $(i,3)$, this $2$-point graph has to be an element of the perturbative expansion of $G_{WW}$.
\end{enumerate}
The same decomposition have to be true for all diagrams contributing to $\sigma_{VV}^{(i)}$, therefore, summing over all the diagrams, we reconstruct the two points functions $G_{VV}$ and $G_{WW}$ :
\begin{equation}
\vcenter{\hbox{\includegraphics[scale=1.2]{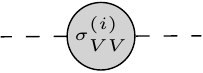} }}=\vcenter{\hbox{\includegraphics[scale=0.9]{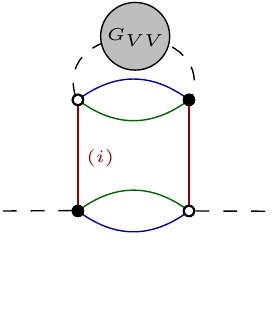} }}+\vcenter{\hbox{\includegraphics[scale=0.9]{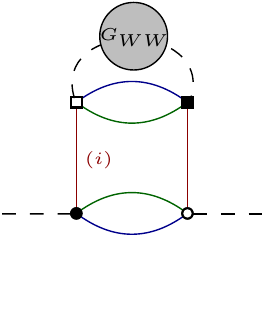} }}\,.\label{closedgraphical}
\end{equation}
This graphical equation does not take into account the symmetry factors. There are two ways to hook the effective function $G_{VV}$ to the boundary vertex $(i,1)$, whereas only a single way to hook the effective function $G_{WW}$ to the boundary $(i,3)$. Taking into account the factor $2$, and translating into formula the equation \eqref{closedgraphical}, we proved the proposition for $\sigma_{VV}^{(i)}$. The argument hold in the same way for $\sigma_{WW}^{(i)}$, exchanging bull and square nodes, and the proposition is proved. 

\begin{flushright}
$\square$
\end{flushright}

\noindent
\textit{Proof of proposition \ref{coro1}.} Let us denote as $\bar{V}_4$ the number of boundary vertices of type 4. Note that because of proposition \ref{cormelons}, all these boundaries have to be identical, and then, all of the type $(i,4)$ for the same $i$. For $2N\neq 4$, we know that the statement is true for $\bar{V}_4=\bar{V}=1$, $\bar{V}$ being the total number of external vertices. But following the previous proof, it is easy to see that the statement have to be true for all $\bar{V}$. Indeed, the arguments about the the closure of all the internal effective vertices hold.\\

\noindent
Now, let us consider the case $\bar{V}_4=\bar{V}=2$. As show in the proof of the proposition \ref{closed}, the closure of the effective vertices require an even number of bull and square nodes. Let $\mathcal{G}$ a $1PI$ leading order Feynman graph having $\bar{V}_4=2$ and $\mathcal{T}_{\mathcal{G}}$ the corresponding tree in the intermediate field representation. From proposition \ref{cormelons}, there must be a path $\mathcal{P}_i$, that we call \textit{skeleton}, made of edges of color $i$ in $\mathcal{T}_{\mathcal{G}}$ between the loop vertices sharing the two vertices of type $4$. Let $v_1$ and $v_2$ these two end loop vertices and $\{\ell_n\}$ the set of edges of color $i$ building the path $\mathcal{P}_i$. Let $\ell_1=(v_1,v)$ the first intermediate field edge, hooked to $v_1$. 

\noindent
\begin{itemize}
\item If $v=v_1$, the path $\mathcal{P}_i$ has a length equal to zero, and the two boundary vertices of type $4$ are hooked to the same loop vertex. Together, they provide an even number of bull and square nodes, and the closure of the effective vertex is allowed. Note that we can not hook another vertex of type $(i,4)$ to this effective vertex because of Proposition \ref{closed}. Indeed, because of the tree structure of the graph, the connected sub-tree having such a vertex has root build a $2$-point graph with boundary vertex of type $(i,4)$.

\item If $v\neq v_1$, the previous argument hold for $v_1$ if and only if the colored edge $\ell_1$ corresponds to a vertex of type $(i,4)$. A second colored edge of type $4$ have to be hooked to $v$ to ensure its closure. But from proposition \ref{closed} once again, this hooked edge cannot be the root of a sub-connected tree, because this tree would correspond to a $2$-point function with type $4$ boundary vertex. As a result, either $v=v_2$, either the required colored edge of type $4$ have to be an element of the path $\mathcal{P}_i$.

\end{itemize}
Recursively, one has then proved that intermediate field representation of leading order graphs with $N=N_4=2$ are $2$-point trees with a mono-colored path build of vertices of type $4$ only. Figure \ref{contrex} provide two examples. 

\begin{center}
\includegraphics[scale=1]{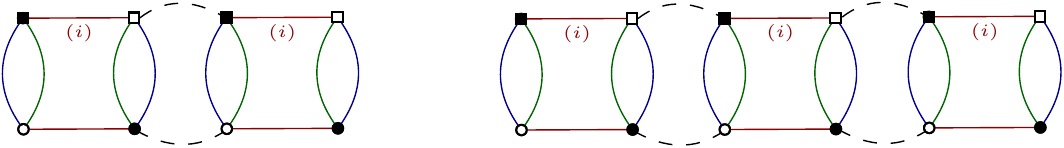} 
\captionof{figure}{Two leading order non-vanishing configurations having two vertices of type $(i,4)$ as boundaries.} \label{contrex}
\end{center}

\noindent
From such a leading order graph, we can break some tadpoles along the boundary of the long external faces of color $i$ to get $1PI$ Feynman graphs with $N_4=2$, $N>2$. \\

\noindent
Now, let us consider a graph with arbitrary $N_4=N$ boundary vertices, all of the type $(i,4)$ for the same $i$. From the previous argument, each of them can be the boundary vertex of a $2$-point sub-tree having a monocolored path building of vertices of type $(i,4)$. Because of the proposition \ref{cormelons}, there must be exist a path of color $i$, $\mathcal{P}_{v,v^\prime}^{(i)}$ between each pair $(v,v^\prime)$ of boundary loop-vertex. These paths are made of segments of color $i$, and we distinguish two family of such a segments: The type $\mathcal{P}_i$ made of a colored edge of type $(i,4)$, and the type $\mathcal{P}_{i}^{\sharp}$, made of colored edge of type $(i,n)$ for $n\neq 4$. The set of all these paths build a monocolored tree, the skeleton of the graph, that we denote as $\mathfrak{S}(\mathcal{T}_{\mathcal{G}})$. In this tree, we can investigate the possibility to close each loop vertex. The obstruction does not come from the paths of type $\mathcal{P}_{i}^{\sharp}$, therefore, we can contract all the them following the contraction rule :
\begin{itemize}
\item We delete the edge connecting the loop vertices $v_1$ and $v_2$.
\item We glue together the boundary loop-vertices.
\end{itemize}
Because we delete an even number of bull and square nodes per loop vertex, we do not change the closure constraint per vertex, and the remaining tree, $\bar{\mathfrak{S}}(\mathcal{T}_{\mathcal{G}})$,  has only paths of type $\mathcal{P}_i$. $N_4$ of them are hooked to the external loop vertex. Let us denote as $\mathcal{C}$ the total number of bull nodes in the tree $\bar{\mathfrak{S}}(\mathcal{T}_{\mathcal{G}})$. Denoting as $\mathcal{L}$ denoting the number of internal paths (i.e. not connected to external loop vertices), it is easy to cheek that $\mathcal{C}=N_4+2\mathcal{L}$, $\mathcal{L}$ because each internal path contribute twice. Therefore $\mathcal{C}$ is even and the closure is guarantee if and only if $N_4$ is even. \\

\noindent
Finally, from leading order $1PI$ graph with $N_4$ external edges, we can delete some tadpole to increase the number of boundary vertices of type $(i,n)$, for $n\neq 4$. 

\begin{flushright}
$\square$
\end{flushright}

\begin{remark}
Note that the configurations corresponding to the $2$-point trees as pictured on Figure \ref{contrex} can be easily summed using the same technique explained in [Samary-Lahoche]. The two diagrams pictured on Figure \ref{contrex} correspond to skeleton of length zero and one respectively. Note that some vertices could be added on each dotted edges, and order by order they build nothing but effective $2$-point functions of type $G_{VV}$ and $G_{WW}$. As a result, summing over the length of the skeleton, one get:
\begin{align}
\nonumber\pi_2^{(1,4)}(p_1,p_1^\prime)&=c_2+c_2^2\sum_{\vec{p}}\delta_{\vec{p}_\bot\vec{p}_\bot^\prime} G_{VV}(\vec{p}\,)G_{WW}(\vec{p}\,^\prime)+c_2^3\left(\sum_{\vec{p}}\delta_{\vec{p}_\bot\vec{p}_\bot^\prime} G_{VV}(\vec{p}\,)G_{WW}(\vec{p}\,^\prime)\right)^2+\cdots\\
&=\frac{c_2}{1-c_2\sum_{\vec{p}}\delta_{\vec{p}_\bot\vec{p}_\bot^\prime} G_{VV}(\vec{p}\,)G_{WW}(\vec{p}\,^\prime)}\,,
\end{align}
where $\vec{p}_\bot :=(p_2,\cdots,p_d)\in \mathbb{Z}^{d-1}$. 
\end{remark}

\section{Useful computations}\label{appC}
In this appendix we provide some explicit computations involved in the text. 

\subsection{Computation of $J_n$}
The integrals $J_n$ occurring in the flow equations are defined as:
\begin{equation}
J_n:= \int d^2x \vert \vec{x} \,\vert^n \theta(R-\vert x_1 \vert -\vert x_2 \vert)\,.
\end{equation}
Rescaling the $x$ variables as $y=x/R$, we get:
\begin{equation}
J_n= R^{n+2} \int d^2 y \vert \vec{y}\, \vert^n \theta(1-\vert y_1 \vert -\vert y_2 \vert)\,.
\end{equation}
Rewriting the Heaviside distribution $\theta(x)$ as
\begin{equation}
\theta(1-\vert y_1 \vert -\vert y_2 \vert)=\int_{0}^1 dz \delta(z-\vert y_1 \vert -\vert y_2 \vert)\,,
\end{equation}
we obtain, using the properties of the Dirac distribution and up to the rescaling $y\to z y$:
\begin{equation}
J_n=R^{n+2} \int_0^1 z^{n+1} dz\,\int d^2y\, \delta(1-\vert y_1 \vert -\vert y_2 \vert) =2^2 \frac{R^{n+2}}{n+2}\,\int_{(\mathbb{R}^+)^2} d^2y\, \delta(1-\vert y_1 \vert -\vert y_2 \vert)\,.
\end{equation}
The last integral may be trivially computed, 
\begin{equation}
\int_{[0,1]^2} d^2y\, \delta(1-\vert y_1 \vert -\vert y_2 \vert)=\int_0^1 dy [\theta(1-y)-\theta(-y)]=\int_0^1dy=1\,.
\end{equation}
As a result:
\begin{equation}
J_n=2^2 \frac{R^{n+2}}{n+2}\,.
\end{equation}

\subsection{Computation of $\mathcal{A}_{3,IJK}$,   $\Delta_{n,I}$  and $\Delta_{3,IJK}$}
In this section we provide the computation of the quantities $\mathcal{A}_{3,IJK}$,   $\Delta_{n,I}$  and $\Delta_{n,IJK}$ occurring in the computation of the flow equations, section \ref{solving}. \\

We recall the definitions:
\begin{equation}
\mathcal{A}_{3,IJK}:= \sum_{\vec{p}_\bot} G_I(\vec{p}_\bot)G_J(\vec{p}_\bot)G_K(\vec{p}_\bot)\,,
\end{equation}
\begin{equation}
\Delta_{n,IJK}:= \sum_{\vec{p}_\bot} \frac{\partial r_{k,II}}{\partial \vert p_1\vert}(\vec{p}_\bot) G_I(\vec{p}_\bot)G_J(\vec{p}_\bot)G_K(\vec{p}_\bot)\,,
\end{equation}
\begin{equation}
\Delta_{n,I}:= \sum_{\vec{p}_\bot}\frac{\partial r_{k,II}}{\partial \vert p_1\vert}(\vec{p}_\bot) G_I^n(\vec{p}_\bot)\,.
\end{equation}

\noindent
$\bullet$ \textbf{Computation of $\mathcal{A}_{3,IJK}$.}\\
Using the explicit expression of $\Gamma_k^{(2)}$ and $r_{k,II}$, $\mathcal{A}_{3,IJK}$ splits into two contributions, for $\vert \vec{p} \,\vert \leq k$ and $\vert \vec{p}\,\vert \geq k$:
\begin{equation}
\mathcal{A}_{3,IJK}=: \mathcal{A}_{3,IJK}^{\leq}+\mathcal{A}_{3,IJK}^{\geq}\,,
\end{equation}
with:
\begin{equation}
\mathcal{A}_{3,IJK}^{\leq}:=\left[\prod_{L=(I,J,K)}\frac{1}{Z_{LL} k+m_{LL}}\right]\,  \sum_{\vec{p}_\bot}\theta(k-\vert \vec{p}_\bot\,\vert )\label{sum1}
\end{equation}
and
\begin{equation}
\mathcal{A}_{3,IJK}^{\geq}:=\sum_{\vec{p}_\bot}\left[\prod_{L=(I,J,K)}\frac{1}{Z_{II}\vert \vec{p}_\bot\vert+m_{LL}}\right]\theta(\vert \vec{p}_\bot\,\vert -k ) \label{sum2}
\end{equation}
To do the computations, we exploit the fact that we are in the deep UV limit, and we use of the same integral approximation used for compute $J_n$. In particular:
\begin{equation}
\sum_{\vec{p}_\bot}\theta(k-\vert \vec{p}_\bot\,\vert ) \to J_0\,,
\end{equation}
and 
\begin{equation}
\mathcal{A}_{3,IJK}^{\leq}=2 {k^{-1}}\left[\prod_{L=(I,J,K)}\frac{Z_{LL}^{-1}}{1+\bar{m}_{LL}}\right]\,,
\end{equation}
where we introduced the dimensionless renormalized mass $\bar{m}_{II}:=Z^{-1}_{II}k^{-1} m_{II}$. In the same way, the integral version of \eqref{sum2} writes as:
\begin{equation}
\mathcal{A}_{3,IJK}^{\geq}=k^{-1}\int d^2x\left[\prod_{L=(I,J,K)}\frac{Z_{LL}^{-1}}{\vert \vec{x}\,\vert+\bar{m}_{LL}}\right]\theta(\vert \vec{x}\,\vert -1 ) \,.
\end{equation}
Using the same strategy as for computing $J_n$, we introduce the integral representation of $\theta$ with Dirac function:
\begin{align}
\mathcal{A}_{3,IJK}^{\geq}&=k^{-1}\int d^2x \int_{1}^{+\infty} dz\left[\prod_{L=(I,J,K)}\frac{Z_{LL}^{-1}}{z+\bar{m}_{LL}}\right]\delta(\vert \vec{x}\,\vert -z ) \\
&= k^{-1} \left[\int_{1}^{+\infty} dz\,z\prod_{L=(I,J,K)}\frac{Z_{LL}^{-1}}{z+\bar{m}_{LL}}\right]\int d^2x\delta(\vert \vec{x}\,\vert -1 )\,.
\end{align}
The last factor is equal to $4$, and to compute the bracket, we observe that at least to indices have to be equals (because we have only two indices). Then, we have to evaluate the integral:
\begin{equation}
\int_{1}^{+\infty} dz\,z \frac{1}{(z+\bar{m}_{II})^2}\frac{1}{z+\bar{m}_{JJ}}=\int_{1}^{+\infty} dz\,\left(1-\frac{m_{II}}{z+\bar{m}_{II}}\right)\frac{1}{z+\bar{m}_{II}}\frac{1}{z+\bar{m}_{JJ}}\,.
\end{equation}
The first term of the right hand side may be easily computed:
\begin{align}
\nonumber\int_{1}^{+\infty} dz\,\frac{1}{z+\bar{m}_{II}}\frac{1}{z+\bar{m}_{JJ}}&= \int_{1}^{+\infty} dz\, \frac{1}{\bar{m}_{JJ}-\bar{m}_{II}}\left(\frac{1}{z+\bar{m}_{II}}-\frac{1}{z+\bar{m}_{JJ}}\right)\\
&=\frac{1}{\bar{m}_{JJ}-\bar{m}_{II}} \ln \left(\frac{1+\bar{m}_{JJ}}{1+\bar{m}_{II}}\right)\,.
\end{align}
For the last term, observe that :
\begin{equation}
-m_{II} \int_{1}^{+\infty} dz\, \frac{1}{(z+\bar{m}_{II})^2}\frac{1}{z+\bar{m}_{JJ}}=m_{II} \int_{1}^{+\infty} dz\, \frac{\partial}{\partial m_{II}} \frac{1}{z+\bar{m}_{II}}\frac{1}{z+\bar{m}_{JJ}}\,,
\end{equation}
and because the integrated function is absolutely convergent, the integral and the derivative may be exchanged:
\begin{equation}
m_{II} \int_{1}^{+\infty} dz\, \frac{\partial}{\partial m_{II}} \frac{1}{z+\bar{m}_{II}}\frac{1}{z+\bar{m}_{JJ}}=m_{II}\frac{\partial}{\partial m_{II}} \int_{1}^{+\infty} dz\,  \frac{1}{z+\bar{m}_{II}}\frac{1}{z+\bar{m}_{JJ}}\,.
\end{equation}
Explicitly:
\begin{equation}
-m_{II} \int_{1}^{+\infty} dz\, \frac{1}{(z+\bar{m}_{II})^2}\frac{1}{z+\bar{m}_{JJ}}=\frac{\bar{m}_{II}}{\bar{m}_{JJ}-\bar{m}_{II}} \left(\frac{\ln \left(\frac{1+\bar{m}_{JJ}}{1+\bar{m}_{II}}\right)}{\bar{m}_{JJ}-\bar{m}_{II} }-\frac{1}{1+\bar{m}_{II}}\right)\,,
\end{equation}
and:
\begin{equation}
\mathcal{A}_{3,IJK}^{\leq}=4k^{-1}Z_{II}^{-2}Z_{JJ}^{-1}\frac{1}{\bar{m}_{JJ}-\bar{m}_{II}} \left(\frac{\bar{m}_{JJ}}{\bar{m}_{JJ}-\bar{m}_{II} }\ln \left(\frac{1+\bar{m}_{JJ}}{1+\bar{m}_{II}}\right)-\frac{\bar{m}_{II}}{1+\bar{m}_{II}}\right)\,.
\end{equation}
The complete dimensionless function $\bar{\mathcal{A}}_{3,IJK}$ is then given by:
\begin{align}
\bar{\mathcal{A}}_{3,IIJ}=\frac{2}{(1+\bar{m}_{II})^2}\frac{1}{1+\bar{m}_{JJ}}+\frac{4}{\bar{m}_{JJ}-\bar{m}_{II}} \left(\frac{\bar{m}_{JJ}}{\bar{m}_{JJ}-\bar{m}_{II} }\ln \left(\frac{1+\bar{m}_{JJ}}{1+\bar{m}_{II}}\right)-\frac{\bar{m}_{II}}{1+\bar{m}_{II}}\right)\,.
\end{align}
Note that it is easy to cheek the continuity in the limit $\bar{m}_{II}\to\bar{m}_{JJ}$. For this case:
\begin{equation}
\bar{\mathcal{A}}_{3,III}=\frac{2}{(1+\bar{m}_{II})^3}+4\left(-\frac{1}{1+\bar{m}_{II}}+2\frac{1}{(1+\bar{m}_{II})^2}\right)
\end{equation}

\noindent
$\bullet$ \textbf{Computation of $\Delta_{n,I}$  and $\Delta_{3,IJK}$.}\\
From the definition of $r_{k,II}$ we get:
\begin{equation}
\frac{\partial r_{k,II}}{\partial \vert p_1\vert}(\vec{p}_\bot)=-Z_{II}\theta(k-\vert \vec{p}_\bot\vert)\,,
\end{equation}
and :
\begin{equation}
\Delta_{n,I}=-\frac{1}{Z_{II}^{n-1}}\frac{1}{k^n} \left(\frac{1}{1+\bar{m}_{II}}\right)^n \sum_{\vec{p}_\bot}\theta(k-\vert \vec{p}_\bot\vert) \to-\frac{1}{Z_{II}^{n-1}}\frac{1}{k^n} \left(\frac{1}{1+\bar{m}_{II}}\right)^n J_0 \,,
\end{equation}
explicitly:
\begin{equation}
\Delta_{n,I}=-2\frac{1}{Z_{II}^{n-1}}\frac{1}{k^{n-2}} \left(\frac{1}{1+\bar{m}_{II}}\right)^n\,.
\end{equation}
In the same way:
\begin{equation}
\Delta_{3,IIJ}=-2k^{-1}Z_{II}^{-1}Z_{JJ}^{-1} \left(\frac{1}{1+\bar{m}_{II}}\right)^2 \left(\frac{1}{1+\bar{m}_{JJ}}\right)\,.
\end{equation}

%\begin{equation}
%\int_{1}^{+\infty} dz\,z \frac{1}{(z+\bar{m}_{II})^3}=\int_{1}^{+\infty} dz\,\left(1-\frac{m_{II}}{z+\bar{m}_{II}}\right)\frac{1}{(z+\bar{m}_{II})^2}
%\end{equation}

%\begin{equation}
%\frac{1}{1+\bar{m}_{II}}-2\frac{\bar{m}_{II}}{(1+\bar{m}_{II})^2}
%\end{equation}

%\begin{equation}
%-\frac{1}{1+\bar{m}_{II}}+2\frac{1}{(1+\bar{m}_{II})^2}
%\end{equation}

\end{document}